\documentclass[12pt,draftclsnofoot,onecolumn]{IEEEtran}

\usepackage{amsfonts}
\usepackage{cite}
\usepackage{graphicx}
\usepackage{amssymb}
\usepackage{subfigure}
\usepackage{array}
\usepackage{color}
\usepackage{amsmath}
\usepackage{algorithmic}
\usepackage{algorithm}
\usepackage{cases}
\usepackage{bm}
\usepackage{mathrsfs}
\usepackage{epsfig}
\usepackage{psfrag}
\usepackage{url}
\usepackage{setspace}

\usepackage{etoolbox}
\makeatletter
\patchcmd{\@makecaption}
  {\scshape}
  {}
  {}
  {}
\makeatletter
\patchcmd{\@makecaption}
  {\\}
  {.\ }
  {}
  {}
\makeatother

\newcommand{\eqla}{\stackrel{(a)}{=}}
\newcommand{\eqlb}{\stackrel{(b)}{=}}
\newcommand{\eqlc}{\stackrel{(c)}{=}}

\newtheorem{Thm}{Theorem}
\newtheorem{Lem}{Lemma}
\newtheorem{Cor}{Corollary}

\newtheorem{Prob}{Problem}

\IEEEoverridecommandlockouts

\begin{document}

\title{\textcolor{black}{ML and MAP Device Activity Detections for Grant-Free \textcolor{black}{Massive} Access in Multi-Cell Networks}}

\author{
\IEEEauthorblockN{Dongdong Jiang, {\em Student Member, IEEE} and Ying Cui, {\em Member, IEEE}}
\thanks{D. Jiang and Y. Cui are with the Department of  Electronic Engineering, Shanghai Jiao Tong University, China. This paper \textcolor{black}{was} presented in part at IEEE WCNC 2020~\cite{Jiang20WCNC} and IEEE SPAWC 2020~\cite{Jiang20SPAWC}.
}}

\maketitle

% \vspace{-20mm}
\begin{abstract}
Device activity detection is one main challenge in grant-free massive access, which is recently proposed to support \textcolor{black}{massive} machine-type communications (mMTC). Existing solutions for device activity detection fail to consider inter-cell interference generated by massive IoT devices or important prior information on device activities and \textcolor{black}{inter-cell interference}. In this paper, given different numbers of observations and network parameters, we consider both non-cooperative device activity detection and cooperative device activity detection in a \textcolor{black}{multi-cell} network, consisting of many access points (APs) and IoT devices. Under each activity detection mechanism, we consider the joint maximum likelihood (ML) estimation and joint maximum a posterior probability (MAP) estimation of both device activities and interference powers,  utilizing tools from probability, stochastic geometry\textcolor{black}{,} and optimization. Each estimation problem is a challenging non-convex problem, and a coordinate descent algorithm is proposed to obtain a stationary point. \textcolor{black}{Each proposed joint ML} estimation extends the existing one \textcolor{black}{for a single-cell network} by considering the estimation of interference powers, together with the estimation of device activities. \textcolor{black}{Each proposed joint MAP} estimation further enhances the \textcolor{black}{corresponding joint ML} estimation by exploiting prior distributions of device activities and interference powers. The proposed \textcolor{black}{joint} ML estimation and \textcolor{black}{joint} MAP estimation under cooperative detection outperform the respective ones under non-cooperative detection at the costs of increasing backhaul burden, knowledge of network parameters, and computational complexities. Numerical results show the substantial gains of the proposed designs over well-known existing designs and reveal the importance of explicit consideration of inter-cell interference, the value of prior information, and the advantage of AP cooperation in device activity detection.
\end{abstract}

\begin{IEEEkeywords}
 Massive machine-type communications (mMTC), grant-free \textcolor{black}{massive} access, device activity detection, \textcolor{black}{inter-cell interference},  maximum likelihood (ML) estimation, maximum a posterior probability (MAP) estimation.
\end{IEEEkeywords}

% \clearpage

\section{Introduction}

Driven by the proliferation of Internet of Things (IoT), massive machine-type communication (mMTC) has been identified as one of the three generic services in the fifth generation (5G) cellular technologies \cite{Popovski17Mag,Liu18Mag,Bockelmann16Mag,Chen18TCOM}.
Massive access is a critical and challenging task for supporting mMTC. Although a large number of devices are associated with a single access point (AP), only a small number of devices are active at a time\textcolor{black}{,} and a small amount of data is transmitted from each active device. The traditional grant-based random access with orthogonal sequences is no longer effective for mMTC due to the heavy access overhead.
% huge number of devices and the limited radio resources. %channel coherence time. %fact that the number of potential devices often greatly exceeds the number of time-frequency dimensions.
% Instead,
\textcolor{black}{Therefore,} grant-free massive access is recently proposed as an \textcolor{black}{essential} technique for supporting massive access. In most grant-free massive access schemes, each device is assigned a specific pilot sequence, all active devices send their pilot sequences, and %to their associated APs.
each AP detects the activities of its associated devices~\cite{Liu18Mag,Bockelmann16Mag}. %each active device directly transmits the pilot and data to the AP without waiting for the grant \cite{Liu18Mag,Bockelmann16Mag}.
% In grant-free random access,
% In mMTC, an access point (AP) connects to a large number of devices, among which only a small fraction are active over any communication resource due to the sporadic traffic~\cite{Shirvanimoghaddam17Mag}. Under such circumstances,
A \textcolor{black}{primary} challenge at each AP is to
%provide massive uplink random access for a large number of devices.
% One of the main goals of the random access procedure is for the network to
identify the set of active devices in the presence of an excessive number of potential devices, as it is not possible to assign  mutually orthogonal pilot sequences to \textcolor{black}{all} devices within
a cell.\footnote{\textcolor{black}{Device activity detection itself is a fundamental problem in grant-free massive access. For applications where active devices do not send data, only device activity detection is required. For applications where active devices have very few data to transmit, data can be embedded into pilots~\cite{Senel18TCOM}, and joint activity and data detection (which can be easily extended from device activity detection~\cite{Jiang20SPAWC,Yu19ICC}) is required. For applications where active devices have many of data to transmit, joint activity detection and channel estimation \cite{Liu18TSP} or separate activity detection and channel estimation (with conventional channel estimation methods given detected device activities~\cite{Cui20JSAC}) can be conducted.}} %in order to enable resource allocation in the subsequent transmissions.
% Many well-known protocols on massive random access are based on the ALOHA [7]-[10]. In ALOHA, each transmitter sends the data as one packet repeatedly, and the receiver decodes the packet if there is no collision. Note that most of the works abstract the physical channel model into a collision channel model without fading and noise.
% % Different from the works mentioned above, this paper studies random access with physical channel model containing both fading and noise.
% Mutually orthogonal pilot sequences adopted in ALOHA cannot efficiently support the massive connectivity in mMTC.
% % and designing an uncoordinated pilot collision resolution protocol, [13] investigates the user activity detection in multi-cell massive MIMO systems and analyzes the collision probability.

% \textcolor{black}{[xx]-[xx] adopts the orthogonal pilot sequence for users in the grant-free random access.}

Due to inherent sparse device activities in mMTC, device activity detection can be formulated as compressed sensing (CS) problems and solved by many CS-based algorithms.
In~\cite{Xu15ICC}, the authors consider device activity detection and channel estimation %in uplink CRAN systems
and propose a modified Bayesian CS algorithm, which exploits the active device sparsity and chunk sparsity feature of the channel matrix.
In~\cite{Bockelmann13}, the authors consider joint activity and data detection %in a sensor network with star topology,
and apply the greedy group orthogonal matching pursuit (GOMP) algorithm, which exploits block-sparsity information of the detected data.
% \textcolor{black}{In \cite{Choi19TCOM}, the authors apply power-domain NOMA to CS-based random access, %determine the power levels for successful SIC with a high probability,
% and employ maximum a posteriori probability (MAP) detection for device activity detection.}
In \cite{Zhang18TVT}, the authors propose a message passing-based block sparse Bayesian learning (MP-BSBL) algorithm for device activity detection and channel estimation, which has much lower computational complexity compared to the  block orthogonal matching pursuit (BOMP) algorithm.
% \textcolor{black}{Joint estimations of device activity, channel states, and the device data are performed via BiG-AMP and Turbo-BiG-AMP Algorithms respectively in \cite{Ding19TWC}, which exploit the sparsity embedded in the device signals and distributions of signals and channel states.}
In \cite{Liu18TSP} and \cite{Chen18TSP}, the authors consider joint device activity detection and channel estimation %in a simple single-cell network,
and propose approximate message passing (AMP) algorithms, which exploit the statistical channel information.
In \cite{Shao19IoTJ} and \cite{Senel18TCOM}, the authors %consider a three-phase transmission protocol consisting of device detection and channel estimation, uplink transmission and downlink transmission,
employ efficient AMP-based algorithms for device activity detection and channel estimation and analyze %and optimize the uplink and downlink
achievable rates. %which is calculated based on the estimated channel,
% and maximize the overall performance by optimizing the length of each phase.
% \textcolor{black}{Reference \cite{Senel18TCOM} employ AMP-based algorithms for device activity detection and data decoding.}
Reference \cite{Chen19TWC} adopts the AMP-based algorithm for non-cooperative and cooperative device activity detections in a multi-cell network.

Recently, maximum likelihood (ML) estimation-based device activity detection designs are proposed and analyzed in \cite{Caire18ISIT,Alexander19,Yu19ICC,Chen19Asilomar}. Specifically, in \cite{Caire18ISIT}, the authors %study the activity detection in a massive MIMO setup where the AP has multiple antennas, and
formulate the device activity detection as an ML estimation problem %in which %and propose a coordinate descent method to solve the problem.
% the received signals at multiple antennas affect the detection results via their empirical covariance matrix,
and propose a coordinate descent algorithm to solve the non-convex estimation problem.
Reference \cite{Alexander19} employs the ML-based approach for the data detection in an unsourced massive random access, where each device transmits a codeword from the same codebook\textcolor{black}{,} and the AP aims to detect the transmitted codewords rather than the device activities.
In \cite{Yu19ICC}, the authors
%propose a
extend the ML-based approach to joint device activity and data detection and analyze the \textcolor{black}{estimation error distribution}. %in which each device is assigned with multiple pilot sequences and the transmitted data is embedded in the selection of the transmitted pilot sequence.
% The coordinate descent method in \cite{Caire18ISIT} is adopted to solve the joint activity and data detection problem.
% The results in \cite{Yu19ICC} show that compared with AMP-based methods, the covariance-based approach can exploit the multiple antennas at the AP more effectively, especially when the length of pilot sequence is short.
In \cite{Chen19Asilomar}, the authors adopt the ML-based approach %the ML estimation in \cite{Caire18ISIT}
for device activity detection, %propose a necessary condition on the Fisher information matrix such that the estimation error  tends to zero in the massive MIMO regime,
and analyze the \textcolor{black}{estimation error distribution}.
It is shown that the ML-based approach significantly outperforms the AMP-based algorithms in activity detection accuracy, especially when the length of pilot sequences is short and the number of antenna\textcolor{black}{s} at each AP is moderate or large \cite{Caire18ISIT,Yu19ICC}, at the cost of computational complexity increase.
%\textcolor{black}{Note that \cite{Caire18ISIT,Alexander19,Yu19ICC,Chen19Asilomar} do not consider interference generated by other devices, and hence may not provide desirable detection performance in practical mMTC where interference from massive IoT devices cannot be ignored.} In addition,

On the other hand, deep learning-based approaches are proposed for device activity detection problems \cite{Liu19GLOBECOM,Zhang19TVT,Cui20JSAC,Ye19TII}.
Specifically,  in \cite{Liu19GLOBECOM}, the authors
% propose a deep learning based false alarm likelihood estimator
employ a deep neural network (DNN) \textcolor{black}{to identify a device that} is most likely to be falsely alarmed under the AMP-based algorithms to improve
% as a false alarm likelihood estimator to assist conventional AMP-based algorithms for
activity detection and channel estimation \textcolor{black}{performance}.
% by learning the features of the false alarm events in the AMP-based detection.
In \cite{Zhang19TVT}, the authors propose a DNN\textcolor{black}{-}aided MP-BSBL algorithm for device activity detection and channel estimation, which transfers the iterative message passing process of MP-BSBL in \cite{Zhang18TVT} from a factor graph to a DNN, \textcolor{black}{mainly to  alleviate the
convergence problem of the MP-BSBL algorithm}. %whose weights are imposed on the messages and trained to minimize the estimation error.
\textcolor{black}{In\cite{Cui20JSAC}}, the authors use auto-encoder in deep learning %consider device activity detection and propose a data-driven approach
to jointly design pilot sequences and activity detection (or channel estimation) methods, which can exploit properties of sparsity patterns \textcolor{black}{to a certain extent. \textcolor{black}{However, the} proposed model-driven approaches in \cite{Cui20JSAC} rely on methods that cannot effectively utilize general correlation in device activities}.
In \cite{Ye19TII}, the authors establish a DNN model for grant-free nonorthogonal multiple access (NOMA) based on \textcolor{black}{a} deep variational auto-encoder whose decoder jointly \textcolor{black}{detects} device activities and transmitted symbols.
% Li18SPL,Zhang20icassp,Li20WCNC

% Note that the CS-based algorithms in \cite{Xu15ICC,Bockelmann13} have high computational complexity and cannot be applied in practical mMTC with a large number of devices. The MP-BSBL algorithm in~\cite{Zhang18TVT} has \textcolor{black}{no convergence guarantee} due to the densely-connected factor graph.
% In addition,
% % Note that
% the AMP-based methods \textcolor{black}{in \cite{Liu18TSP,Chen18TSP,Shao19IoTJ,Senel18TCOM}} cannot effectively exploit the feature of common support of the sparse signals at multiple antennas.

% Notice that the ML-based algorithms in \cite{Caire18ISIT,Alexander19,Yu19ICC,Chen19Asilomar} do not consider interference or the prior knowledge on device activities. How to take into account the impacts of interference and prior knowledge in \textcolor{black}{an optimal way to maximally} improve device activity detection remains an open problem.

Note that \cite{Xu15ICC,Bockelmann13,Zhang18TVT,Liu18TSP,Chen18TSP,Shao19IoTJ,Senel18TCOM,Liu19GLOBECOM,Zhang19TVT,Ye19TII,Caire18ISIT,Alexander19,Yu19ICC,Chen19Asilomar,Cui20JSAC} consider device activity detection in single-cell networks without \textcolor{black}{inter-cell} interference; %generated by inter-cell devices;
\cite{Chen19TWC} considers \textcolor{black}{inter-cell interference} in non-cooperative device activity detection \textcolor{black}{for multi-cell networks}, but does not take \textcolor{black}{inter-cell interference} into account in cooperative device activity detection \textcolor{black}{for multi-cell networks}. %the cooperative device activity detection in \cite{Chen19TWC} does not consider interference and the non-cooperative device activity detection in \cite{Chen19TWC} does not properly treat the inter-cell interference.
Hence, the resulting algorithms may not provide \textcolor{black}{a} desirable detection performance in practical mMTC with nonnegligible interference from massive IoT devices in other cells.
In addition, notice that the ML-based algorithms in \cite{Caire18ISIT,Alexander19,Yu19ICC,Chen19Asilomar} do not consider prior knowledge on sparsity patterns of device activities; the CS-based algorithms in \cite{Xu15ICC,Bockelmann13,Zhang18TVT,Liu18TSP,Chen18TSP,Shao19IoTJ,Senel18TCOM,Chen19TWC} \textcolor{black}{and the deep learning-based approaches in \cite{Liu19GLOBECOM,Zhang19TVT,Cui20JSAC,Ye19TII}} \textcolor{black}{only exploit the active probability for independently and identically distributed (i.i.d.) device activities, \textcolor{black}{specific} simple sparsity patterns, \textcolor{black}{(}such as group sparsity\textcolor{black}{)}, or limited correlation of device activities}. Hence, these algorithms may not achieve promising detection performance when device activities possess arbitrary sparsity patterns. %Although the deep learning-based algorithms in \cite{Cui20JSAC} are effective in exploiting general sparsity patterns,
\textcolor{black}{Last, notice that the algorithms in~\cite{Liu19GLOBECOM,Zhang19TVT,Cui20JSAC,Ye19TII}}  cannot efficiently adapt to different network setups or provide (theoretical) performance guarantees.

In summary, how to %optimally
systematically and rigorously %\footnote{Note that our primary goal is to improve detection accuracy rather than reducing detection complexity. The obtained optimization-based designs will provide important benchmarks for practical low complexity designs.}
take into account \textcolor{black}{inter-cell interference} and prior knowledge on sparsity patterns to maximally improve device activity detection \textcolor{black}{in multi-cell networks} remains an open problem.
%Recently, \cite{Chen19TWC} considers device activity detection in a multi-cell network and adopts the AMP-based algorithm for non-cooperative and cooperative device activity detections. %Specifically, for non-cooperative design, each AP detects activities of its associated users and treats inter-cell interference as noise; for cooperative design, each AP detects all devices in the network and forward the results to the central unit which makes the final decision on device activity.
% Note that besides the limitation of the AMP-based algorithm, \cite{Chen19TWC} considers device activity detection in a finite network and does not properly treat the inter-cell interference.% the non-cooperative design in \cite{Chen19TWC} does not properly treat the inter-cell interference and the cooperative design needs to detect activities of all devices in a finite network.
In this paper, we would like to shed some light on this problem.
% aim to tackle the above issues.
In particular, given different numbers of observations and network parameters, we consider non-cooperative device activity detection and cooperative device activity detection, both in the presence of \textcolor{black}{inter-cell interference}, in a \textcolor{black}{multi-cell} network consisting of many APs and IoT devices. Under each activity detection mechanism, we investigate two scenarios, with and without prior distributions on device activities and interference powers. The main contributions of this paper are listed as follows.

\begin{itemize}

\item When prior distributions are not available, we consider the joint ML estimation of both the device activities and interference powers under each activity detection mechanism. The challenges \textcolor{black}{of} incorporating interference lie in  the modeling of inter-cell
interference in grant-free massive access  %the analysis of the p.d.f. of the interference,
and the joint estimation of the device activities and interference powers. %which involves solving more complex non-convex problems.
% In particular,
\textcolor{black}{Under each activity detection mechanism}, by carefully approximating the interference powers, we first obtain a tractable expression for the likelihood of observations in the presence of inter-cell interference. Then, we formulate the joint ML estimation problem, which is a challenging non-convex problem. By making good use of the problem structure, we propose a coordinate descent algorithm \textcolor{black}{that} %\textcolor{black}{has a closed-form optimal solution to each coordinate descent optimization under the non-cooperative device activity detection and can be shown to}
converges to a stationary point.
\textcolor{black}{Each proposed joint} ML estimation successfully extends the existing ML estimation \textcolor{black}{for a single-cell network}~\cite{Caire18ISIT} to \textcolor{black}{a multi-cell network}.%the practical scenario with \textcolor{black}{inter-cell interference}. %\textcolor{magenta}{Under the existing ML estimation, more devices are estimated to be active when interference exists, than under the proposed ML estimation}.

\item When prior distributions are known, we consider the joint MAP estimation of both the device activities and interference powers under each activity detection mechanism. Specifically, we adopt the multivariate Bernoulli (MVB) model~\cite{NIPS2011_4209} to capture a general distribution of (possibly correlated) random device activities. We also present some
typical instances for the MVB model. To our knowledge, general correlation among device activities has not yet been theoretically investigated or effectively utilized via neural networks for device activity detection in grant-free massive access \cite{Liu18TSP,Chen18TSP,Shao19IoTJ,Senel18TCOM,Chen19TWC,Caire18ISIT,Alexander19,Yu19ICC,Chen19Asilomar}. \textcolor{black}{Under each activity detection mechanism}, using tools from stochastic geometry, we \textcolor{black}{derive} a tractable expression for the distributions of the interference powers. Based on the prior distributions of the device activities and interference powers together with the \textcolor{black}{conditional distribution} of the observations \textcolor{black}{in the presence of inter-cell interference}, we formulate the joint MAP estimation problem, which is a more challenging non-convex problem. By exploiting the problem structure, we propose a coordinate descent algorithm \textcolor{black}{that} %\textcolor{black}{has a closed-form optimal solution to each coordinate descent optimization under the non-cooperative device activity detection and can be shown to}
converges to a stationary point.
% \textcolor{black}{A closed-form optimal solution is obtained for each coordinate descent optimization under the non-cooperative device activity detection}.
\textcolor{black}{Each proposed joint} MAP estimation further enhances the \textcolor{black}{corresponding joint} ML estimation by taking the prior information on the device activities and interference powers into consideration. We also show that the influence of the prior information reduces as the number of antennas increases.

\item Finally, we show the substantial gains of the proposed designs over well-known existing designs \textcolor{black}{by numerical results}. The numerical results also demonstrate the importance of explicit consideration of \textcolor{black}{inter-cell interference}, the value of prior information, and the advantage of AP cooperation in device activity detection.

\end{itemize}

The rest of this paper is organized as follows. Section II describes the system model for grant-free massive access and introduces the non-cooperative and cooperative activity detection mechanisms. Section III considers the joint ML and MAP estimations of device activities and interference powers \textcolor{black}{under} non-cooperative \textcolor{black}{detection}. Section IV considers the joint ML and MAP estimations of device activities and interference powers \textcolor{black}{under} cooperative \textcolor{black}{detection}. Numerical results are provided in Section V. Finally, Section VI concludes this paper.

\section{System Model}

\begin{figure}[t]
\begin{center}
 \includegraphics[width=5.5cm]{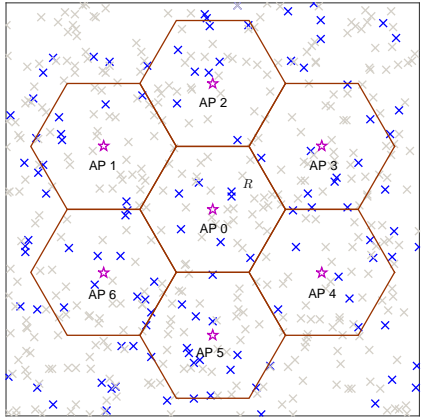}
  \end{center}
  \vspace{-2mm}
  \caption{\small{System model. The red stars represent APs. The black and gray crosses represent active and inactive devices, respectively.}}
  \vspace{-2mm}  %$N=40$, $R=88$, $\lambda=0.002$, $p_a=0.2$.
\label{fig:system_model}
\end{figure}

% \begin{figure}[t]
% \begin{center}
% \subfigure[\small{Non-cooperative device activity detection}]
% {\resizebox{6.8cm}{!}{\includegraphics{Fig_Nov23_system_model_individual.eps}}}\quad\quad\quad
% \subfigure[\small{Cooperative device activity detection}]
% {\resizebox{6.8cm}{!}{\includegraphics{Fig_Nov23_system_model_coop.eps}}}
% \end{center}
% % \vspace{-2mm}
% \caption{\small{System model. The red stars represent APs. The black and gray crosses represent active and inactive devices, respectively.}}
% % \vspace{-4mm}
% \label{fig:system_model}
% \end{figure}

% \subsection{Network Model}
As shown in Fig.~\ref{fig:system_model}, we consider %a massive  access scenario arising from mMTC in
a \textcolor{black}{multi-cell} network which consists of $M$-antenna APs and single-antenna devices. %\footnote{\textcolor{black}{As in \cite{Xu15ICC,Liu18TSP,Chen18TSP,Shao19IoTJ,Senel18TCOM,Chen19TWC}, we consider the scenario where the locations of devices do not change over time.}}
The locations of APs are distributed according to \textcolor{black}{the} hexagonal grid model with the side length of each hexagonal cell equal to $R$. %\footnote{\textcolor{black}{ For tractability and for ease of exposition, we use the hexagonal grid model. Under such model, we can obtain first-order insights and reveal impacts of inter-cell interference on device activity detection in multi-cell networks.}.}
%For the ease of illustration, w
The APs and their cells are indexed by $j$ %Let $\mathcal J$ denote
and the set of indices of APs is denoted as $\mathcal J\triangleq \{0,1,\cdots\}$. \textcolor{black}{The devices remain stationary or move slowly over time}. %and the cell of AP $j$ is also indexed with $j$.
% \textcolor{black}{We focus on the typical cell. There are $N_0$ devices in the typical cell. Let $\Phi_0$ denote the set of $N_0$ devices}.
%The area size of each cell is given by $S=\frac{3\sqrt{3}}{2}R^2$.
%Without loss of generality (w.l.o.g.), the AP centered at the origin is denoted as AP $0$, and its $6$ neighbor APs are indexed with $1,2,\cdots,6$, respectively.
% There are $N_j$ devices in cell $j$.
The devices are indexed by $i$, %Let $\mathcal I$ denote
and the set of indices of devices is denoted as $\mathcal I\triangleq \{1,2,\cdots\}$. Let $a_i\in\{0,1\}$ denote the activity state of device $i$, where $a_i=1$ indicates that device $i$ is active, and $a_i=0$ otherwise.
% Let $\Phi_j$ denote
Denote $\Phi_j$ as the set of indices of the devices in cell $j$. %and denote $\Phi\triangleq \cup_{j\in\mathbb N^+}\Phi_j$.
Let $N_j\triangleq |\Phi_j|$ denote the number of devices in cell $j$.
Denote $\mathbf a_j\triangleq (a_i)_{i\in\Phi_j}\in\{0,1\}^{N_j}$ as the activity vector of devices in cell $j$.
%, which follows the Poisson distribution with mean $S\lambda$. %, i.e., $\Pr(N_j=n)=\frac{(S\lambda)^n}{n!}\exp(-S\lambda)$.
% We assume that the devices in cell $j$ are indexed after those in cell $j-1$. Therefore, $\Phi_j\triangleq\{N_{j-1}+1,\cdots,N_j\}$, where $N_j\triangleq\sum_{k=0}^{j}n_k$ is the number of devices in the first $j$ cells.
\textcolor{black}{We consider both large-scale fading and small-scale fading.}
Let $d_{i,j}$ denote the distance between device $i$ and AP $j$.
\textcolor{black}{We consider a narrow-band system~\cite{Liu18TSP,Chen18TSP,Shao19IoTJ,Senel18TCOM}. \textcolor{black}{We} adopt the commonly used power-law path loss model for large-scale random networks~\cite{FTNhaenggi09,haenggi2012stochastic,Andrews11TCOM}, i.e.,}
% For large-scale fading, %consider power-law path loss model \textcolor{black}{and take the path loss intercept to be one} \cite{haenggi2012stochastic}, i.e.,
transmitted signals with distance $d$ are attenuated with a factor $d^{-\alpha}$, where $\alpha\geq 2$ is the path loss exponent~\cite{haenggi2012stochastic}.\footnote{\textcolor{black}{As in \cite{FTNhaenggi09,haenggi2012stochastic,Andrews11TCOM} which study large-scale random networks, we do not consider shadowing for tractability. The results in this paper can be readily extended to incorporate the exact shadowing effect of a device to be detected and approximate shadowing effect of an interfering device.}}	 %\textcolor{black}{Note that as in \cite{Shao19IoTJ,Senel18TCOM,Liu18TSP,Chen18TSP}, we consider a narrow-band system, and the variation of pathloss due to carrier frequency is negligible.}
Let $\gamma_{i,j}\triangleq d_{i,j}^{-\alpha}$ denote the path loss between device $i$ and AP $j$. %\footnote{\textcolor{black}{As in \cite{Shao19IoTJ,Liu18TSP,Chen18TSP}, we consider a narrow-band system, where the variation of path loss due to carrier frequency is negligible, and the power-law path loss model, which is tractable for large-scale random networks with infinitely many nodes.}} %Denote $\bm \gamma_j\triangleq (\gamma_{i,j})_{i\in\Phi_j}$.
%Assume that $\bm \gamma_j$ is perfectly known at AP $j$~\cite{Xu15ICC}.
For small-scale fading, \textcolor{black}{we} consider \textcolor{black}{the} block fading channel model, i.e., the channel is static in each coherence block and changes across blocks in an i.i.d. manner. %Each coherence block contains $L$ signal dimensions.
% Consider block fading channel model between devices and APs, i.e., the channel is static in each coherence block and changes across blocks. %Each coherence block contains $L$ signal dimensions.
Let $\mathbf h_{i,j}\in\mathbb C^M$ denote the \textcolor{black}{small-scale fading coefficient} between device $i$ and AP $j$. We assume Rayleigh fading \textcolor{black}{for small-scale fading, i.e.,}  %all entries of
$\mathbf h_{i,j}$, $i\in\mathcal I$ and $j\in\mathcal J$ are i.i.d. according to $\mathcal{CN}(\mathbf 0,\mathbf I_M)$. %complex Gaussian random variables, each with zero mean and unit variance, i.e., following $\mathcal{CN}(\mathbf 0,\mathbf I_M)$.

We consider a massive access scenario
arising from mMTC, where each cell contains a large number of devices, and very small of them are
% Due to the sporadic traffic of mMTC,
% only a small fraction is
active in each coherence block. That is, for all $j\in\mathcal J$, $\sum_{i\in\Phi_j}a_i\ll N_j$.
% we assume that
% each device accesses the channel with a small
% probability %$p_a\ll 1$ %in an i.i.d. manner
% in each coherence block.
% Denote $\mathbf a_j\triangleq (a_i)_{i\in\Phi_j}$.
% \textcolor{black}{Denote $\mathcal L\triangleq \{1,2,\cdots,L\}$.}
We adopt a grant-free \textcolor{black}{massive} access scheme \cite{Caire18ISIT,Alexander19,Yu19ICC,Chen19Asilomar}. %and devote a coherence block to device activity detection~\cite{Caire18ISIT}. For the purpose of activity detection,
\textcolor{black}{Specifically, each device $i$ is assigned a specific pilot sequence $\mathbf p_i=(p_{i,\ell})_{\ell\in\mathcal L}\in \mathbb C^L$ of length $L$, where $\mathcal L\triangleq \{1,2,\cdots,L\}$. Note that $L$ is much smaller than the number of devices in each cell}.
% Denote $\mathcal L\triangleq \{1,2,\cdots,L\}$.
Let $\mathbf P_j\triangleq (\mathbf p_i)_{i\in\Phi_j}\in\mathbb C^{L\times N_j}$ denote the $L\times N_j$ matrix of the pilot sequences of the devices in cell $j$.
% In massive access, the number of devices in each cell is much larger than the length of the pilot sequence,
As $L\ll N_j$, $j\in\mathcal J$, it is not possible to assign mutually orthogonal pilot sequences to the devices within a cell.
As in\textcolor{black}{~\cite{Xu15ICC,Liu18TSP,Chen18TSP,Shao19IoTJ,Chen19TWC,Yu19ICC}}, we assume that the pilot sequences for all devices are generated in an i.i.d. manner according to %complexed Gaussian distribution with zero mean and unit variance, i.e.,
$\mathcal{CN}(\mathbf 0,\mathbf I_L)$. \textcolor{black}{By noting that a Gaussian random variable is continuous, the probability of assigning different devices the same pilot sequence is zero.}
%Assume that $\mathbf p_i$, $i\in\Phi_j$ are known at AP $j$.
In each coherence block, all active devices synchronously send their pilot sequences\textcolor{black}{~\cite{Liu18TSP,Chen19TWC,Yu19ICC}}, and %to their associated APs.
each AP aims to detect the activities of its associated devices (i.e., the devices in its own cell).
Let $\mathbf Y_j\in\mathbb C^{L\times M}$ denote the received signal over \textcolor{black}{the} $L$ signal dimensions and $M$ antennas at AP $j$. Then, we have
\begin{align}
\mathbf  Y_j &= \sum_{i\in \mathcal I}a_i \textcolor{black}{\gamma_{i,j}^{\frac{1}{2}}}\mathbf p_i \mathbf h_{i,j}^T+\mathbf Z_j,\quad j\in\mathcal J,\notag
\end{align}
where $\mathbf Z_j\in\mathbb C^{L\times M}$ is the additive white Gaussian noise (AWGN) at AP $j$ with each element following $\mathcal{CN}(0,\mathbf \delta^2)$.
In this paper, w.l.o.g., we focus on the device activity detection at a typical AP located at the origin. The typical AP is  denoted as AP $0$, and its six neighbor APs are indexed with $1,2,\cdots,6$, respectively.
% The receive signal at AP $j$ is given by
% \begin{align}
% \mathbf Y_j = \sum_{i\in \Phi_j}a_id_{i,j}^{-\frac{\alpha}{2}}\mathbf p_i \mathbf h_{i,j}^T+\sum_{i\in\Phi\setminus \Phi_j}a_id_{i,j}^{-\frac{\alpha}{2}}\mathbf p_i \mathbf h_{i,j}^T+\mathbf Z_j,\notag
% \end{align}
% where $\mathbf Z_j$ is the additive white Gaussian noise (AWGN) at AP $j$ with each element following $\mathcal{CN}(0,\mathbf \delta^2)$.
% Note that the first term is the received signal from devices in $\Phi_j$ and the second term is the received signal from devices in other cells.
% The inter-cell interference caused by non-orthogonal pilot sequences will degenerate the device activity detection accuracy.
We consider two types of activity detection mechanisms, i.e., non-cooperative device activity detection and cooperative device activity detection. \textcolor{black}{For ease of exposition, we assume that the large-scale fading powers are known for non-cooperative and cooperative activity detections.}\footnote{In the case of stationary devices, path losses (and shadowing effects) of the devices to be detected can be easily obtained.
In the case of slowly moving devices, %devices have small or moderate speed,
path losses (and shadowing effects) of the devices to be detected can be estimated. \textcolor{black}{The results for non-cooperative device activity detection in this paper can be readily extended to the case with unknown large-scale fading powers as discussed in \cite{Caire18ISIT}. We leave the investigation of cooperative device activity detection with unknown large-scale fading powers to our future work}.}

\begin{itemize}
\item {\em Non-cooperative Device Activity Detection}: %As shown in Fig.~\ref{fig:system_model}~(a), we assume that $\overline{\Phi}_0$ contains the devices in an enlarged hexagonal cell with the side length equal to $\overline{R}\geq R$. That is, $\overline{\Phi}_0$ contains the devices associated with the typical AP as well as the devices within distance $\overline{R}-R$ from the boundary of the typical cell.
Let $\bm\gamma_{0}\triangleq (\gamma_{i,0})_{i\in\Phi_0}\in\mathbb R^{N_0}$ denote the path losses between \textcolor{black}{the} devices in cell $0$ and AP \textcolor{black}{$0$}.
Under non-cooperative device activity detection, %the pilot sequence and location information of devices in $\Phi_0$ (i.e.,
% given $\mathbf P_0$ and $\bm \gamma_{0}$, %are assumed to be perfectly known at the typical AP.
% Let $\mathbf Y_0\in\mathbb C^{L\times M}$ denote the received signal over $L$ signal dimensions and $M$ antennas at the typical AP.
% For the non-cooperative device activity detection,
AP \textcolor{black}{$0$} has knowledge of $\mathbf P_0$ and $\bm \gamma_{0}$ \cite{Liu18TSP,Chen18TSP,Shao19IoTJ,Senel18TCOM} and would like to detect the activities of \textcolor{black}{the} devices in $\Phi_0$ from the received signal $\mathbf Y_0$. %\textcolor{black}{or the sample covariance $\widehat{\mathbf \Sigma}_{\mathbf Y_0} \triangleq \frac{1}{M}\mathbf Y_0\mathbf Y_0^H$}.%has the knowledge of the pilot sequence and location information of devices in $\Phi_0$ (i.e., $\mathbf P$ and $\mathbf \Gamma_{0}$), and
% performs activity detection from $\mathbf Y_0$.
%For device activity detection without AP cooperation, the typical AP has the knowledge of the pilot sequence and location information of devices in $\overline{\Phi}_0$ (i.e., $\mathbf P$ and $\mathbf \Gamma_{0}$), and performs activity detection solely based on its received signal (i.e.,$\mathbf Y_0$).

\item {\em Cooperative Device Activity Detection}: %As shown in Fig.~\ref{fig:system_model}~(b), %Under cooperative device activity detection,
% we assume that $\overline{\Phi}_0$ contains the devices associated with the typical AP as well as the devices connected to its six neighbor APs, i.e., $\overline{\Phi}_0=\cup_{j=0}^6 \Phi_j$.
Denote $\overline{\Phi}_0=\cup_{j=0}^6 \Phi_j$ as the set of indices of the devices in cell \textcolor{black}{$0$} as well as its six neighbor cells. Denote $\overline{N}_0\triangleq |\overline{\Phi}_0|=\sum_{j=0}^6N_j$. %as the number of devices in $\overline{\Phi}_0$.
Let $\overline{\mathbf P}_0\triangleq (\mathbf P_j)_{j\in\{0,1,\cdots,6\}}\in\mathbb C^{L\times \overline{N}_0}$ denote the $L\times \overline{N}_0$ matrix of the pilot sequences of the devices in $\overline{\Phi}_0$.
Let $\overline{\bm\gamma}_{j}\triangleq (\gamma_{i,j})_{i\in\overline{\Phi}_0}\in\mathbb R^{\overline N_0}$ denote the path losses between the devices in $\overline{\Phi}_0$ and AP $j$.
%we assume that each AP has the knowledge of the pilot sequence and location information of its associated devices as well as the devices connected to its $6$ neighbor APs.
Under cooperative device activity detection,
% the pilot sequence and location information of devices in $\overline{\Phi}_0$ (i.e.,
% Let $\mathbf Y_j\in\mathbb C^{L\times M}$ denote the received signal over $L$ signal dimensions and $M$ antennas at AP $j$.
each AP $j\in\{1,2,\cdots,6\}$ transmits its received signal $\mathbf Y_j$ \textcolor{black}{or the sample covariance $\widehat{\mathbf \Sigma}_{\mathbf Y_j} \triangleq \frac{1}{M}\mathbf Y_j\mathbf Y_j^H$ (a sufficient statistics for estimating device activities, which will be seen shortly)} to AP $0$ via an error-free backhaul link.\footnote{As in~\cite{Chen19TWC}, for tractability, we consider an error-free backhaul link in the analysis and optimization. The resulting detection performance provides an upper bound for that in practical networks. \textcolor{black}{Although aggregating the detection results from the six neighbor APs yields a smaller backhaul burden, we do not utilize these detection results for cooperative device activity detection mainly due to two reasons. Firstly, without extra observations, each AP may not detect the activities of the devices in the seven cells with a satisfactory accuracy. Secondly, it is unknown how to effectively utilize the detection results for one device from the six neighbor APs which have different accuracies.}%\textcolor{black}{Later, we shall see that the sample covariance $\widehat{\mathbf \Sigma}_{\mathbf Y_j}$ is a sufficient statistics for estimating device activities. This involves each neighbor AP transmitting $\frac{L(L+1)}{2}$ complex numbers to AP $0$ in the cooperative device activity detection, as $\widehat{\mathbf \Sigma}_{\mathbf Y_j}$ is a Hermitian matrix.}
}
% Note that the impact of limited backhaul links on the device activity detection is also investigated in this paper. %in Section~\ref{subsec:limited_backhaul_cap}.
% For cooperative device activity detection,
% Given $\overline{\mathbf P}_0$ and $\overline{\bm \gamma}_{j}$, $j\in\{0,1,\cdots,6\}$, %are assumed to be perfectly known at the typical AP,
AP \textcolor{black}{$0$} has knowledge of $\overline{\mathbf P}_0$ and $\overline{\bm \gamma}_{j}$, $j\in\{0,1,\cdots,6\}$ \cite{Chen19TWC} and would like to detect the activities of the devices in $\overline{\Phi}_0$ (i.e., the devices in its cell and its six neighbor cells)
% its associated devices (i.e., devices in $\Phi_0$) in combining with the devices in its six neighbor cells (i.e., devices in $\cup_{j\in\{1,2,\cdots,6\}}\Phi_j$) %has the knowledge of the pilot sequence and location information of devices in $\overline{\Phi}_0$ (i.e., $\mathbf P$ and $\mathbf \Gamma_{j}$, $j\in\{0,1,\cdots,6\}$), and
% performs activity detection
from $\overline{\mathbf Y}_0\triangleq [\mathbf Y_0,\mathbf Y_1,\cdots,\mathbf Y_6]\in\mathbb C^{L\times 7M}$ or \textcolor{black}{$\widehat{\mathbf \Sigma}_{\mathbf Y_j}$, $j\in\{0,1,2\cdots,6\}$}.
Note that each AP detects activities of the devices in its six neighbor cells rather than simply treating them as interference, for the purpose of improving the accuracy for detecting the devices in $\Phi_0$.
\end{itemize}

% \textcolor{black}{Note that as in \cite{Xu15ICC,Chen19TWC}, we assume that the lar}
Later, we shall see that with extra knowledge and observations, cooperative device activity detection can achieve high detection accuracy for \textcolor{black}{the} device activities in $\Phi_0$, at the cost of \textcolor{black}{computational} complexity increase, compared to non-cooperative device activity detection.

\section{Non-cooperative Device Activity Detection}\label{sec:signal_AP_detection}

In this section, we consider non-cooperative device activity detection, where given $\mathbf P_0$ and $\bm \gamma_0$, AP \textcolor{black}{$0$} detects the activities of the devices in $\Phi_0$ from $\mathbf Y_0$. Then, $\mathbf Y_0$ can be rewritten as
\begin{align}
\mathbf  Y_0
&=\mathbf P_0\mathbf A_0\bm \Gamma_{0}^{\frac{1}{2}}\mathbf H_{0}^T+\sum_{i\in\mathcal I\setminus \Phi_0}a_i\gamma_{i,0}^{\frac{1}{2}}\mathbf p_i \mathbf h_{i,0}^T+\mathbf Z_0, \label{eqn:receive_signal}
\end{align}
where $\mathbf A_0\triangleq{\rm diag}(\mathbf a_0)$, %with $\mathbf a_0\triangleq (a_i)_{i\in\Phi_0}$ representing the activity vector of devices in $\Phi_0$, %with $\mathbf a_j\triangleq (a_i)_{i\in\Phi_j}$,
$\mathbf \Gamma_{0}\triangleq{\rm diag}(\bm\gamma_{0})$, %with $\bm\gamma_{0}\triangleq (\gamma_{i,0})_{i\in\Phi_0}$ representing the path loss vector between devices in $\Phi_0$ and the typical AP,  %with $\bm \gamma_j\triangleq (\gamma_{i,j})_{i\in\Phi_j}$ and $\gamma_{i,j}\triangleq d_{i,j}^{-\alpha}$,
and $\mathbf H_{0}\triangleq (\mathbf h_{i,0})_{i\in\Phi_0}\in\mathbb C^{M\times N_0}$. %is the $M\times N_0$ channel matrix between devices in $\Phi_0$ and the typical AP.
% $\widetilde{\mathbf Y}=\widetilde{\mathbf P}\widetilde{\bm \Gamma}^{\frac{1}{2}}\widetilde{\mathbf H}$ is the received signal from other cells, and $\mathbf Z$ is the additive white Gaussian noise with $\mathbf Z_{:,i}\sim\mathcal{CN}(0,\sigma^2\mathbf I)$.
% and $\mathbf Z_0$ is the additive white Gaussian noise (AWGN) at the typical AP with each element following $\mathcal{CN}(0,\mathbf \delta^2)$. %$\mathbf Z_j\triangleq [\mathbf z_j(1),\cdots,\mathbf z_j(L)]^T$.
Note that the first term in~\eqref{eqn:receive_signal} is the received signal from the devices in $\Phi_0$, and the second term is the received \textcolor{black}{inter-cell interference} from the other devices.
% \textcolor{black}{In contrast with \cite{Caire18ISIT}, interference is explicitly considered in device activity detection.}

\begin{figure}[t]
\begin{center}
 \includegraphics[width=7cm]{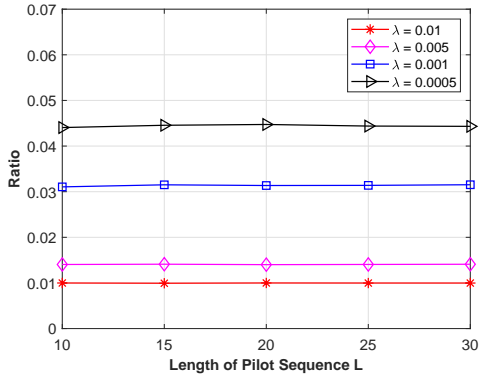}
  \end{center}
  \vspace{-2mm}
  \caption{\small{Average ratio \textcolor{black}{$\mathbb E\left[\frac{\tilde{X}_{i,j}}{\tilde{X}_{i,i}}\right]$, $i,j\in\mathcal L$ and $j\neq i$, where $\tilde{X}_{i,j}$ denotes the element in the $i$-th row and the $j$-th column of $\widetilde{\mathbf X}$.}
  % between the Euclidean norms of \textcolor{black}{the} diagonal and non-diagonal elements of $\widetilde{\mathbf X}$. %Assume that the locations of
  Active interfering devices in $\mathcal I\setminus\Phi_0$ are assumed to follow a homogeneous Poisson point process (PPP) with density $\lambda$. $R=200$ and $\alpha=3$. \textcolor{black}{Note that the non-diagonal elements of $\widetilde{\mathbf X}$ are i.i.d., and the diagonal elements of $\widetilde{\mathbf X}$ are i.i.d. The expectation is evaluated via Monte Carlo simulations.}}}
  \vspace{-2mm}
\label{fig:diagonally_dominant}
\end{figure}

Let $\mathbf y_{0,m}$ denote the $m$-th column of $\mathbf Y_0$.
Under Rayleigh fading and AWGN, %It is known that
\textcolor{black}{given device activities, large-scale fading, and pilot sequences,}
$\mathbf y_{0,m}$, $m\in\{1,2,\cdots,M\}$ are i.i.d. according to
%As channel vectors are spatially white Gaussian, the columns of $\mathbf Y_0$ are i.i.d. Gaussian vectors with
$\mathcal{CN}(\mathbf 0,\mathbf P_0\mathbf A_0\bm \Gamma_{0}\mathbf P_0^H+\widetilde{\mathbf X}+\delta^2\mathbf I_L)$, where $\widetilde{\mathbf X}\triangleq \sum\limits_{i\in \mathcal I\setminus\Phi_0}a_i\gamma_{i,0}\mathbf p_i\mathbf p_i^H \in \mathbb C^{L\times L}$. %for given $\mathbf p_i$, $i\in\Phi$.
Here, $\mathbf P_0\mathbf A_0\bm \Gamma_{0}\mathbf P_0^H$, $\widetilde{\mathbf X}$, and $\delta^2\mathbf I_L$ are the covariance matrices of the received signal, inter-cell interference, and noise at AP \textcolor{black}{$0$}, respectively. Besides $\mathbf a_0$, $\widetilde{\mathbf X}$ is also unknown and has to be estimated. %On one hand,
Notice that estimation of the $L\times L$ matrix $\widetilde{\mathbf X}$ involves \textcolor{black}{very high computational complexity}, especially when $L$ is moderate or large, \textcolor{black}{and will yield optimization problems that are not tractable}. In addition, note that
% are independently and randomly generated according to $\mathcal {CN}(\mathbf 0,\mathbf I_L)$,
% we can see that
$\widetilde{\mathbf X}$ is diagonally dominant \textcolor{black}{(as shown in Fig.~\ref{fig:diagonally_dominant}), %and the diagonal entries of $\widetilde{\mathbf X}$ are independently distributed,
since} pilot sequences are generated from i.i.d. \textcolor{black}{$\mathcal {CN}(\mathbf 0,\mathbf I_L)$}~\cite{Chen19Asilomar}. Therefore, we approximate
% Due to the lack of information on the covariance matrix of interference (i.e., $\sum_{i\in \mathcal I\setminus\Phi_0}a_i\gamma_{i,0}\mathbf p_i\mathbf p_i^H$) at the typical AP, we cannot accurately characterize the distribution of $\mathbf y_{0,m}$. %the distribution of $\mathbf y_{0,m}$ is approximated as follows.
% As pilot sequences are generated from i.i.d. complexed Gaussian distribution, we can see that the diagonal entries of $\sum_{i\in \mathcal I\setminus\Phi_0}a_i\gamma_{i,0}\mathbf p_i\mathbf p_i^H$ are independently distributed.
% By approximating
$\widetilde{\mathbf X}$ with $\mathbf X\triangleq  {\rm diag}(\mathbf x)$, where $\mathbf x\triangleq(x_{\ell})_{\ell\in\mathcal L}\in [0,\infty)^L$\cite{Chen19TWC,Andrews07WC,Choi19TCOM}.
\textcolor{black}{Fig.~\ref{fig:diagonally_dominant} demonstrates that the approximation is reasonable and has a negligible error when there are massive interfering devices.}
%\textcolor{black}{Approximating $\widetilde{\mathbf X}$ with a diagonal matrix can maintain the main characteristics of the covariance matrix of inter-cell interference.}
Later, we shall see that allowing the entries of $\mathbf x$ to be different facilitates the coordinate descent optimization in device activity detection. %\footnote{
%\textcolor{black}{From the simulation results, we can see that the proposed designs obtained under the approximation achieve significant performance gains over the baseline detection schemes, which highlights the benefit of the diagonal matrix approximation}. %}
Note that $\mathbf x$ can be interpreted as the interference powers at the $L$ signal dimensions. %and different entries of $\mathbf x$ can take different values.
% It is worth noting
In addition, rewriting $\mathbf X$ as $\sum_{\ell\in\mathcal L}x_{\ell}\mathbf e_\ell\mathbf e_\ell^H$, where $\mathbf e_\ell$ is the $\ell$-th standard basis which has a $1$ as its $\ell$-th entry and $0$s elsewhere, the \textcolor{black}{inter-cell interference} can be viewed as from $L$ active devices with pilots $\mathbf e_\ell$, $\ell\in\mathcal L$ and \textcolor{black}{path} losses $x_\ell$, $\ell\in\mathcal L$.
% \textcolor{black}{Fig.~\ref{fig:diagonally_dominant} plots the average ratio between the Euclidean norms of diagonal and non-diagonal elements of matrix $\widetilde{\mathbf X}$ versus pilot length $L$. From Fig.~\ref{fig:diagonally_dominant}, we can see that the average ratio is quite small, and does not scale with $L$, demonstrating that our approximation is reasonable.}
Under the approximation of $\widetilde{\mathbf X}$, the distribution of $\mathbf y_{0,m}$ is \textcolor{black}{approximated} by
% \vspace{-2mm}
\begin{align}
\mathbf y_{0,m}\sim \mathcal{CN}\left(\mathbf 0,\mathbf P_0\mathbf A_0\bm \Gamma_{0}\mathbf P_0^H+\mathbf X+\delta^2\mathbf I_L\right), \quad m\in\{1,2,\cdots, M\}.\label{eqn:distribution_Y}
\end{align}
% where
% \begin{align}
% \mathbf \Sigma \triangleq \mathbf P_0\mathbf A_0\bm \Gamma_{0}\mathbf P_0^H+\mathbf X+\delta^2\mathbf I_L.\notag
% \end{align}
%$\bm \Sigma\triangleq \mathbf P\bm \Gamma\mathbf P^H+\left(X+\sigma^2\right)\mathbf I_D$ and .
Based on \eqref{eqn:distribution_Y} and the fact that $\mathbf y_{0,m}$, $m\in\{1,2,\cdots, M\}$ are i.i.d., the likelihood of $\mathbf  Y_0$ %, given $\mathbf a_0$ and $\mathbf  x$,
is given by
% \vspace{-2mm}
\begin{align}
&f_{\mathbf a_0,\mathbf  x}(\mathbf Y_0)\propto\frac{\exp\left(-{\rm tr}\left(\left(\mathbf P_0\mathbf A_0\bm \Gamma_{0}\mathbf P_0^H+\mathbf X+\delta^2\mathbf I_L\right)^{-1}\mathbf Y_0\mathbf Y_0^H\right)\right)}{\vert(\mathbf P_0\mathbf A_0\bm \Gamma_{0}\mathbf P_0^H+\mathbf X+\delta^2\mathbf I_L)\vert^M},\label{eqn:likelihood_no_coop}
\end{align}
where $\propto$ means ``proportional to'', $\vert\cdot\vert$ is the determinant of a matrix, and ${\rm tr}(\cdot)$ is the trace of a matrix. Note that the constant coefficient is omitted for notation simplicity.
\textcolor{black}{From~\eqref{eqn:likelihood_no_coop}, we know that $f_{\mathbf a_0,\mathbf  x}(\mathbf Y_0)$ depends on $\mathbf Y_0$ only through the sample covariance matrix $\widehat{\mathbf \Sigma}_{\mathbf Y_0}$.
Thus, $\widehat{\mathbf \Sigma}_{\mathbf Y_0}$ is a sufficient statistics for estimating $\mathbf a_0$ and $\mathbf x$.}
Based on~\eqref{eqn:likelihood_no_coop}, in the following, we consider the joint ML estimation and joint MAP estimation of $N_0$ device activities $\mathbf a_0$ and $L$ interference powers $\mathbf x$, respectively. %first derive likelihood function and posterior distribution

%Assume that $\bm \gamma_j$ is perfectly known at AP $j$~\cite{Xu15ICC}.
% Assume that $\mathbf p_i$, $i\in\Phi_j$ are known at AP $j$.
% In this section, we assume that $\overline{\Phi}_j$ contains the devices in an enlarged hexagonal cell with the side length equal to $\overline{R}\geq R$, as shown in Fig.~\ref{fig:system_model}. That is, $\overline{\Phi}_j$ contains the devices associated with AP $j$ as well as the devices that are within distance $\overline{R}-R$ from the cell edge of AP $j$. Note that when $\overline{R}= R$, $\overline{\Phi}_j$ equals to $\Phi_j$ and %the considered device activity detection reduces the case where
% each AP estimates the activities of its associated devices only.
% For device activity detection without AP cooperation, the typical AP has the knowledge of the pilot sequence and location information of devices in $\overline{\Phi}_0$ (i.e., $\mathbf P$ and $\mathbf \Gamma_{0}$), and performs activity detection solely based on its received signal (i.e.,$\mathbf Y_0$).

% \subsection{Likelihood Function and Posterior Distribution}

% % In this part, we derive the likelihood function of $\mathbf Y_0$ given $\mathbf a$ and $\mathbf x$, and the posterior distribution of $\mathbf a$ and $\mathbf x$ given an observation of $\mathbf Y_0$, which will be used in activity detection in subsequent parts.

% % First, we obtain the likelihood of $\mathbf Y_0$, given $\mathbf a$ and $\mathbf x$.

% % \subsection{Optimization Problems}
\vspace{-2mm}
\subsection{Joint ML Estimation of Device Activities and Interference Powers}

% \subsubsection{Likelihood Function}

% \subsubsection{ML Estimation}
In this part,
we
% consider the case where the interference power $\mathbf x$ is deterministic and unknown and make
perform the joint ML estimation of $\mathbf a_0$ and $\mathbf x$.
The maximization of the likelihood \textcolor{black}{$f_{\mathbf a_0,\mathbf  x}(\mathbf Y_0)$} is equivalent to the minimization of the negative log-likelihood $-\log \textcolor{black}{f_{\mathbf a_0,\mathbf  x}(\mathbf Y_0)}\\ \propto f_{\rm ML}(\mathbf a_0, \mathbf x)$, where
%The log-likelihood of $\mathbf Y_0$ given $\mathbf a$ and $\mathbf x$ is
% \vspace{-2mm}
\begin{align}
% -\log f(\mathbf Y_0|\mathbf a_0,\mathbf  x)
% %&=-\frac{1}{M} \log\left(p(\mathbf Y_0|\mathbf a,\mathbf x)\right)\notag\\
% \propto&
f_{\rm ML}(\mathbf a_0, \mathbf x) \triangleq \log|\mathbf P_0\mathbf A_0\mathbf \Gamma_{0}\mathbf P_0^H+\mathbf X+\delta^2\mathbf I_L|+{\rm tr}((\mathbf P_0\mathbf A_0\mathbf \Gamma_{0}\mathbf P_0^H+\mathbf X +\delta^2\mathbf I_L)^{-1}\widehat{\mathbf \Sigma}_{\mathbf Y_0}),\label{eqn:f_ml}
\end{align}
with %$h(\bm \gamma,\mathbf x)\triangleq \log|\mathbf P\mathbf \Gamma\mathbf P^H+\mathbf X+\sigma^2\mathbf I_L|$, $q(\bm \gamma,\mathbf x)={\rm tr}((\mathbf P\mathbf \Gamma\mathbf P^H+\mathbf X +\sigma^2\mathbf I_L)^{-1}\widehat{\mathbf \Sigma}_{\mathbf Y})$,
$\widehat{\mathbf \Sigma}_{\mathbf Y_0} = \frac{1}{M}\mathbf Y_0\mathbf Y_0^H$.
% Therefore, the joint ML estimate of $\mathbf a$ and $\mathbf x$ can be found by solving the following optimization problem
By omitting the constant term in the negative log-likelihood function, the joint ML estimation of $\mathbf a_0$ and $\mathbf x$ without AP cooperation \textcolor{black}{is formulated as follows.}
\begin{Prob}[Joint ML Estimation without AP Cooperation]\label{Prob:ML}
% \vspace{-2mm}
\begin{align}
\min_{\mathbf a_0,\mathbf x} &\quad f_{\rm ML}(\mathbf a_0, \mathbf x)\notag\\
s.t. &\quad  1\geq a_i\geq 0,\quad i\in \Phi_0,\label{eqn:single_a}\\
&\quad x_{\ell}\geq 0,\quad \ell\in\mathcal L.\label{eqn:single_x}
\end{align}
Let $(\mathbf a_0^*,\mathbf x^*)$ denote an optimal solution of Problem~\ref{Prob:ML}.
\end{Prob}

\textcolor{black}{Note that in this paper,
% as in~\cite{Caire18ISIT} and \cite{Yu19ICC},
binary condition $a_i\in\{0,1\}$ is relaxed to continuous condition $a_i\in[0,1]$ in each estimation problem, and binary detection results are obtained by performing thresholding after solving the estimation problem as in~\cite{Caire18ISIT} and \cite{Yu19ICC}.}
Different from the ML estimation in \cite{Caire18ISIT} and \cite{Yu19ICC}, which only focuses on estimating $\mathbf a_0$ in a single-cell network without \textcolor{black}{inter-cell interference}, Problem~\ref{Prob:ML} considers the joint ML estimation of $\mathbf a_0$ and $\mathbf x$ \textcolor{black}{in the multi-cell network with inter-cell interference}.
% Due to the existence of the extra nonnegative diagonal matrix $\mathbf X$ in $f_{\rm ML}(\mathbf a_0, \mathbf x)$, $\mathbf a_0^*$ obtained from the proposed joint ML estimation is likely to have fewer zero elements than the optimal solution of the ML estimation in \cite{Caire18ISIT}, when $\mathbf Y_0$ contains interference.
% Let $\mathbf e_i$ denote the $i$-th standard basis which has a $1$ as its $i$-th entry and $0$s elsewhere. Comparing $\mathbf X=\sum_{\ell\in\mathcal L}x_{\ell}\mathbf e_\ell\mathbf e_\ell^H$ with $\mathbf P_0\mathbf A_0\mathbf \Gamma_{0}\mathbf P_0^H=\sum_{i\in\overline{\Phi}_0 }a_i\gamma_{i,0}\mathbf p_i\mathbf p_i^H$, we can see that the estimation of $x_{\ell}$ is equivalent to estimating the activity of an additional device with pilot sequence $\mathbf e_\ell$ and unit distance from the typical AP. Therefore, the joint estimation of $\mathbf a_0$ and $\mathbf x$ can be regarded as estimating activities of $ N_0+L$ devices. %with pilot matrix $[\mathbf P,\mathbf I_L]$ and path loss vector $[\bm \gamma_{0}^T,\mathbf 1_{1,L}]^T$.
As $\log|\mathbf P_0\mathbf A_0\mathbf \Gamma_{0}\mathbf P_0^H+\mathbf X+\delta^2\mathbf I_L|$ is a concave function of $\mathbf a_0$ and $\mathbf x$, and ${\rm tr}((\mathbf P_0\mathbf A_0\mathbf \Gamma_{0}\mathbf P_0^H+\mathbf X+\delta^2\mathbf I_L)^{-1}\widehat{\mathbf \Sigma}_{\mathbf Y_0})$ is a convex function of $\mathbf a_0$ and $\mathbf x$, %and $\frac{(x-\mu)^2}{2\sigma^2}$ is a convex function of $x$,
$f_{\rm ML}(\mathbf a_0, \mathbf x)$ is a difference of convex (DC) function. Combining with the fact that the inequality constraints are linear, Problem~\ref{Prob:ML} is a DC programming problem, which is a subcategory of non-convex problems. Note that obtaining a stationary point is the classic goal for solving a non-convex problem.
% Suppose that the pilot sequence $\{\mathbf p_i\}$, $i\in\mathcal N$ are such that cone$(\mathbf p_{\mathcal N})$ coincides with the cone of $L\times L$ positive semi-definite matrices $\mathcal S_{L}^{+}$.
In the following, we extend the coordinate descent method  for the case without \textcolor{black}{inter-cell interference} in~\cite{Caire18ISIT} to obtain a stationary point of Problem~\ref{Prob:ML} for the case with \textcolor{black}{inter-cell interference}.
As a closed-form optimal solution can be obtained for the optimization of each coordinate, the coordinate descent method is more computationally efficient than standard methods for DC programming, such as convex-concave procedure, \textcolor{black}{where the convex approximate problem in each iteration cannot be solved analytically}.%\footnote{\textcolor{black}{Later, we shall see that the complexity for calculating the closed-form solution in each iteration of the coordinate descent algorithm is $\mathcal O(N_0L^2+L^3)$. The complexity for solving one approximate convex problem in each iteration of the convex-concave procedure with an interior point method is $O((N_0+L)^3)$ \cite{Boyd04book}.}}

\textcolor{black}{In each iteration of the proposed coordinate descent algorithm, all coordinates are updated once}. At each step of \textcolor{black}{one iteration}, %the proposed coordinate descent algorithm,
we optimize $f_{\rm ML}(\mathbf a_0,\mathbf x)$ with respect to one of the coordinates in $\{a_i$: $i\in\Phi_0\}\cup\{x_{\ell}:\ell\in\mathcal L\}$. %Let $\mathbf e_i$ denote the $i$-th standard basis which has a $1$ as its $i$-th entry and $0$s elsewhere.
Specifically, given $\mathbf a_0$ and $\mathbf x$ obtained in the previous step, the coordinate descent optimization with respect to $a_i$ is equivalent to the optimization of the increment $d$ in $a_i$:
\begin{align}
\min_{1-a_i\geq d\geq -a_i}\ f_{\rm ML}(\mathbf a_0 + d\mathbf e_i, \mathbf x),\label{eqn:ML_a}
\end{align}
and the coordinate descent optimization with respect to $x_{\ell}$  is equivalent to the optimization of the increment $d$ in $x_{\ell}$:
\begin{align}
\min_{d\geq -x_{\ell}}\ f_{\rm ML}(\mathbf a_0,\mathbf x + d\mathbf e_\ell).\label{eqn:ML_x}
\end{align}

Based on structural properties of the coordinate descent optimization problems in~\eqref{eqn:ML_a} and \eqref{eqn:ML_x}, we can derive their closed-form optimal solutions.\footnote{\textcolor{black}{Without the approximation of covariance matrix of inter-cell interference, there are $L^2$ variables related to interference that have to be optimized, and the corresponding optimization have more complex structures which do not allow analytical solutions for the coordinate descent optimization problems.}}
\begin{Thm}[Optimal Solutions of Coordinate Descent Optimizations in~\eqref{eqn:ML_a} and \eqref{eqn:ML_x}]\label{Thm:Step_one_AP}
Given $\mathbf a_0$ and $\mathbf x$ obtained in the previous step,
the optimal solution of the coordinate optimization with respect to \textcolor{black}{the increment in} $a_i$ in~\eqref{eqn:ML_a} is given by
\begin{align}
\min\left\{\max\left\{\frac{\mathbf p_{i}^H\mathbf \Sigma^{-1}\widehat{\mathbf \Sigma}_{\mathbf Y_0}\mathbf \Sigma^{-1}\mathbf p_{i}-\mathbf p_{i}^H\mathbf \Sigma^{-1}\mathbf p_{i}}{\gamma_{i,0}(\mathbf p_{i}^H\mathbf \Sigma^{-1}\mathbf p_{i})^2},-a_{i}\right\},1-a_i\right\},\label{eqn:d_ML_a}
\end{align}
and the optimal solution of \textcolor{black}{the} coordinate optimization with respect to \textcolor{black}{the increment in} $x_{\ell}$ in \eqref{eqn:ML_x} is given by
\begin{align}
\max\left\{\frac{\mathbf e_{\ell}^H\mathbf \Sigma^{-1}\widehat{\mathbf \Sigma}_{\mathbf Y_0}\mathbf \Sigma^{-1}\mathbf e_{\ell}-\mathbf e_{\ell}^H\mathbf \Sigma^{-1}\mathbf e_{\ell}}{(\mathbf e_{\ell}^H\mathbf \Sigma^{-1}\mathbf e_{\ell})^2},-x_{\ell}\right\}.\label{eqn:d_ML_x}
\end{align}
Here, $\mathbf \Sigma \triangleq \mathbf P_0\mathbf A_0\bm \Gamma_{0}\mathbf P_0^H+\mathbf X+\delta^2\mathbf I_L$ is determined by $\mathbf a_0$ and $\mathbf x$.
\end{Thm}
\begin{IEEEproof}
Please refer to Appendix A.
\end{IEEEproof}

\begin{table}[t]
\caption{\textcolor{black}{Computational complexity of each iteration of an iterative algorithm. %the proposed estimation algorithms and the existing algorithm in~\cite{Caire18ISIT}.
}}\label{tab:computation_complexity}
\begin{center}
\vspace{-4mm}
\begin{scriptsize}
\textcolor{black}{\begin{tabular}{c|c}%!{\vrule width 1.5pt}
\hline
Estimation algorithm &Computational complexity of each iteration\rule{0pt}{3mm}\\
\hline
joint ML estimation in non-cooperative  mechanism &  $\mathcal O(N_0L^2+L^3)$\\
\hline
joint MAP estimation in non-cooperative  mechanism  & $\mathcal O(N_02^{N_0}+N_0L^2+L^3)$\\
\hline
joint ML estimation in cooperative  mechanism & $\mathcal O(\overline{N}_0L^2+L^3)$\\
\hline
joint MAP estimation in cooperative  mechanism &  $\mathcal O(\sum_{j=0}^6N_j2^{N_j}+\overline{N}_0L^2+L^3)$\\
\hline
ML estimation in \cite{Caire18ISIT} &  $\mathcal O(NL^2)$\\
\hline
\end{tabular}}
\end{scriptsize}
\end{center}
\vspace{-8mm}
\end{table}

The details of the coordinate descent algorithm for solving Problem~\ref{Prob:ML} are summarized in Algorithm~1. Specifically, in Steps $4-6$,  each coordinate of $\mathbf a_0$ is updated. In Steps $9-11$, each coordinate of $\mathbf x$ is updated.
\textcolor{black}{Unlike} the coordinate descent algorithm \textcolor{black}{for the ML estimation} in \cite{Caire18ISIT} which only updates the coordinates of $\mathbf a_0$, the coordinate updates in Algorithm~\ref{alg:ML_descend} for the \textcolor{black}{joint} ML estimation are with respect to both $\mathbf a_0$ and $\mathbf x$.
In addition, as in~\cite{Yu19ICC}, we update $\mathbf \Sigma^{-1}$ instead \textcolor{black}{of} $\mathbf \Sigma$ in each coordinate descent optimization (i.e., Steps $6$ and $11$), which avoids the calculation of matrix inversion and improves the computation efficiency
(the proof for the update of $\mathbf \Sigma^{-1}$ can be found in Appendix~A).
% Note that the computational complexities for solving the coordinate optimizations in \eqref{eqn:d_ML_a} and \eqref{eqn:d_ML_x} per iteration are $O(N_0L^2+L^3)$. %, which is the same as the computational complexity per iteration for the ML detection in~\cite{Yu19ICC}.
\textcolor{black}{As shown in Table~\ref{tab:computation_complexity},} the computational complexity of each iteration of Algorithm~\ref{alg:ML_descend} for the \textcolor{black}{joint} ML estimation %$O(N_0L^2+L^3)$
is higher than the one for the ML estimation in \cite{Caire18ISIT} %\textcolor{black}{$O(N_0L^2)$},
due to the \textcolor{black}{extra estimation of $\mathbf x_0$}. %of interference power.
% The complexity of the proposed algorithm is lower than the AMP algorithms in \cite{Shao19IoTJ,Senel18TCOM,Liu18TSP,Chen18TSP} when $M$ is extremely large.
As $f_{\rm ML}(\mathbf a_0, \mathbf x)$ is continuously differentiable, and each of the coordinate optimizations in \eqref{eqn:ML_a} and \eqref{eqn:ML_x} has a unique optimal solution,
% the coordinate descent optimizations are solved optimally,
by\textcolor{black}{\cite[Proposition 2.7.1]{Bertsekas99}}, we know that Algorithm~\ref{alg:ML_descend} for the \textcolor{black}{joint} ML estimation converges to a stationary point of Problem~\ref{Prob:ML}, as the number of iterations goes to infinity.\footnote{\textcolor{black}{When different initial points are set, Algorithm 1 may converge to different stationary points. From numerical results, we find that the stationary points corresponding to \textcolor{black}{the} initial point \textcolor{black}{$\mathbf a_0=\mathbf 0$, $\mathbf x=\mathbf 0$} usually provide good detection performance \textcolor{black}{in most setups}.}}

\begin{algorithm} \caption{Coordinate Descent Algorithm without AP Cooperation}
\small{\begin{algorithmic}[1]
\STATE Initialize $\mathbf \Sigma^{-1}=\frac{1}{\delta^2} \mathbf I_L$, $\mathbf a_0=\mathbf 0$, $\mathbf x=\mathbf 0$.
\STATE \textbf{repeat}
\FOR {$i\in\Phi_0$}
\STATE {\textbf ML:} Calculate $d$ according to \eqref{eqn:d_ML_a}.
\STATE {\textbf MAP:} Calculate $d$ according to \eqref{eqn:d_MAP_a}.
\STATE Update $a_{i}=a_{i}+d$ and $\mathbf \Sigma^{-1} = \mathbf \Sigma^{-1}-\frac{d\gamma_{i,0}\mathbf \Sigma^{-1}\mathbf p_{i}\mathbf p_{i}^H\mathbf \Sigma^{-1}}{1+d\gamma_{i,0}\mathbf p_{i}^H\mathbf \Sigma^{-1}\mathbf p_{i}}$.
\ENDFOR
\FOR {$\ell\in\mathcal L$}
\STATE {\textbf ML:} Calculate $d$ according to \eqref{eqn:d_ML_x}.
\STATE {\textbf MAP:} Calculate $d$ according to \eqref{eqn:d_MAP_x}.
\STATE Update $x_{\ell}=x_{\ell}+d$ and $\mathbf \Sigma^{-1} = \mathbf \Sigma^{-1}-\frac{d\mathbf \Sigma^{-1}\mathbf e_{\ell}\mathbf e_{\ell}^H\mathbf \Sigma^{-1}}{1+d\mathbf e_{\ell}^H\mathbf \Sigma^{-1}\mathbf e_{\ell}}$.
\ENDFOR
\STATE \textbf{until} $(\mathbf a_0,\mathbf x)$ satisfies some stopping criterion.
\end{algorithmic}}\normalsize\label{alg:ML_descend}
\end{algorithm}
% \vspace{-2mm}

\subsection{Joint MAP Estimation of Device Activities and Interference Powers}

In this part, we assume that %the distributions of
$\mathbf a_0$ and $\mathbf x$ are random and %their distributions are known to the typical AP. In this case,
perform
% consider the case where the interference power is random and make
the joint MAP estimation of $\mathbf a_0$ and $\mathbf x$.

\subsubsection{Prior Distributions}\label{Sec:prior_dist}

We assume that $\mathbf a_0$ and $\mathbf x$ are independently distributed. Note that this is a weak assumption, as it only requires that the device activities in cell \textcolor{black}{$0$} are independent of those in the other cells.
% \textcolor{black}{In this paper, we focus on the estimation of device activity in the typical cell.} %(i.e., $\mathbf a_0$).
% \textcolor{black}{Let $\mathbf a_j\triangleq (a_i)_{i\in\Phi_j}$, $j\in\mathbb N$ denote the activity vector of devices in cell $j$}.
% \textcolor{black}{For ease of analysis, we assume that devices in $\Phi\setminus \Phi_0$ access the channel with probability $p_a$ in an i.i.d. manner. The probability mass function of $\mathbf a_j$, $j\in\mathbb N^+$ is
% \begin{align}
% p(\mathbf a_j)=\exp\left(\log \frac{p_a}{1-p_a}\sum_{i\in\Phi_j}a_i+N_j\log(1-p_a)\right),\quad j\in\mathbb N^+.\label{eqn:pmf_a_out}
% \end{align}
First, we introduce a general prior distribution of the Bernoulli random vector $\mathbf a_0$.  %We assume that the distribution of $\mathbf a_{j_1}$ is independent with that of $\mathbf a_{j_2}$, $j_2\neq j_1$.
\textcolor{black}{Unlike} \cite{Chen19TWC} where devices in a cell are assumed to access the channel in an i.i.d. manner,  we
allow for correlation among the activities of \textcolor{black}{the} devices in cell \textcolor{black}{$0$}. In particular, % consider that devices in $\Phi_0$ can access the channel in a more general manner such that coupling effects between $a_i$, $i\in\Phi_0$ are allowed.
%which assume that devices access the channel in an i.i.d. manner,
we adopt the multivariate Bernoulli (MVB) model for $\mathbf a_0$~\cite{NIPS2011_4209}. %and assume that device activity in a cell is independent with those in other cells.
% Let $\overline{\Psi}_j$ denote the power set of $\Phi_0$ and denote $\Psi_0=\overline{\Psi}_0\setminus \{\emptyset\}$. Let $\omega\subseteq \Phi_0$ denote an element in $\Psi_0$.
% The MVB model of device activity in $\Phi_0$ is given as follows:
The probability mass function (p.m.f.) of $\mathbf a_0$ under the MVB model is given by
\begin{align}
p_0\left(\mathbf a_0\right)=\exp\left(\sum_{\omega\in\Psi_0}\left(c_{\omega}\prod_{i\in\omega}a_i\right)+b_0\right),\label{eqn:pmf_a_typical}
\end{align}
where $\Psi_0$ denotes the set of nonempty subsets of $\Phi_0$,
$c_{\omega}$ is the coefficient  reflecting the correlation among $a_i$, $i\in\omega$, and
$b_0\triangleq -\log(\sum_{\mathbf a_0\in\{0,1\}^{N_0}}\exp(\sum_{\omega\in\Psi_0}(c_{\omega}\prod_{i\in\omega}a_i)))$ is the normalization factor.
Note that $c_{\omega}$, $\omega\in\Psi_0$ %reveal the coupling effect between $a_i$, $i\in\Phi_j$ and
can be estimated based on the historical device activity data using existing methods~\cite{NIPS2011_4209}. In addition, given the p.m.f. of $\mathbf a_0$ in any form, the coefficients $c_{\omega}$, $\omega\in\Psi_0$ can be calculated according to~\cite[Lemma~$2.1$]{NIPS2011_4209}. %In this paper, we assume that $c_{\omega}$, $\omega\in\Psi_j$ are known and can take different values to reveal different kinds of coupling effects.
When $c_{\omega}=0$ for all $|\omega|>2$, the MVB model reduces to the Ising model~\cite{banerjee2008model}. When $c_{\omega}=0$ for all $|\omega|>1$, the MVB model reduces to the independent model with $\Pr(a_i=1)= \frac{\exp\left(c_{\{i\}}\right)}{\exp\left(c_{\{i\}}\right)+1}$, $i\in\Phi_0$.
When $c_{\omega}=0$ for all $|\omega|>1$ and $c_{\omega}=c$ for all $|\omega|=1$, the MVB model reduces to the i.i.d. model in \cite{Chen19TWC} with $\Pr(a_i=1)= \frac{\exp\left(c\right)}{\exp\left(c\right)+1}$, $i\in\Phi_0$.%\footnote{\textcolor{black}{If the typical AP has only partial information of $\mathbf a_0$, such as the marginal distribution of $\mathbf a_0$, it can also be incorporated into the joint MAP estimation and hence improve the estimation accuracy. The numerical results in Section~\ref{sec:simulation} will verify this point.}}

\textcolor{black}{To further illustrate the MVB model, we present two instances of $p_0(\mathbf a_0)$ under group device activity. %, which exhibit correlations in $\mathbf a$.
In both instances, the devices in $\Phi_0$ are divided into $K$ groups and the device activities in different groups are independent. For all $k\in\mathcal K\triangleq\{1,2,\cdots,K\}$, let $\mathcal G_k\subseteq \Phi_0$ denote the set of indices of the devices in the $k$-th group. %and denote $\mathbf a_k\triangleq (a_i)_{i\in\mathcal N_k}$.
Note that $\cup_{k\in\mathcal K}\mathcal G_k=\Phi_0$ and $\mathcal G_k\cap\mathcal G_{k'}=\emptyset$ for  $k,k'\in\mathcal K$, $k\neq k^{'}$. According to\textcolor{black}{\cite[Theorem 2.1]{NIPS2011_4209}}, we have
\begin{align}
c_{\omega}=0,\quad  \nexists\ k\in\mathcal K \text{ such that }\omega\subseteq \mathcal G_k. \label{eqn:c_cor_group}
\end{align}
It remains to specify $c_{\omega}$, $\omega\subseteq \mathcal G_k$, $k\in\mathcal K$ in the two instances.
\begin{itemize}
	\item {\em First Instance}:
	% In the first instance,
	Each group contains two devices, i.e., $|\mathcal G_k|=2$, $k\in\mathcal K$; every two devices in a group are correlated with correlation coefficient $\eta$; $\Pr[a_i=1]=p_a$, $i\in\Phi_0$.
% each device is active with probability $p_a$.
Consequently, for two devices $i_1$ and $i_2$ in one group, $\Pr(a_{i_1}=1,a_{i_2}=1)=\eta p_a +(1-\eta)p_a^2$, $\Pr(a_{i_1}=0,a_{i_2}=1)=\Pr(a_{i_1}=1,a_{i_2}=0)=(1-\eta)(p_a-p_a^2)$ and $\Pr(a_{i_1}=0,a_{i_2}=0)=1+(\eta-2)p_a+(1-\eta)p_a^2$. %Note that when $\eta = 0$, it reduces to the i.i.d. case.
According to ~\cite[Lemma~$2.1$]{NIPS2011_4209}, we know that %have the following result.
% know that when
% $\exists$ $k\in\{1,2,\cdots,K\}$ such that $\omega\in\mathcal N_k$,
% \begin{Lem}
$p_0(\mathbf a_0)$ satisfies \eqref{eqn:pmf_a_typical} with $c_{\omega}$ satisfying \eqref{eqn:c_cor_group} and
\begin{align}
c_{\omega} =&
\begin{cases}
\frac{(\eta p_a +(1-\eta)p_a^2)(1+(\eta-2)p_a+(1-\eta)p_a^2)}{(1-\eta)^2(p_a-p_a^2)^2}\quad & |\omega| = 2\\
\frac{(1-\eta)(p_a-p_a^2)}{1+(\eta-2)p_a+(1-\eta)p_a^2} &|\omega|=1
\end{cases},\quad \omega\subseteq \mathcal G_k,k\in\mathcal K.\notag
\end{align}
% \end{Lem}
% \subsubsection{Group Device Activity}
\item {\em Second Instance}:
% In the second instance,
The activity states of devices in a group are the same, i.e., $a_i$, $i\in\mathcal G_k$ are the same for all $k\in\mathcal K$,  implying $\sum_{i\in\mathcal G_k}a_i\in\{0,|\mathcal G_k|\}$, $k\in\mathcal K$; %Let $\mathcal N_k$ denote the set of indices of the devices in the $k$-th group and denote $\mathbf a_k\triangleq (a_i)_{i\in\mathcal N_k}$. %We have $\sum_{i\in\mathcal N_k}a_i\in\{0,|\mathcal N_k|\}$.
$\Pr[\cap_{i\in\mathcal G_k}(a_i=1)]=p_k$, $k\in\mathcal K$. %the devices in group $k$ are active with probability $p_k$.
%For tractability, set $\Pr(\mathbf a_k)=\epsilon$, where $\epsilon$ is an arbitrarily small number, for $\mathbf a_k$ with $\sum_{i\in\mathcal N_k}a_i\in\{2,3,\cdots,|\mathcal N_k|-1\}$.
According to~\cite[Lemma~$2.1$]{NIPS2011_4209}, we know that %have the following result.
% know that when $\exists$ $k\in\{1,2,\cdots,K\}$ such that $\omega\in\mathcal N_k$,
% \begin{Lem}
$p_0(\mathbf a_0)$ approaches \eqref{eqn:pmf_a_typical} with $c_{\omega}$ satisfying \eqref{eqn:c_cor_group} and
\begin{align}
c_{\omega} = &
\begin{cases}
(-1)^{|\omega|}\log(\frac{1-p_k}{\epsilon}) \quad &|\omega|<|\mathcal G_k|\\
\log(\frac{p_k}{1-p_k}) &|\omega|=|\mathcal G_k|, |\omega|\text{ is odd}\\
\log(\frac{p_k(1-p_k)}{\epsilon^2}) &|\omega|=|\mathcal G_k|, |\omega|\text{ is even}\\
\end{cases},\quad \omega\subseteq \mathcal G_k,k\in\mathcal K,\notag
\end{align}
where $\epsilon>0$, as $\epsilon\to 0$. %is an arbitrarily small number.
Note that in this instance, $p_0(\mathbf a_0)$  can be well approximated by~\eqref{eqn:pmf_a_typical} with a small $\epsilon$. %Later, we shall see that this approximation is not only accurate but also tractable.
% \end{Lem}
\end{itemize}}

% Based on the likelihood function of $\mathbf Y$ and p.m.f. of $\mathbf a_k$, the MAP estimation of $\mathbf a_k$, $k\in\{0,1,\cdots,K\}$ can be conducted as introduced in section~\ref{subsec:map_estimation}.

Next, we derive a prior distribution of $\mathbf x$. %which will be used for the derivation of the posterior distribution of $\mathbf a$ and $\mathbf x$. %the activity detection in the following.
Under non-cooperative device activity detection, the locations of the active interfering devices in $\mathcal I\setminus\Phi_0$ are assumed to follow a homogeneous Poisson point process (PPP) with density $\lambda$, which is a widely adopted model for large-scale wireless networks~\cite{haenggi2012stochastic}.
As pilot sequences are generated from i.i.d. $\mathcal {CN}(\mathbf 0,\mathbf I_L)$, %the diagonal entries of $\widetilde{\mathbf X}$ are i.i.d..
we assume that $x_{\ell}$, $\ell\in\mathcal L$ are i.i.d. with the same distribution as $\sum_{i\in\mathcal I\setminus\Phi_0} a_i\gamma_{i,0}$.
Therefore,
% Note that
$x_{\ell}$ is a power-law shot noise, whose exact distribution is still \textcolor{black}{unknown}~\cite{haenggi2012stochastic}. As in~\cite{Aljuaid10TVT}, \textcolor{black}{we  approximate the probability density function (p.d.f.) of $x_{\ell}$ with a Gaussian distribution using moment matching}. %denoted as $g(x_{\ell})$.
Note that the Gaussian approximation is accurate when the cell size (i.e., $R$) is large
\cite{Aljuaid10TVT,Hasan07TWC}. %and the density of active nodes out the typical cell is high.
Based on the above assumptions and techniques from stochastic geometry, we have the following results.
\begin{Lem}[Approximated Distribution of $\mathbf x$]\label{Lem:distribution_X}
%The mean of $x$ is $\mathbb E(x)=\frac{2\pi\lambda p_a R^{2-\alpha}}{\alpha-2}$ and the variance of $x$ is ${\rm Var}(x)=\frac{\pi\lambda p_a R^{2-2\alpha}}{\alpha-1}$.
% Using Gaussian distribution to approximate the distribution of $x_\ell$,
The p.d.f. of $\mathbf x$ is approximated by
\begin{align}
g(\mathbf x) = \frac{1}{(\sqrt{2\pi}\sigma)^L}\exp\left(-\frac{\sum_{\ell\in\mathcal L}(x_{\ell}-\mu)^2}{2\sigma^2}\right),\notag
\end{align}
where $\mu = 12\lambda \int_{\frac{\sqrt{3}}{2}R}^{\infty}
\int_{0}^{\frac{\sqrt{3}}{3}x} (x^2+y^2)^{-\frac{\alpha}{2}}{\rm d }y{\rm d}x$ and $\sigma^2 =12\lambda \int_{\frac{\sqrt{3}}{2}R}^{\infty}
\int_{0}^{\frac{\sqrt{3}}{3}x} (x^2+y^2)^{-\alpha}{\rm d }y{\rm d}x$.
% \begin{align}
% &\mu = 12\lambda \int_{\frac{\sqrt{3}}{2}R}^{\infty}
% \int_{0}^{\frac{\sqrt{3}}{3}x} (x^2+y^2)^{-\frac{\alpha}{2}}{\rm d }y{\rm d}x,\notag\\
% % \end{align}
% % \begin{align}
% &\sigma^2 =12\lambda \int_{\frac{\sqrt{3}}{2}R}^{\infty}
% \int_{0}^{\frac{\sqrt{3}}{3}x} (x^2+y^2)^{-\alpha}{\rm d }y{\rm d}x\notag.
% \end{align}
\end{Lem}
\begin{IEEEproof}
Please refer to Appendix B.
\end{IEEEproof}

\begin{figure}[t]
\begin{center}
 \includegraphics[width=7cm]{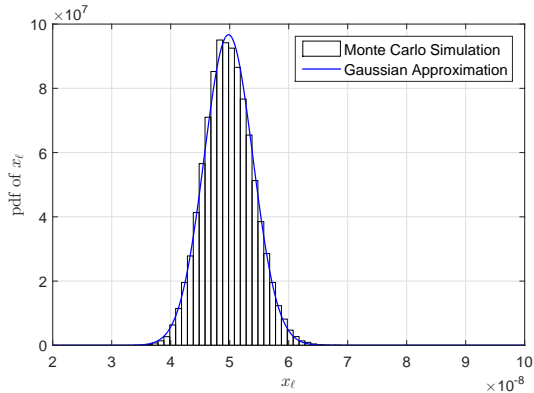}
  \end{center}
  \vspace{-2mm}
  \caption{\small{Comparison between the p.d.f. of $x_\ell$ and its corresponding Gaussian approximation. $R=200$, $\lambda=0.0005$ and $\alpha=4$.}}
  \vspace{-2mm}
\label{fig:gamma_approximation}
\end{figure}

The integral expressions of $\mu$ and $\sigma^2$ in Lemma~\ref{Lem:distribution_X} are for the case where cell \textcolor{black}{$0$} is modeled as a hexagon with side length $R$.\footnote{\textcolor{black}{If the locations of APs follow a homogeneous PPP, the boundary of a cell can be an arbitrary polyhedron.  Lemma 1 can be readily extended. But the integral domains \textcolor{black}{for calculating $\mu$ and $\sigma^2$} rely on the particular shape of cell \textcolor{black}{$0$} and may not be concisely expressed.}}
If cell \textcolor{black}{$0$} is modeled as a disk with radius $R$, $\mu$ and $\sigma^2$ have closed-form expressions, i.e., $\mu=\frac{2\pi\lambda R^{2-\alpha}}{\alpha-2}$ and $\sigma^2= \frac{\pi\lambda R^{2-2\alpha}}{\alpha-1}$, and the following results for non-cooperative device activity detection still hold.
% Note that when approximating the area of the enlarged hexagonal cell to a disk with the same area size, we have $\mu\approx\frac{2\pi\lambda R^{2-\alpha}}{\alpha-2}\left(\frac{3\sqrt{3}}{2\pi}\right)^{1-\frac{\alpha}{2}}$ and $\sigma^2\approx \frac{\pi\lambda R^{2-2\alpha}}{\alpha-1}\left(\frac{3\sqrt{3}}{2\pi}\right)^{1-{\alpha}}$.
Fig.~\ref{fig:gamma_approximation} plots the histogram of $x_{\ell}$ \textcolor{black}{(}which reflects the shape of the p.d.f. of $x_{\ell}$\textcolor{black}{) and} the Gaussian distribution with the same mean and variance. %\textcolor{black}{and the distribution of the diagonal elements of $\mathbf {\widetilde{X}}$}.
From Fig.~\ref{fig:gamma_approximation}, we can see that the Gaussian distribution is a good approximation of the exact p.d.f. of $x_{\ell}$, %under the considered simulation setup,
which verifies Lemma~\ref{Lem:distribution_X}. %\textcolor{black}{In addition, the p.d.f. of $x_\ell$ can well reflect the trend of distribution of the diagonal elements of $\mathbf {\widetilde{X}}$}.

% The pdf of $\gamma_i$ for $i\in[N]$ is given by
% \begin{align}
% f_{\gamma_i}(x) = \frac{2}{\alpha R^{2}}x^{-\frac{2}{\alpha}-1}.\notag
% \end{align}

\subsubsection{\textcolor{black}{Joint} MAP Estimation}

% Next, we derive the joint posterior distribution of $\mathbf a_0$ and $\mathbf x$, given $\mathbf Y_0$.
% Recall that \textcolor{black}{$a_i$, $i\in\Phi\setminus\Phi_0$,}
% % the components of $\mathbf a$
% are i.i.d. Bernoulli random variables \textcolor{black}{and the MVB distribution of $\mathbf a_0$ is given in
% % Combining with the distribution of $\mathbf a_0$
% in \eqref{eqn:pmf_a_typical}. Therefore, we have}
% \begin{align}
% p(\mathbf a)=\exp\left(\sum_{\omega\in\Psi_0}\left(c_{\omega}\prod_{i\in\omega}a_i\right)+b_0+\log \frac{p_a}{1-p_a}\sum_{i\in\overline{\Phi}_0\setminus \Phi_0}a_i+(\overline{N}_0-N_0)\log(1-p_a)\right).\notag
% \end{align}
Based on the \textcolor{black}{conditional density of $\mathbf Y_0$ given $\mathbf a_0$ and $\mathbf x$ (identical to the likelihood of $\mathbf Y_0$ in the \textcolor{black}{joint} ML estimation)} and the prior distributions of $\mathbf a_0$ and $\mathbf x$,
\textcolor{black}{the conditional joint density of
% the joint posterior distribution of
$\mathbf a_0$ and $\mathbf x$, given $\mathbf Y_0$, is given by}
% \vspace{-2mm}
\begin{align}
&\textcolor{black}{f_{\mathbf a_0,\mathbf  x|\mathbf Y_0}(\mathbf a_0,\mathbf  x,\mathbf Y_0)
\propto f_{\mathbf a_0,\mathbf  x}(\mathbf Y_0)}p_0(\mathbf a_0)g(\mathbf x)\notag\\
% \end{align}
% \begin{align}
&\propto \frac{\exp\left(-{\rm tr}\left(\left(\mathbf P_0\mathbf A_0\mathbf \Gamma_{0}\mathbf P_0^H+\mathbf X+\delta^2\mathbf I_L\right)^{-1}\mathbf Y_0\mathbf Y_0^H\right)-\sum_{\ell\in\mathcal L}\frac{(x_{\ell}-\mu)^2}{2\sigma^2}+\sum_{\omega\in\Psi_0}\left(c_{\omega}\prod_{i\in\omega}a_i\right)
\right)}{\vert(\mathbf P_0\mathbf A_0\bm \Gamma_{0}\mathbf P_0^H+\mathbf X+\delta^2\mathbf I_L)\vert^M}.\notag%\exp\left(-\sum\limits_{\ell\in\mathcal L}\frac{(x_{\ell}-\mu)^2}{2\sigma^2}\right)\notag\\
% &\quad\times \exp\left(\sum_{\omega\in\Psi_0}\left(c_{\omega}\prod_{i\in\omega}a_i\right)+\log \frac{p_a}{1-p_a}\sum_{i\in\overline{\Phi}_0\setminus \Phi_0}a_i\right).\notag
\end{align}
% where $\propto$ means ``proportional to''.
% Recall that $a_i$ is a Bernoulli random variable and
% \begin{align}
% P(\mathbf a)=p_a^{\sum_{i\in\Phi_0}a_i}(1-p_a)^{N_0-\sum_{i\in\Phi_0}a_i}.\notag
% \end{align}
The maximization of the \textcolor{black}{conditional joint density $f_{\mathbf a_0,\mathbf  x|\mathbf Y_0}(\mathbf a_0,\mathbf  x,\mathbf Y_0)$} is equivalent to the minimization of the negative logarithm of the \textcolor{black}{conditional joint density} $-\log \textcolor{black}{f_{\mathbf a_0,\mathbf  x|\mathbf Y_0}(\mathbf a_0,\mathbf  x,\mathbf Y_0)}\propto f_{\rm MAP}(\mathbf a_0,\mathbf x)$, where
% Define
% $f_{\rm MAP}(\mathbf a,\mathbf x)$, where
%the MAP estimator of $\mathbf a$ and $\mathbf x$ after simplification is
\begin{align}
% -\log f(\mathbf a_0,\mathbf  x|\mathbf Y_0)
% &
f_{\rm MAP}(\mathbf a_0,\mathbf x)\triangleq f_{\rm ML}(\mathbf a_0, \mathbf x)
%&=-\frac{1}{M} \log\left(p(\mathbf Y_0|\mathbf a,\mathbf x)\right)-\frac{1}{M}\sum_{\ell=1}^L\log g(x_\ell)-\frac{1}{M}\Pr(\mathbf a)\notag\\
% =  \log|\mathbf P\mathbf A\mathbf \Gamma_0\mathbf P^H+\mathbf X+\sigma^2\mathbf I_L|\notag\\
% &\quad+{\rm tr}((\mathbf P\mathbf A\mathbf \Gamma_0\mathbf P^H+\mathbf X +\sigma^2\mathbf I_L)^{-1}\widehat{\mathbf \Sigma}_{\mathbf Y})\notag\\
%&\quad
+ \frac{1}{2M\sigma^2}\sum_{\ell\in\mathcal L}(x_{\ell}-\mu)^2-\frac{1}{M}\sum_{\omega\in\Psi_0}\left(c_{\omega}\prod_{i\in\omega}a_i\right).\label{eqn:f_map}
\end{align}
Note that $\frac{1}{2M\sigma^2}\sum_{\ell\in\mathcal L}(x_{\ell}-\mu)^2$ is from the p.d.f. of $\mathbf x$ %and corresponds to the prior knowledge of the distribution of interference
and $-\frac{1}{M}\sum_{\omega\in\Psi_0}(c_{\omega}\prod_{i\in\omega}a_i)$ is from the p.m.f. of $\mathbf a_0$. %corresponds to the prior knowledge of the distribution of device activity.
The joint MAP estimation of $\mathbf a_0$ and $\mathbf x$ without AP cooperation can be formulated as follows. %found by solving the following optimization problem.
\begin{Prob}[Joint MAP Estimation without AP Cooperation]\label{Prob:MAP}
% \vspace{-3mm}
\begin{align}
\min_{\mathbf a_0,\mathbf x} &\quad f_{\rm MAP}(\mathbf a_0,\mathbf x)\notag\\
%\log|\mathbf P\mathbf \Gamma\mathbf P^H+(\sigma^2+x)\mathbf I_D| +{\rm tr}((\mathbf P\mathbf \Gamma\mathbf P^H+(\sigma^2+x)\mathbf I_D)^{-1}\widehat{\mathbf \Sigma}_{\mathbf Y})+ \frac{(x-\mu)^2}{2\sigma^2}\notag\\
s.t. &\quad\eqref{eqn:single_a}, \eqref{eqn:single_x}.\notag%&\quad 1\geq a_i\geq 0,\quad i\in\Phi_0,\notag\\
% &\quad x_{\ell} \geq 0,\quad \ell\in\mathcal L.\notag
\end{align}
Let $(\mathbf a_0^{\dagger},\mathbf x^\dagger)$ denote an optimal solution of Problem~\ref{Prob:MAP}.
\end{Prob}

% As $\frac{1}{2M\sigma^2}\sum_{\ell\in\mathcal L}(x_{\ell}-\mu)^2$ is a convex function of $\mathbf x$ and $-\frac{1}{M}\log \frac{p_a}{1-p_a}\sum_{i\in\overline{\Phi}_0}a_i$ is a liner function of $\mathbf a$,
By comparing $f_{\rm MAP}(\mathbf a_0,\mathbf x)$ with $f_{\rm ML}(\mathbf a_0,\mathbf x)$, we can draw the following conclusions. The incorporation of prior distribution $g(\mathbf x)$ pushes the estimate of $x_{\ell}$ towards its mean $\mu$ for all $\ell\in\mathcal L$.
The incorporation of the prior distribution $p_0(\mathbf a_0)$ pushes the estimate of $\mathbf a_0$ to \textcolor{black}{the activity states} with high probabilities. As $f_{\rm MAP}(\mathbf a_0,\mathbf x)-f_{\rm ML}(\mathbf a_0,\mathbf x)$ decreases with $M$, the impacts of the prior distributions of $\mathbf a_0$ and $\mathbf x$ reduce as $M$ increases.
%Since $p_a\ll 1$, the incorporation of prior knowledge on device activity pushes the estimate of $a_i$, $i\in\overline{\Phi}_0\setminus\Phi_0$ towards $0$.
% Note that compared with $f_{\rm ML}(\mathbf a,\mathbf x)$, $f_{\rm MAP}(\mathbf a,\mathbf x)$ has an additional term $\frac{1}{2M\sigma^2}\sum_{\ell\in\mathcal L}(x_{\ell}-\mu)^2-\frac{1}{M}\sum_{\omega\in\Psi_0}(c_{\omega}\prod_{i\in\omega}a_i)$, which diminishes to $0$ as $M\to \infty$. Therefore, we have the following result.
% \begin{Lem}[Property of Problem~\ref{Prob:MAP}]
As $M\to \infty$, $f_{\rm MAP}(\mathbf a_0,\mathbf x)\to f_{\rm ML}(\mathbf a_0,\mathbf x)$, Problem~\ref{Prob:MAP} reduces to Problem~\ref{Prob:ML}, and $(\mathbf a_0^\dagger,\mathbf x^\dagger)$ becomes $(\mathbf a_0^*,\mathbf x^*)$.
% \end{Lem}

As $f_{\rm ML}(\mathbf a_0, \mathbf x)$ is a DC function and $-\frac{1}{M}\sum_{\omega\in\Psi_0}(c_{\omega}\prod_{i\in\omega}a_i)$ is a non-convex function,
we can see that Problem~\ref{Prob:MAP} is a challenging non-convex problem with a complicated objective function.
We adopt the coordinate descent method to obtain a stationary point to Problem~\ref{Prob:MAP}. %The coordinate descend optimization of $\mathbf a_j$ in Problem~\ref{Prob:MAP} is the same as that in Problem~\ref{Prob:ML}. %The coordinate descend optimization of $x_\ell$ is conducted as follows.
Specifically, given $\mathbf a_0$ and $\mathbf x$ obtained in the previous step, the coordinate descent optimization with respect to $a_i$, $i\in\Phi_0$ is equivalent to the optimization of the increment $d$ in $a_i$:
% \vspace{-2mm}
\begin{align}
\min_{1-a_i\geq d\geq -a_i} \ f_{\rm MAP}(\mathbf a_0 + d\mathbf e_i, \mathbf x),\label{eqn:MAP_a}
\end{align}
% For all
% \textcolor{black}{Let $d^*$ denote the optimal solution of the optimization in \eqref{eqn:MAP_a}.}
and the coordinate descent optimization with respect to $x_{\ell}$, $\ell\in\mathcal L$ is equivalent to the optimization of the increment $d$ in $x_\ell$:
% \vspace{-2mm}
\begin{align}
\min_{d\geq -x_{\ell}} \ f_{\rm MAP}(\mathbf a_0,\mathbf x+ d\mathbf e_\ell).\label{eqn:MAP_x}
\end{align}
Define
\begin{align}
&f_{a,i}(d,\mathbf a_0,\mathbf x)\triangleq\log(1+d\gamma_{i,0}\mathbf p_i^H\mathbf \Sigma^{-1}\mathbf p_i)-\frac{d\gamma_{i,0}\mathbf p_i^H\mathbf \Sigma^{-1}\widehat{\mathbf \Sigma}_{\mathbf Y_0}\mathbf \Sigma^{-1}\mathbf p_i}{1+d\gamma_{i,0}\mathbf p_i^H\mathbf \Sigma^{-1}\mathbf p_i}
-\frac{d}{M}\sum_{\omega\in\Psi_0:i\in\omega}\left(c_{\omega}\prod_{i^{'}\in\omega,i^{'}\neq i}a_{i^{'}}\right),\notag\\
% \end{align}
% \begin{align}
&f_{x,\ell}(d,\mathbf a_0,\mathbf x)\triangleq \log(1+d\mathbf e_\ell^H\mathbf \Sigma^{-1}\mathbf e_\ell)-\frac{d\mathbf e_\ell^H\mathbf \Sigma^{-1}\widehat{\mathbf \Sigma}_{\mathbf Y_0}\mathbf \Sigma^{-1}\mathbf e_\ell}{1+d\mathbf e_\ell^H\mathbf \Sigma^{-1}\mathbf e_\ell}+\frac{(x_{\ell}-\mu+d)^2}{2M\sigma^2},\notag\\
% \end{align}
% \begin{align}
% \end{align}
% \begin{align}
%&f_j(d,\mathbf s,\mathbf \Sigma)\triangleq \log(1+d\mathbf s^H\mathbf \Sigma^{-1}\mathbf s)-\frac{d\mathbf s^H\mathbf \Sigma^{-1}\widehat{\mathbf \Sigma}_{\mathbf Y_j}\mathbf \Sigma^{-1}\mathbf s}{1+d\mathbf s^H\mathbf \Sigma^{-1}\mathbf s},\notag\\
% &h_{a,i}(d,\mathbf a_0,\mathbf x)\triangleq \frac{\gamma_{i,0}\mathbf p_i^H\mathbf \Sigma^{-1}\mathbf p_i}{1+d\gamma_{i,0}\mathbf p_i^H\mathbf \Sigma^{-1}\mathbf p_i}-\frac{\gamma_{i,0}\mathbf p_i^H\mathbf \Sigma^{-1}\widehat{\mathbf \Sigma}_{\mathbf Y_0}\mathbf \Sigma^{-1}\mathbf p_i}{(1+d\gamma_{i,0}\mathbf p_i^H\mathbf \Sigma^{-1}\mathbf p_i)^2}
% -\frac{1}{M}\sum_{\omega\in\Psi_0:i\in\omega}\left(c_{\omega}\prod_{i^{'}\in\omega,i^{'}\neq i}a_{i^{'}}\right),\notag\\
&h_{x,\ell}(d,\mathbf a_0,\mathbf x)\triangleq \frac{\mathbf e_\ell^H\mathbf \Sigma^{-1}\mathbf e_\ell}{1+d\mathbf e_\ell^H\mathbf \Sigma^{-1}\mathbf e_\ell}-\frac{\mathbf e_\ell^H\mathbf \Sigma^{-1}\widehat{\mathbf \Sigma}_{\mathbf Y_0}\mathbf \Sigma^{-1}\mathbf e_\ell}{(1+d\mathbf e_\ell^H\mathbf \Sigma^{-1}\mathbf e_\ell)^2}
+\frac{d+x_{\ell}-\mu}{M\sigma^2}.\notag
%&f_j^{'}(d,\mathbf s,\mathbf \Sigma)\triangleq \frac{\mathbf s^H\mathbf \Sigma^{-1}\mathbf s}{1+d\mathbf s^H\mathbf \Sigma^{-1}\mathbf s}-\frac{\mathbf s^H\mathbf \Sigma^{-1}\widehat{\mathbf \Sigma}_{\mathbf Y_j}\mathbf \Sigma^{-1}\mathbf s}{(1+d\mathbf s^H\mathbf \Sigma^{-1}\mathbf s)^2}.\notag
\end{align}
We write $f_{a,i}(d,\mathbf a_0,\mathbf x)$, $f_{x,\ell}(d,\mathbf a_0,\mathbf x)$, and $h_{x,\ell}(d,\mathbf a_0,\mathbf x)$ as functions of $\mathbf a_0$ and $\mathbf x$, as $\mathbf \Sigma$ is a function of $\mathbf a_0$ and $\mathbf x$.
Note that $h_{x,\ell}(d,\mathbf a_0,\mathbf x)$ is the derivative function of $f_{x,\ell}(d,\mathbf a_0,\mathbf x)$ with respect to $d$.
% Denote $\mathcal A_{i}(\mathbf a_0,\mathbf x)\triangleq \{d\in[-a_i,1-a_i]:h_{a,i}(d,\mathbf a_0,\mathbf x)=0\}$ as the set of roots of equation $h_{a,i}(d,\mathbf a_0,\mathbf x)=0$ that lie in $[-a_i,1-a_i]$, for given $\mathbf a_0$ and $\mathbf x$.
Denote $\mathcal X_{\ell}(\mathbf a_0,\mathbf x)\triangleq \{d\geq-x_\ell:h_{x,\ell}(d,\mathbf a_0,\mathbf x)=0\}$ as the set of roots of equation $h_{x,\ell}(d,\mathbf a_0,\mathbf x)=0$ that are no smaller than $-x_\ell$, for given $\mathbf a_0$ and $\mathbf x$.
% We write $f_{x,\ell}(d,\mathbf a_0,\mathbf x)$ and $h_{x,\ell}(d,\mathbf a_0,\mathbf x)$ as functions of $\mathbf a_0$ and $\mathbf x$, as $\mathbf \Sigma$ is a function of $\mathbf a_0$ and $\mathbf x$.
% Note that $h_{x,\ell}(d,\mathbf a_0,\mathbf x)$ is the derivative function of $f_{x,\ell}(d,\mathbf a_0,\mathbf x)$ with respect to $d$.
% % Denote $\mathcal A_{i}(\mathbf a,\mathbf x)\triangleq \{d\geq-a_i:h_{a,i}(d,\mathbf a,\mathbf x)=0\}$ as the set of roots of equation $h_{a,i}(d,\mathbf a,\mathbf x)=0$ that are no smaller than $-a_i$.
% Denote $\mathcal X_{\ell}(\mathbf a_0,\mathbf x)\triangleq \{d\geq-x_{\ell}:h_{x,\ell}(d,\mathbf a_0,\mathbf x)=0\}$ as the set of roots of equation $h_{x,\ell}(d,\mathbf a_0,\mathbf x)=0$ that are no smaller than $-x_{\ell}$, for given $\mathbf a_0$ and $\mathbf x$.
Based on structural properties of the coordinate descent optimization problems in \eqref{eqn:MAP_a} and \eqref{eqn:MAP_x}, we have the following results.
%In the coordinate descend optimization with respect to $a_i$, we have the following optimization problem:
%\begin{Prob}[Coordinate Descend Optimization with Respect to $a_i$]\label{Prob:MAP_gamma}
%\begin{align}
%\min_{d} &\quad f_{\rm MAL}(\mathbf a + d\mathbf e_i, \mathbf x)\notag\\
%s.t. &\quad  d\geq -a_i.\notag
%\end{align}
%\end{Prob}
%Similarly, in the coordinate descend optimization with respect to $x_\ell$, we have
%\begin{Prob}[Coordinate Descend Optimization with Respect to $x_\ell$]\label{Prob:MAP_x}
%\begin{align}
%\min_{d} &\quad f_{\rm MAP}(\mathbf a ,\mathbf x+ d\mathbf e_\ell)\notag\\
%s.t. &\quad  d\geq -x_\ell.\notag
%\end{align}
%\end{Prob}
%
%Based on the structure properties of coordinate optimization problems, we can derive the closed-form solution to  Problem~\ref{Prob:MAP_x} as follows.
\textcolor{black}{\begin{Thm}[Optimal Solutions of Coordinate Descent Optimizations in~\eqref{eqn:MAP_a} and \eqref{eqn:MAP_x}]\label{Thm:Step_APs}
Given $\mathbf a_0$ and $\mathbf x$ obtained in the previous step, the optimal solution of the coordinate optimization with respect to \textcolor{black}{the increment in}  $a_i$ in~\eqref{eqn:MAP_a} is given by
\begin{align}
% d^*=
\begin{cases}
\min\left\{\max\left\{s_i(\mathbf a_0,\mathbf x),-a_i\right\}, 1-a_i\right\},& C_i\leq 0\\
\arg\min_{d\in\{s_i(\mathbf a_0,\mathbf x),-a_i+1\}} f_{a,i}(d,\mathbf a_0,\mathbf x), & \textcolor{black}{0 < C_i < \frac{\gamma_{i,0}(\mathbf p_i^H\mathbf \Sigma^{-1}\mathbf p_i)^2}{4\mathbf p_i^H\mathbf \Sigma^{-1}\widehat{\mathbf \Sigma}_{\mathbf Y_0}\mathbf \Sigma^{-1}\mathbf p_i}} \\
-a_i+1,& \textcolor{black}{C_i\geq \frac{\gamma_{i,0}(\mathbf p_i^H\mathbf \Sigma^{-1}\mathbf p_i)^2}{4\mathbf p_i^H\mathbf \Sigma^{-1}\widehat{\mathbf \Sigma}_{\mathbf Y_0}\mathbf \Sigma^{-1}\mathbf p_i}}
\end{cases},\label{eqn:d_MAP_a}
\end{align}
where \textcolor{black}{$s_i(\mathbf a_0,\mathbf x) \triangleq \frac{1}{2C_i}\left(1-\sqrt{1-\frac{4C_i\mathbf p_i^H\mathbf \Sigma^{-1}\widehat{\mathbf \Sigma}_{\mathbf Y_0}\mathbf \Sigma^{-1}\mathbf p_i}{\gamma_{i,0}(\mathbf p_i^H\mathbf \Sigma^{-1}\mathbf p_i)^2}}\right)-\frac{1}{\gamma_{i,0}\mathbf p_i^H\mathbf \Sigma^{-1}\mathbf p_i}$} and $C_i\triangleq \frac{1}{M}\sum_{\omega\in\Psi_0:i\in\omega}\big( c_{\omega}\\ \times\prod_{i^{'}\in\omega,i^{'}\neq i}a_{i^{'}}\big)$, %$\Delta_i \triangleq 1-\frac{4C_i\mathbf p_i^H\mathbf \Sigma^{-1}\widehat{\mathbf \Sigma}_{\mathbf Y_0}\mathbf \Sigma^{-1}\mathbf p_i}{\gamma_{i}(\mathbf p_i^H\mathbf \Sigma^{-1}\mathbf p_i)^2}$,
and the optimal solution of the coordinate optimization with respect to \textcolor{black}{the increment in} $x_{\ell}$ in~\eqref{eqn:MAP_x} is given by
% \vspace{-4mm}
\begin{align}
\mathop{\arg\min}\limits_{d\in\mathcal X_{\ell}(\mathbf a_0,\mathbf x)\cup\{-x_{\ell}\}}
f_{x,\ell}(d,\mathbf a_0,\mathbf x).\label{eqn:d_MAP_x}
\end{align}
\end{Thm}}
\begin{IEEEproof}
\textcolor{black}{Please refer to Appendix C.}
\end{IEEEproof}

% The roots of equation $h_{x,\ell}(d,\mathbf a_0,\mathbf x)=0$ can be obtained by solving a cubic equation with one variable, which has closed-form solutions.
The roots of equation $h_{x,\ell}(d,\mathbf a_0,\mathbf x)=0$ can be obtained \textcolor{black}{in closed form} by solving
a cubic equation with one variable.
Thus, the coordinate descent optimizations can be efficiently solved.
From Theorem~\ref{Thm:Step_APs}, we can see that in the coordinate descent optimizations, prior information on $\mathbf a_0$ and $\mathbf x$ affects the updates of $a_i$, $i\in\Phi_0$ and $x_{\ell}$, $\ell\in\mathcal L$, respectively.
By Theorem~\ref{Thm:Step_APs}, we obtain the \textcolor{black}{closed-form} optimal solution of the coordinate optimization with respect to $a_i$ in~\eqref{eqn:d_MAP_a} in the i.i.d. case where $a_i$, $i\in\Phi_0$ are i.i.d. with $\Pr(a_i=1)=p_a$.

\begin{Cor}[Optimal Solutions of Coordinate Descent Optimizations in~\eqref{eqn:MAP_a} in i.i.d. Case]\label{Cor:opt_a_iid_no_coop}
Given $\mathbf a_0$ and $\mathbf x$ obtained in the previous step, the optimal solution of the coordinate optimization with respect to $a_i$ in~\eqref{eqn:MAP_a} is given by
\small{\begin{align}
\min\Bigg\{&\max\Bigg\{\frac{M}{2\log(\frac{p_a}{1-p_a})}\Bigg(1-\sqrt{1-\frac{\frac{4}{M}\log(\frac{p_a}{1-p_a})\mathbf p_i^H\mathbf \Sigma^{-1}\widehat{\mathbf \Sigma}_{\mathbf Y_0}\mathbf \Sigma^{-1}\mathbf p_i}{\gamma_{i,0}(\mathbf p_i^H\mathbf \Sigma^{-1}\mathbf p_i)^2}}\Bigg)-\frac{1}{\gamma_{i,0}\mathbf p_i^H\mathbf \Sigma^{-1}\mathbf p_i},-a_i\Bigg\},%\notag\\&\quad
 1-a_i\Bigg\}.\label{eqn:d_MAP_a_iid}
\end{align}}\normalsize
\end{Cor}
% \vspace{-4mm}
\begin{IEEEproof}
Please refer to Appendix D.
\end{IEEEproof}

From Corollary~\ref{Cor:opt_a_iid_no_coop}, we can see that as $M\to \infty$ or $p_a\to 0.5$, % $\frac{1}{M}\log(\frac{p_a}{1-p_a})\to 0$,
the optimal solution in \eqref{eqn:d_MAP_a_iid} reduces to the optimal solution  in \eqref{eqn:d_ML_a}. In addition, as $p_a\to 0$, %$\frac{M}{2\log(\frac{p_a}{1-p_a})}\left(1-\sqrt{1-\frac{\frac{4\gamma_{i,0}}{M}\log(\frac{p_a}{1-p_a})\mathbf p_i^H\mathbf \Sigma^{-1}\widehat{\mathbf \Sigma}_{\mathbf Y_0}\mathbf \Sigma^{-1}\mathbf p_i}{(\gamma_{i,0}\mathbf p_i^H\mathbf \Sigma^{-1}\mathbf p_i)^2}}\right)\to 0$, and
the optimal solution in \eqref{eqn:d_MAP_a_iid} becomes $-a_i$, and hence \textcolor{black}{the updated} $a_i$ converges to $0$.
The details of the coordinate descent algorithm for solving Problem~\ref{Prob:MAP} are also summarized in Algorithm~\ref{alg:ML_descend}.
% Note that
\textcolor{black}{As shown in Table~\ref{tab:computation_complexity},} the computational complexities for solving the coordinate optimizations in \eqref{eqn:MAP_a} and \eqref{eqn:MAP_x} per iteration %are $O(N_02^{N_0}+N_0L^2+L^3)$, and
are higher than those for solving the coordinate optimizations in \eqref{eqn:ML_a} and \eqref{eqn:ML_x}, as the objective functions incorporating the prior distributions of $\mathbf a_0$ and $\mathbf x$ are more complex. %in the worst case, where an $N_0$-order interaction exists in $\mathbf a_0$.
In the group activity cases given by the first and second instances in Section~\ref{Sec:prior_dist}, the computational complexities \textcolor{black}{for solving the coordinate optimizations in \eqref{eqn:MAP_a} and \eqref{eqn:MAP_x}} per iteration are $O(N_0L^2+L^3)$ and $O(\sum_{k\in\mathcal K}|\mathcal G_k|2^{|\mathcal G_k|}+N_0L^2+L^3)$, respectively.
%In the group activity case given by the second instance in Section~\ref{Sec:prior_dist}, the computational complexity per iteration is $O(N_0L^2+L^3+\sum_{j=0}^6|\mathcal N_j|^22^{|\mathcal N_j|})$. %when $|\mathcal N_k|=\frac{N_0}{K}$ for $k\in\mathcal K$.
In addition, as $\mathbf a_0$ is a sparse vector, the actual  computational complexity is much lower. % than that in the worst-case.
As $f_{\rm MAP}(\mathbf a_0,\mathbf x)$ is continuously differentiable, we know that
Algorithm~\ref{alg:ML_descend} for the \textcolor{black}{joint} MAP estimation converges to a stationary point of Problem~\ref{Prob:MAP} under a mild condition that each of the coordinate optimizations in \eqref{eqn:MAP_a} and \eqref{eqn:MAP_x} has a unique optimal solution~\cite[Proposition 2.7.1]{Bertsekas99}.

\vspace{-1mm}
\section{Cooperative Device Activity Detection}\label{sec:signal_APs_detection}
\vspace{-1mm}

In this section, we consider cooperative device activity detection, where given $\overline{\mathbf P}_0$ and $\overline{\bm \gamma}_{j}$, $j\in\{0,1,\cdots,6\}$, %are assumed to be perfectly known at the typical AP,
AP \textcolor{black}{$0$} detects the activities of the devices in $\overline{\Phi}_0$
% the typical AP performs device activity detection
from $\overline{\mathbf Y}_0$. %in order to improve the estimation accuracy of its associated devices in $\Phi_0$.
Then,
% from perspective of the typical AP, %the received signal at AP $j\in\{0,1,\cdots,6\}$
$\mathbf Y_j$ can be rewritten as
% \vspace{-1mm}
\begin{align}
&\mathbf  Y_j = \overline{\mathbf P}_0\overline{\mathbf A}_0\overline{\mathbf  \Gamma}_{j}^{\frac{1}{2}}\overline{\mathbf H}_{j}^T+\sum_{i\in\mathcal I\setminus \overline{\Phi}_0}a_i\gamma_{i,j}^{\frac{1}{2}}\mathbf p_i \mathbf h_{i,j}^T+\mathbf Z_j, \quad j\in\{0,1,\cdots,6\},\label{eqn:receive_signal_cooper}
\end{align}
% \vspace{-1mm}
where $\overline{\mathbf A}_0\triangleq{\rm diag}(\overline{\mathbf a}_0)$ with $\overline{\mathbf a}_0\triangleq (a_i)_{i\in\overline{\Phi}_0}$, %representing the activity vector of devices in $\overline{\Phi}_0$,
$\overline{\mathbf \Gamma}_{j}\triangleq{\rm diag}(\overline{\bm\gamma}_{j})$, %with $\bm\gamma_{j}\triangleq (\gamma_{i,j})_{i\in\overline{\Phi}_0}$ representing the path loss vector between devices in $\overline{\Phi}_0$ and AP $j$,
\textcolor{black}{and} $\overline{\mathbf H}_{j}\triangleq (\mathbf h_{i,j})_{i\in\overline{\Phi}_0}\in\mathbb C^{M\times \overline{N}_0}$.
Note that the first term in \eqref{eqn:receive_signal_cooper} is the received signal from the devices in $\overline{\Phi}_0$ and the second term is the received \textcolor{black}{inter-cell interference} %\footnote{\textcolor{black}{In cooperative device activity detection, inter-cell interference is referred to as the interference generated by devices in $\overline{\Phi}_0$.}}
from the other devices. \textcolor{black}{By comparing \eqref{eqn:receive_signal_cooper} with \eqref{eqn:receive_signal}, we see that cooperative \textcolor{black}{device} activity detection deals with less interference than non-cooperative \textcolor{black}{device} activity detection.}

% \subsection{Likelihood Function and Posterior Distribution}

% We consider the ideal case and $\overline{\mathbf Y}_0 = [\mathbf Y_0,\mathbf Y_1,\cdots,\mathbf Y_6]$.
Let $\mathbf y_{j,m}$ denote the $m$-th column of $\mathbf Y_j$.
 % It is known that
Under Rayleigh fading and AWGN, $\mathbf Y_j$, $j\in\{0,1,\cdots,6\}$ are independent, and for all $j\in\{0,1,\cdots,6\}$, $\mathbf y_{j,m}$, $m\in\{1,\cdots, M\}$ are i.i.d. according to $\mathcal{CN}(\mathbf 0,\overline{\mathbf P}_0\overline{\mathbf A}_0\overline{\bm \Gamma}_{j}\overline{\mathbf P}_0^H+\sum_{i\in \mathcal I\setminus\overline{\Phi}_0}a_i\gamma_{i,j}\mathbf p_i\mathbf p_i^H+\delta^2\mathbf I_L)$. %\textcolor{black}{where $\mathbf{\widetilde{X}_j}\triangleq $}.
% Due to the lack of information on the covariance matrix of interference (i.e., $\sum_{i\in \Phi\setminus\overline{\Phi}_0}a_i\gamma_{i,j}\mathbf p_i\mathbf p_i^H$) at the typical AP, we cannot accurately characterize the distribution of $\mathbf y_{j,m}$.
% As in Section~\ref{sec:signal_AP_detection},
Similarly, for tractability, we approximate
% $\sum_{i\in \mathcal I\setminus\overline{\Phi}_0}a_i\gamma_{i,j}\mathbf p_i\mathbf p_i^H$
$\sum_{i\in \mathcal I\setminus\overline{\Phi}_0}a_i\gamma_{i,j}\mathbf p_i\mathbf p_i^H$ with $\mathbf X_j \triangleq  {\rm diag}(\mathbf x_j)$, where $\mathbf x_j\triangleq (x_{j,\ell})_{\ell\in\mathcal L}\in [0,\infty)^L$.
Note that given extra information on $\mathbf P_j$, $j\in\{1,2,\cdots,6\}$, we approximate fewer terms of the covariance of $\mathbf y_{0,m}$, $m\in\{1,2,\cdots,M\}$ than in Section~\ref{sec:signal_AP_detection}.
Under the approximation, the distribution of $\mathbf y_{j,m}$ is \textcolor{black}{approximated} by
\begin{align}
&\mathbf y_{j,m}\sim \mathcal{CN}(\mathbf 0,\overline{\mathbf P}_0\overline{\mathbf A}_0\overline{\bm \Gamma}_{j}\overline{\mathbf P}_0^H+\mathbf X_j +\delta^2\mathbf I_L),\quad j\in\{0,1,\cdots,6\} ,m\in\{1,2,\cdots,M\}.\label{eqn:distribution_yj_cooper}
\end{align}
% where
% \begin{align}
% \mathbf \Sigma_j\triangleq .\notag
% \end{align}
Based on \eqref{eqn:distribution_yj_cooper} and the fact that $\mathbf y_{j,m}$, $m\in\{1,2,\cdots, M\}$ are i.i.d.,
the likelihood of $\mathbf Y_j$ is given by
\begin{align}
& \textcolor{black}{\bar{f}_{j,\overline{\mathbf a}_0,\mathbf x_j}(\mathbf Y_j)}
\propto \frac{\exp\left(-{\rm tr}\left(\left(\overline{\mathbf P}_0\overline{\mathbf A}_0\overline{\bm \Gamma}_{j}\overline{\mathbf P}_0^H+\mathbf X_j+\delta^2\mathbf I_L\right)^{-1}\mathbf Y_j\mathbf Y_j^H\right)\right)}{\vert(\overline{\mathbf P}_0\overline{\mathbf A}_0\overline{\bm \Gamma}_{j}\overline{\mathbf P}_0^H+\mathbf X_j+\delta^2\mathbf I_L)\vert^M},\quad j\in\{0,1,\cdots,6\}.\notag
\end{align}
% From \eqref{eqn:likelihood_coop_j}, we can see that $\overline{\mathbf a}_0$ appears in the expression of $\bar{f}_j(\mathbf Y_j|\overline{\mathbf a}_0,\mathbf x_j)$, $j\in\{0,1,\cdots, 6\}$, and $\mathbf x_j$ only appears in the expression of $\bar{f}(\mathbf Y_j|\overline{\mathbf a}_0,\mathbf x_j)$.
As $\mathbf Y_j$, $j\in\{0,1,\cdots, 6\}$ are independent, the likelihood of $\overline{\mathbf Y}_0$ is given by
% $p\left(\overline{\mathbf Y}_0|\overline{\mathbf a}_0,\overline{\mathbf x}_0\right)=\prod_{j=0}^6p\left(\mathbf Y_j|\overline{\mathbf a}_0,\mathbf x_j\right)$, where
\begin{align}
\textcolor{black}{\bar{f}_{\overline{\mathbf a}_0,\overline{\mathbf x}_0}(\overline{\mathbf Y}_0)} &= \prod_{j=0}^6 \bar{f}_{j,\overline{\mathbf a}_0,\mathbf x_j}(\mathbf Y_j)
\propto \frac{\exp\left(-\sum_{j=0}^6{\rm tr}\left(\left(\overline{\mathbf P}_0\overline{\mathbf A}_0\overline{\bm \Gamma}_{j}\overline{\mathbf P}_0^H+\mathbf X_j+\delta^2\mathbf I_L\right)^{-1}\mathbf Y_j\mathbf Y_j^H\right)\right)}{\prod_{j=0}^6\vert(\overline{\mathbf P}_0\overline{\mathbf A}_0\overline{\bm \Gamma}_{j}\overline{\mathbf P}_0^H+\mathbf X_j+\delta^2\mathbf I_L)\vert^M},\label{eqn:likelihood_coop}
\end{align}
where $\overline{\mathbf x}_0\triangleq [\mathbf x_0^T,\cdots,\mathbf x_6^T]^T$.
Based on~\eqref{eqn:likelihood_coop}, in the following, we consider the joint ML estimation and joint MAP estimation of $\overline{N}_0$ device activities $\overline {\mathbf a}_0$ and $7L$  interference powers $\overline{\mathbf x}_0$, respectively.

\vspace{-1mm}
\subsection{Joint ML Estimation of Device Activities and Interference Powers}

In this part, we
% consider the case where the interference power $\mathbf x$ is deterministic and unknown and make
perform the joint ML estimation of  $\overline{\mathbf a}_0$ and $\overline{\mathbf x}_0$ under AP cooperation.
The maximization of \textcolor{black}{the likelihood $\bar{f}_{\overline{\mathbf a}_0,\overline{\mathbf x}_0}(\overline{\mathbf Y}_0)$} %$\bar{f}(\overline{\mathbf Y}_0|\overline{\mathbf a}_0,\overline{\mathbf x}_0)$
is equivalent to the minimization of the negative log-likelihood $-\log \textcolor{black}{\bar{f}_{\overline{\mathbf a}_0,\overline{\mathbf x}_0}(\overline{\mathbf Y}_0)} \propto \overline{f}_{\rm ML}(\overline{\mathbf a}_0, \overline{\mathbf x}_0)\triangleq \sum_{j=0}^6\overline{f}_{{\rm ML},j}(\overline{\mathbf a}_0, \mathbf x_j)$, %Define
%The log-likelihood of $\mathbf Y_0$ given $\mathbf a$ and $\mathbf x$ is
where %$-\log \bar{f}(\overline{\mathbf Y}_0|\overline{\mathbf a}_0,\overline{\mathbf x}_0)$
\begin{align}
% &-\log \bar{f}(\mathbf Y_j|\overline{\mathbf a}_0,\mathbf x_j)\propto\notag\\
%&=-\frac{1}{M} \log\left(p(\mathbf Y_0|\mathbf a,\mathbf x)\right)\notag\\
% &-\log \bar{f}(\overline{\mathbf Y}_0|\overline{\mathbf a}_0,\overline{\mathbf x}_0)=-\sum_{j=0}^6\log f(\mathbf Y_j|\overline{\mathbf a}_0,\mathbf x_j)\notag\\
&\overline{f}_{{\rm ML},j}(\overline{\mathbf a}_0, \mathbf x_j) \triangleq \log\vert(\overline{\mathbf P}_0\overline{\mathbf A}_0\overline{\bm \Gamma}_{j}\overline{\mathbf P}_0^H+\mathbf X_j+\delta^2\mathbf I_L)\vert+{\rm tr}\left(\left(\overline{\mathbf P}_0\overline{\mathbf A}_0\overline{\bm \Gamma}_{j}\overline{\mathbf P}_0^H+\mathbf X_j+\delta^2\mathbf I_L\right)^{-1}\widehat{\mathbf \Sigma}_{\mathbf Y_j}\right) \notag
\end{align}
with $\widehat{\mathbf \Sigma}_{\mathbf Y_j} = \frac{1}{M}\mathbf Y_j\mathbf Y_j^H$. Note that $\overline{f}_{{\rm ML},j}(\overline{\mathbf a}_0, \mathbf x_j)$ corresponds to the negative log-likelihood function of $\mathbf Y_j$. %and the negative log-likelihood function of $\overline{\mathbf Y}_0$ equals to the sum of the negative log-likelihood function of $\mathbf Y_j$, $j\in\{0,1,\cdots, 6\}$. %We can see that $\overline{\mathbf a}_0$ appears in the expression of $\overline{f}_{\rm ML,j}(\overline{\mathbf a}_0, \mathbf x_j)$, $j\in\{0,1,\cdots, 6\}$, and $\mathbf x_j$ only appears in the expression of $\overline{f}_{\rm ML,j}(\overline{\mathbf a}_0, \mathbf x_j)$.
% where %$h(\bm \gamma,\mathbf x)\triangleq \log|\mathbf P\mathbf \Gamma\mathbf P^H+\mathbf X+\sigma^2\mathbf I_L|$, $q(\bm \gamma,\mathbf x)={\rm tr}((\mathbf P\mathbf \Gamma\mathbf P^H+\mathbf X +\sigma^2\mathbf I_L)^{-1}\widehat{\mathbf \Sigma}_{\mathbf Y})$,
% $\widehat{\mathbf \Sigma}_{\mathbf Y_j} \triangleq \frac{1}{M}\mathbf Y_j\mathbf Y_j^H$.
% Therefore, the joint ML estimate of $\mathbf a$ and $\mathbf x$ can be found by solving the following optimization problem
By omitting the constant term in the negative log-likelihood function, the joint ML estimation of $\overline{\mathbf a}_0$ and $\overline{\mathbf x}_0$ with AP cooperation \textcolor{black}{is formulated as follows.}
\begin{Prob}[Joint ML Estimation with AP Cooperation]\label{Prob:ML_coop}
% \vspace{-1mm}
\begin{align}
\min_{\overline{\mathbf a}_0, \overline{\mathbf x}_0} &\quad \overline{f}_{\rm ML}(\overline{\mathbf a}_0, \overline{\mathbf x}_0)\notag\\
s.t. &\quad  1\geq a_i\geq 0,\quad i\in \overline{\Phi}_0,\label{eqn:coop_cstrt_a}\\
&\quad x_{j,\ell}\geq 0,\quad j\in\{0,1,\cdots, 6\},\ \ell\in\mathcal L.\label{eqn:coop_cstrt_x}
\end{align}
Let $(\overline{\mathbf a}_0^*, \overline{\mathbf x}_0^*)$ denote an optimal solution of Problem~\ref{Prob:ML_coop}.
\end{Prob}

Different from the ML estimation in \cite{Caire18ISIT} and \cite{Yu19ICC}, %which only focuses on estimating $\mathbf a_0$ in a single-cell network without interference, %activities of the associated devices in $\Phi_0$,
Problem~\ref{Prob:ML_coop} considers the joint ML estimation of $\overline{\mathbf a}_0$ and $\overline{\mathbf x}_0$ in the presence of \textcolor{black}{inter-cell interference} and under AP cooperation.
Compared with Problem~\ref{Prob:ML}, Problem~\ref{Prob:ML_coop} makes use of $\overline{f}_{{\rm ML},j}(\overline{\mathbf a}_0, \mathbf x_j)$, $j\in\{1,2,\cdots,6\}$ in the joint ML estimation under AP cooperation and hence is likely to provide device activity detection with higher accuracy. %estimates the activity of devices in a larger device set $\overline{\Phi}_0$.
% In addition, different from Problem~\ref{Prob:ML} in non-cooperative device activity detection, %which uses the received signal $\mathbf Y_0$ at the typical AP,
% Problem~\ref{Prob:ML_coop} That is, in the estimation of activity of devices in $\Phi_0$, the typical AP reconstructs the signals of devices in $\overline{\Phi}_0\setminus \Phi_0$ instead of treating them as interference.
% utilizes the received signal $\mathbf Y_0$ at the typical AP and the received signals $\mathbf Y_j$, $j\in\{1,2,\cdots, 6\}$ at the six neighbor APs in the cooperative device activity detection. %at estimates the activity of devices associated with the typical AP and the devices connected to its six neighbor APs. %That is, in the estimation of activity of devices in $\Phi_0$, the typical AP reconstructs the signal of devices associated with its six neighbor APs, instead of treating them as interference.
Similarly, we can see that $\overline{f}_{\rm ML}(\overline{\mathbf a}_0, \overline{\mathbf x}_0)$ is a DC function, and Problem~\ref{Prob:ML_coop} is a DC programming problem.
We adopt the coordinate descent method to obtain a stationary point of Problem~\ref{Prob:ML_coop}.
Specifically, given $\overline{\mathbf a}_0$ and $\overline{\mathbf x}_0$ obtained in the previous step, the coordinate descent optimization with respect to $a_i$ is equivalent to the optimization of the increment $d$ in $a_i$:
\begin{align}
\min_{1-a_i\geq d\geq -a_i}\ \overline{f}_{\rm ML}(\overline{\mathbf a}_0+d\mathbf e_i, \overline{\mathbf x}_0),\label{eqn:ML_a_coop}
\end{align}
and the coordinate descent optimization with respect to $x_{j,\ell}$ is equivalent to the optimization of the increment $d$ in $x_{j,\ell}$:
\begin{align}
\min_{d\geq -x_{j,\ell}}\ \overline{f}_{{\rm ML},j}(\overline{\mathbf a}_0, \mathbf x_j+d\mathbf e_{\ell}).\label{eqn:ML_x_coop}
\end{align}
Define
\begin{align}
&\overline{f}_{a,i}(d,\overline{\mathbf a}_0, \overline{\mathbf x}_0)\triangleq\sum_{j=0}^6\left(\log(1+d\gamma_{i,j}\mathbf p_i^H\mathbf \Sigma_j^{-1}\mathbf p_i)-\frac{d\gamma_{i,j}\mathbf p_i^H\mathbf \Sigma_j^{-1}\widehat{\mathbf \Sigma}_{\mathbf Y_j}\mathbf \Sigma_j^{-1}\mathbf p_i}{1+d\gamma_{i,j}\mathbf p_i^H\mathbf \Sigma_j^{-1}\mathbf p_i}\right),\notag\\
% \end{align}
% \begin{align}
% &f_{x,j,\ell}(d,\overline{\mathbf a}_0,\mathbf x_j)\triangleq \log(1+d\mathbf e_\ell^H\overline{\mathbf \Sigma}_j^{-1}\mathbf e_\ell)-\frac{d\mathbf e_\ell^H\overline{\mathbf \Sigma}_j^{-1}\widehat{\mathbf \Sigma}_{\mathbf Y_j}\overline{\mathbf \Sigma}_j^{-1}\mathbf e_\ell}{1+d\mathbf e_\ell^H\overline{\mathbf \Sigma}_j^{-1}\mathbf e_\ell}\notag\\
&\overline{h}_{a,i}(d,\overline{\mathbf a}_0, \overline{\mathbf x}_0)\triangleq \sum_{j=0}^6\left(\frac{\gamma_{i,j}\mathbf p_i^H\mathbf \Sigma_j^{-1}\mathbf p_i}{1+d\gamma_{i,j}\mathbf p_i^H\mathbf \Sigma_j^{-1}\mathbf p_i}-\frac{\gamma_{i,j}\mathbf p_i^H\mathbf \Sigma_j^{-1}\widehat{\mathbf \Sigma}_{\mathbf Y_j}\mathbf \Sigma_j^{-1}\mathbf p_i}{(1+d\gamma_{i,j}\mathbf p_i^H\mathbf \Sigma_j^{-1}\mathbf p_i)^2}\right),\notag
% &h_{x,j,\ell}(d,\overline{\mathbf a}_0,\mathbf x_j)\triangleq \frac{\mathbf e_\ell^H\mathbf \Sigma^{-1}\mathbf e_\ell}{1+d\mathbf e_\ell^H\mathbf \Sigma^{-1}\mathbf e_\ell}-\frac{\mathbf e_\ell^H\mathbf \Sigma^{-1}\widehat{\mathbf \Sigma}_{\mathbf Y_0}\mathbf \Sigma^{-1}\mathbf e_\ell}{(1+d\mathbf e_\ell^H\mathbf \Sigma^{-1}\mathbf e_\ell)^2}.\notag
\end{align}
where $\mathbf \Sigma_j\triangleq \overline{\mathbf P}_0\overline{\mathbf A}_0\overline{\bm \Gamma}_{j}\overline{\mathbf P}_0^H+\mathbf X_j +\delta^2\mathbf I_L$ is determined by $\overline{\mathbf a}_0$ and $\mathbf x_j$.
Note that $\overline{h}_{a,i}(d,\overline{\mathbf a}_0, \overline{\mathbf x}_0)$ is the derivative function of $\overline{f}_{a,i}(d,\overline{\mathbf a}_0, \overline{\mathbf x}_0)$ with respect to $d$. Denote $\overline{\mathcal A}_{i}(\overline{\mathbf a}_0, \overline{\mathbf x}_0)\triangleq \{d\in[-a_i,1-a_i] :\overline{h}_{a,i}(d,\overline{\mathbf a}_0, \overline{\mathbf x}_0)=0\}$ as the set of roots of equation $\overline{h}_{a,i}(d,\overline{\mathbf a}_0, \overline{\mathbf x}_0)=0$ that lie in $[-a_i,1-a_i]$, for given $\overline{\mathbf a}_0$ and $\overline{\mathbf x}_0$.
Based on structural properties of the coordinate descent optimization problems in \eqref{eqn:ML_a_coop} and \eqref{eqn:ML_x_coop}, we have the following results.

\begin{Thm}[Optimal Solutions of Coordinate Descent Optimizations in~\eqref{eqn:ML_a_coop} and \eqref{eqn:ML_x_coop}]\label{Thm:Step_one_AP_coop}
Given $\overline{\mathbf a}_0$ and $\overline{\mathbf x}_0$ obtained from in the previous step,
the optimal solution of the coordinate optimization with respect to \textcolor{black}{the increment in} $a_i$ in~\eqref{eqn:ML_a_coop} is given by
\begin{align}
\mathop{\arg\min}\limits_{d\in\overline{\mathcal A}_{i}(\overline{\mathbf a}_0,\overline{\mathbf x}_0)\cup\{-a_i,1-a_i\}}
\overline{f}_{a,i}(d,\overline{\mathbf a}_0, \overline{\mathbf x}_0),\label{eqn:d_ML_a_coop}
\end{align}
and the optimal solution of the coordinate optimization with respect to \textcolor{black}{the increment in} $x_{j,\ell}$ in \eqref{eqn:ML_x_coop} is given by
\begin{align}
\max\left\{\frac{\mathbf e_{\ell}^H\mathbf \Sigma_j^{-1}\widehat{\mathbf \Sigma}_{\mathbf Y_j}\mathbf \Sigma_j^{-1}\mathbf e_{\ell}-\mathbf e_{\ell}^H\mathbf \Sigma_j^{-1}\mathbf e_{\ell}}{(\mathbf e_{\ell}^H\mathbf \Sigma_j^{-1}\mathbf e_{\ell})^2},-x_{j,\ell}\right\}.\label{eqn:d_ML_x_coop}
\end{align}
\end{Thm}
% \begin{IEEEproof}
% Please refer to Appendix B.
% \end{IEEEproof}

The roots of equation $\overline{h}_{a,i}(d,\overline{\mathbf a}_0, \overline{\mathbf x}_0)=0$ can be obtained by solving \textcolor{black}{a univariate polynomial equation of degree\footnote{Note that the degree of the univariate polynomial equation is $2u-1$, where $u$ denotes the number of cooperative APs.}
% Note that when the received signals from $k$ neighbor APs are transmitted to and exploited at the typical AP, the degree of the univariate polynomial equation to be solved is $2k-1$.
13} using mathematical tools, e.g., MATLAB, and hence $\overline{\mathcal A}_{i}(\overline{\mathbf a}_0, \overline{\mathbf x}_0)$ can be easily obtained. Note that the closed-form optimal solution in \eqref{eqn:d_ML_x_coop}  is analogous to that in \eqref{eqn:d_ML_x}. %Compared with the closed-form optimal solution for $x_{\ell}$ in \eqref{eqn:d_ML_x} for the joint ML estimation under non-cooperative device activity detection, the closed-form solution for $x_{j,\ell}$ in \eqref{eqn:d_ML_x_coop} for the joint ML estimation under cooperative device activity detection can be obtained by replacing $\mathbf \Sigma$ and $\widehat{\mathbf \Sigma}_{\mathbf Y_0}$ with $\mathbf \Sigma_j$ and $\widehat{\mathbf \Sigma}_{\mathbf Y_j}$ in \eqref{eqn:d_ML_x}, respectively.
The details of the coordinate descent algorithm for solving Problem~\ref{Prob:ML_coop} are summarized in Algorithm~\ref{alg:ML_descend_coop}.
% Note that
\textcolor{black}{As shown in Table~\ref{tab:computation_complexity},} the computational complexities for solving the coordinate optimizations in \eqref{eqn:ML_a_coop} and \eqref{eqn:ML_x_coop} per iteration %are $O(\overline{N}_0L^2+L^3)$, and
are higher than those for solving the coordinate optimizations in~\eqref{eqn:ML_a} and \eqref{eqn:ML_x} due to the detection of more devices.
Similarly, as $\overline{f}_{{\rm ML}}(\overline{\mathbf a}_0, \overline{\mathbf x}_0)$ is  continuously differentiable and each coordinate optimization in~\eqref{eqn:ML_x_coop} has a unique optimal solution, we can obtain a stationary point of Problem~\ref{Prob:ML_coop} by Algorithm~\ref{alg:ML_descend_coop} for the \textcolor{black}{joint} ML estimation under a mild condition that each coordinate optimization in~\eqref{eqn:ML_a_coop} has a unique optimal solution~\cite[Proposition 2.7.1]{Bertsekas99}.

% \vspace{-2mm}
\begin{algorithm} \caption{Coordinate Descent Algorithm with AP Cooperation}
\small{\begin{algorithmic}[1]
\STATE Initialize $\mathbf \Sigma_j^{-1}=\frac{1}{\delta^2} \mathbf I_L$ for $j\in\{0,1,\cdots, 6\}$, $\overline{\mathbf a}_0=\mathbf 0$, $\overline{\mathbf x}_0=\mathbf 0$.
% \FOR {$n=1,2,\cdots$}
\STATE \textbf{repeat}
\FOR {$i\in\overline{\Phi}_0$}
% \IF {$i\leq N$}
\STATE {\textbf ML:} Calculate $d$ according to \eqref{eqn:d_ML_a_coop}.
\STATE {\textbf MAP:} Calculate $d$ according to \eqref{eqn:d_MAP_a_coop}.
\STATE Update $a_{i}=a_{i}+d$ and $\mathbf \Sigma_j^{-1} = \mathbf \Sigma_j^{-1}-\frac{d\gamma_{i,j}\mathbf \Sigma_j^{-1}\mathbf p_{i}\mathbf p_{i}^H\mathbf \Sigma_j^{-1}}{1+d\gamma_{i,j}\mathbf p_{i}^H\mathbf \Sigma_j^{-1}\mathbf p_{i}}$ for all $j\in\{0,1,\cdots, 6\}$.
% \ELSE
\ENDFOR
\FOR {$j=0$ to $6$}
\FOR {$\ell\in\mathcal L$}
% \STATE $\ell = i-N$.
\STATE {\textbf ML:} Calculate $d$ according to \eqref{eqn:d_ML_x_coop}.
\STATE {\textbf MAP:} Calculate $d$ according to \eqref{eqn:d_MAP_x_coop}.
\STATE Update $x_{j,\ell}=x_{j,\ell}+d$ and $\mathbf \Sigma_j^{-1} = \mathbf \Sigma_j^{-1}-\frac{d\mathbf \Sigma_j^{-1}\mathbf e_{\ell}\mathbf e_{\ell}^H\mathbf \Sigma_j^{-1}}{1+d\mathbf e_{\ell}^H\mathbf \Sigma_j^{-1}\mathbf e_{\ell}}$.
% \ENDIF
\ENDFOR
\ENDFOR
% \ENDFOR
\STATE \textbf{until} $(\overline{\mathbf a}_0, \overline{\mathbf x}_0)$ satisfies some stopping criterion.
\end{algorithmic}}\normalsize\label{alg:ML_descend_coop}
\end{algorithm}

% \vspace{-8mm}
\subsection{Joint MAP Estimation of Device Activities and Interference Powers}

In this part, we assume that $\overline{\mathbf a}_0$ and $\overline{\mathbf x}_0$ are random and perform the joint MAP estimation of $\overline{\mathbf a}_0$ and $\overline{\mathbf x}_0$ under AP cooperation.

\subsubsection{Prior Distribution}

For tractability, we assume that $\mathbf a_j$, $\mathbf x_j$, $j\in\{0,1,\cdots,6\}$ are independently distributed. %Note that this is a weak assumption, as it only requires that the device activities in the typical cell are independent of those in the other cells.
% First, we introduce a general prior distribution of the Bernoulli Random vector $\overline{\mathbf a}_0$.
% We allow for correlation among activities of devices in each cell $j\in\{0,1,\cdots,6\}$. In particular, % consider that devices in $\Phi_0$ can access the channel in a more general manner such that coupling effects between $a_i$, $i\in\Phi_0$ are allowed.
%which assume that devices access the channel in an i.i.d. manner,
Similarly, we adopt the MVB model for $\mathbf a_j$~\cite{NIPS2011_4209}. %and assume that device activity in a cell is independent with those in other cells.
% Let $\overline{\Psi}_j$ denote the power set of $\Phi_0$ and denote $\Psi_0=\overline{\Psi}_0\setminus \{\emptyset\}$. Let $\omega\subseteq \Phi_0$ denote an element in $\Psi_0$.
% The MVB model of device activity in $\Phi_0$ is given as follows:
Then, the p.m.f. of $\mathbf a_j$ under the MVB model is given by
\begin{align}
p_j\left(\mathbf a_j\right)=\exp\left(\sum_{\omega\in\Psi_j}\left(c_{\omega}\prod_{i\in\omega}a_i\right)+b_j\right), \quad j\in\{0,1,\cdots,6\},\label{eqn:pmf_a_j}
\end{align}
where $\Psi_j$ denote the set of nonempty subsets of $\Phi_j$, and
$b_j\triangleq \log(\sum_{\mathbf a_j\in\{0,1\}^{N_j}}\exp(\sum_{\omega\in\Psi_j}(c_{\omega}\prod_{i\in\omega}a_i)))$ is the normalization factor.

% Next, we derive a prior distribution of $\overline{\mathbf x}_0$.
Under cooperative device activity detection, the locations of the active interfering devices in $\mathcal I\setminus\overline{\Phi}_0$ are assumed to follow a homogeneous PPP with density $\lambda$.
As pilot sequences are generated from i.i.d. \textcolor{black}{$\mathcal{CN}(\mathbf 0,\mathbf I_L)$}, the diagonal entries of $\sum_{i\in \mathcal I\setminus\overline{\Phi}_0}a_i\gamma_{i,j}\mathbf p_i\mathbf p_i^H$ are i.i.d.. Similarly, we assume that $x_{j,\ell}$, $\ell\in\mathcal L$ are i.i.d. with the same distribution as $\sum_{i\in\mathcal I\setminus\overline{\Phi}_0} a_i\gamma_{i,j}$. Therefore, $x_{j,\ell}$ is a power-law shot noise, whose exact p.d.f. is still not known and can be approximated with a Gaussian distribution~\cite{Aljuaid10TVT}. %denoted as $g_j(x_{j,\ell})$.
Based on the above assumptions and techniques from stochastic geometry, we have the following results.
\begin{Lem}[Approximated Distribution of  $\mathbf x_j$]\label{Lem:distribution_X_coop}
% Using Gaussian distribution to approximate the distribution of $x_{j,\ell}$,
The p.d.f. of $\mathbf x_j$ is approximated by
\begin{align}
g_j(\mathbf x_j) = \frac{1}{(\sqrt{2\pi}\sigma_j)^L}\exp\left(-\frac{\sum_{\ell\in\mathcal L}(x_{j,\ell}-\mu_j)^2}{2\sigma_j^2}\right),\quad j\in\{0,1,\cdots,6\},\notag
\end{align}
where %$\mu_j$ and $\sigma_j$ is given in
% \begin{figure*}
\begin{align}
&\mu_0 =12\lambda \int_{\frac{\sqrt{3}}{2}R}^{\infty}
\int_{0}^{\frac{\sqrt{3}}{3}x} (x^2+y^2)^{-\frac{\alpha}{2}}{\rm d }y{\rm d}x-12\lambda \int_{\frac{\sqrt{3}R}{2}}^{\frac{3\sqrt{3}R}{2}}\int_0^{U_0(x)}(x^2+y^2)^{-\frac{\alpha}{2}}{\rm d }y{\rm d}x,\notag\\
% \end{align}
% \begin{align}
&\mu_j =\frac{\mu_0}{2}+6\lambda \int_{\frac{\sqrt{3}}{2}R}^{\infty}
\int_{0}^{\frac{\sqrt{3}}{3}x} (x^2+y^2)^{-\frac{\alpha}{2}}{\rm d }y{\rm d}x-2\lambda \int_{\sqrt{3}R}^{\frac{5\sqrt{3}R}{2}}\int_{U_0(x)}^{U_1(x)}
(x^2+y^2)^{-\frac{\alpha}{2}}{\rm d }y{\rm d}x,\notag\\
&\hspace{100mm} j\in\{1,2,\cdots, 6\},\notag\\
% \end{align}
% \begin{align}
&\sigma_0^2 =12\lambda \int_{\frac{\sqrt{3}}{2}R}^{\infty}
\int_{0}^{\frac{\sqrt{3}}{3}x} (x^2+y^2)^{-\alpha}{\rm d }y{\rm d}x-12\lambda \int_{\frac{\sqrt{3}R}{2}}^{\frac{3\sqrt{3}R}{2}}\int_0^{U_0(x)}(x^2+y^2)^{-{\alpha}}{\rm d }y{\rm d}x,\notag\\
% \end{align}
% \begin{align}
&\sigma_j^2 =\frac{\sigma_0^2}{2}+6\lambda \int_{\frac{\sqrt{3}}{2}R}^{\infty}
\int_{0}^{\frac{\sqrt{3}}{3}x} (x^2+y^2)^{-\alpha}{\rm d }y{\rm d}x-2\lambda \int_{\sqrt{3}R}^{\frac{5\sqrt{3}R}{2}}\int_{U_0(x)}^{U_1(x)}
(x^2+y^2)^{-{\alpha}}{\rm d }y{\rm d}x,\notag\\
&\hspace{100mm} j\in\{1,2,\cdots, 6\},\notag\\
% \end{align}
% \begin{align}
&U_0(x) =\begin{cases}
\frac{\sqrt{3}}{3}x,   & \frac{\sqrt{3}}{2}R\leq  x < \sqrt{3}R\\
-\frac{\sqrt{3}}{3}x+2R,  & \sqrt{3}R\leq  x \leq \frac{3\sqrt{3}}{2}R\\
0, & \frac{3\sqrt{3}}{2}R \leq x
\end{cases}, \quad%\notag\\
% \end{align}
% \begin{align}
U_1(x) =\begin{cases}
\frac{\sqrt{3}}{3}x+R,   &   \sqrt{3}R\leq  x < \frac{3\sqrt{3}}{2}R\\
-\frac{\sqrt{3}}{3}x+4R,  & \frac{3\sqrt{3}}{2}R\leq  x \leq 2\sqrt{3}R\\
-\frac{\sqrt{3}}{3}x+3R, & 2\sqrt{3}R\leq  x < \frac{5\sqrt{3}}{2}R
\end{cases}. \notag
\end{align}
% \end{figure*}
\end{Lem}
\begin{IEEEproof}
\textcolor{black}{Lemma~\ref{Lem:distribution_X_coop} can be proved in a similar way to Lemma~\ref{Lem:distribution_X}. We omit the details due to page limitation.}
\end{IEEEproof}

\begin{figure}[t]
\begin{center}
\subfigure[\small{Typical AP}]
{\resizebox{7cm}{!}{\includegraphics{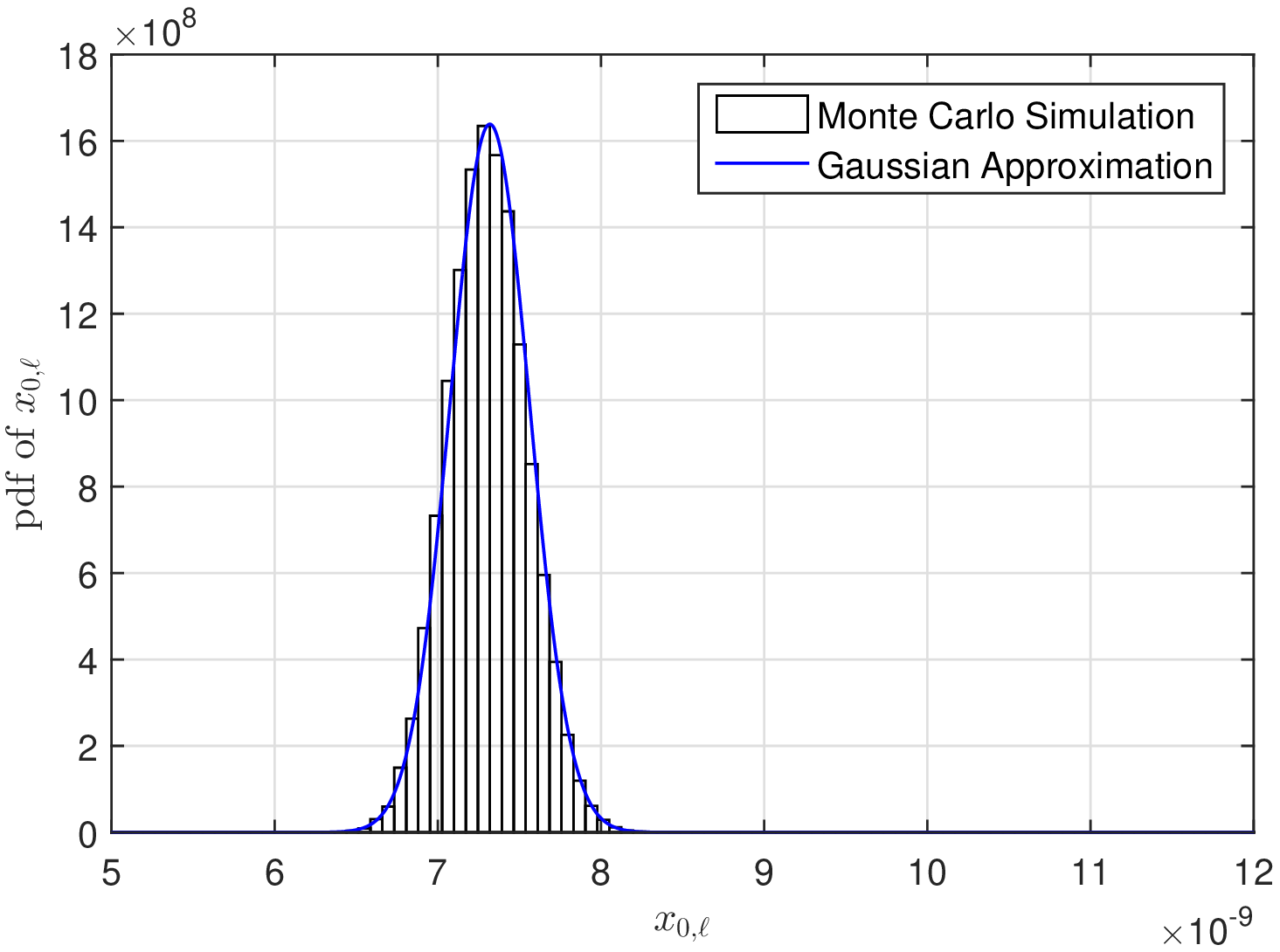}}}\quad\quad
\subfigure[\small{AP $1$}]
{\resizebox{7cm}{!}{\includegraphics{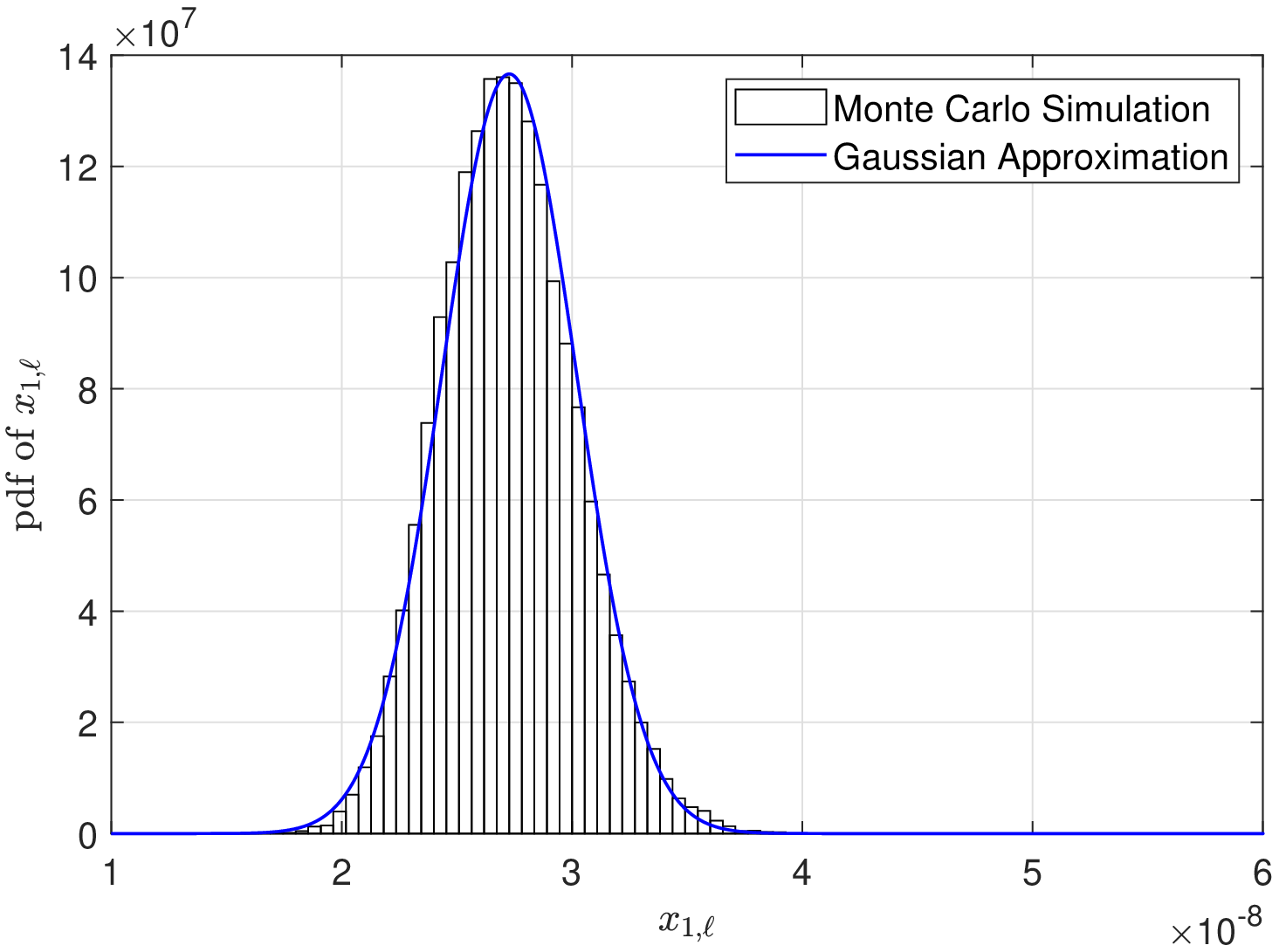}}}
\end{center}
\vspace{-2mm}
\caption{\small{Comparison between the p.d.f. of $x_{j,\ell}$ and its corresponding Gaussian approximation. $R=200$, $\lambda=0.0005$ and $\alpha=4$.}}
\vspace{-2mm}
\label{fig:gamma_approx_coop}
\end{figure}

Fig.~\ref{fig:gamma_approx_coop} plots the histogram of the $x_{j,\ell}$ \textcolor{black}{(}which reflects the shape of the p.d.f. of $x_{j,\ell}$\textcolor{black}{) and} the Gaussian distributions with the same mean and variance. %\textcolor{black}{and the distribution of the diagonal elements of $\mathbf {\widetilde{X}_j}$}.
From Fig.~\ref{fig:gamma_approx_coop}, we can see that the Gaussian distribution is a good approximation of the exact p.d.f. of $x_{j,\ell}$ under the considered simulation setup, which verifies Lemma~\ref{Lem:distribution_X_coop}.  %\textcolor{black}{In addition, the p.d.f. of $x_{j,\ell}$ can well reflect the trend of distribution of the diagonal elements of $\mathbf {\widetilde{X}_j}$}.

\subsubsection{\textcolor{black}{Joint} MAP Estimation}
% Next, we derive the joint posterior distribution of $\overline{\mathbf a}_0$ and $\overline{\mathbf x}_0$, given $\overline{\mathbf Y}_0$.
% Recall that components of $\overline{\mathbf a}_0$ are i.i.d. Bernoulli random variables and
% \begin{align}
% p(\mathbf a)=\exp\left(\log\frac{p_a}{1-p_a}\sum_{i\in\overline{\Phi}_0}a_i +\overline{N}_0\log(1-p_a)\right).\notag
% \end{align}
Based on \textcolor{black}{the conditional distribution of $\overline{\mathbf Y}_0$ given $\overline{\mathbf a}_0$ and $\overline{\mathbf x}_0$} and the distributions of $\mathbf a_j$ and $\mathbf x_j$, $j\in\{0,1,\cdots,6\}$, \textcolor{black}{the conditional joint density of %joint posterior distribution
of $\overline{\mathbf a}_0$ and $\overline{\mathbf x}_0$, given $\overline{\mathbf Y}_0$, is given by}
\begin{align}
&\textcolor{black}{\overline{f}_{\overline{\mathbf a}_0,\overline{\mathbf x}_0|\overline{\mathbf Y}_0}\left(\overline{\mathbf a}_0,\overline{\mathbf x}_0,\overline{\mathbf Y}_0\right)
\propto \overline{f}_{\overline{\mathbf a}_0,\overline{\mathbf x}_0}\left(\overline{\mathbf Y}_0\right)} \left(\prod_{j=0}^6 p_j(\mathbf a_j)\right)\left(\prod_{j=0}^6g_j(\mathbf x_j)\right)\notag\\
% \end{align}
% \begin{align}
&\propto \frac{\exp\left(-\sum\limits_{j=0}^6{\rm tr}\left(\left(\overline{\mathbf P}_0\overline{\mathbf A}_0\overline{\bm \Gamma}_{j}\overline{\mathbf P}_0^H+\mathbf X_j+\delta^2\mathbf I_L\right)^{-1}\mathbf Y_j\mathbf Y_j^H\right)
\right)}{\prod_{j=0}^6\vert(\overline{\mathbf P}_0\overline{\mathbf A}_0\overline{\bm \Gamma}_{j}\overline{\mathbf P}_0^H+\mathbf X_j+\delta^2\mathbf I_L)\vert^M}\exp\left(
-\sum\limits_{j=0}^6\sum\limits_{\ell=1}^L\frac{(x_{j,\ell}-\mu_j)^2}{2\sigma_j^2}
\right)\notag
\end{align}
\begin{align}
&\quad \times \exp\left(\sum_{j=0}^6\sum_{\omega\in\Psi_j}\left(c_{\omega}\prod_{i\in\omega}a_i\right)\right).\notag
\end{align}
% In this part, we assume that $\overline{\mathbf a}_0$ and $\overline{\mathbf x}_0$ are random and their distributions are known to the typical AP. In this case,
% % For activity detection with AP cooperation, AP $0$ performs activity detection of its associated devices based on $\overline{\mathbf Y}_0$. We consider the case where the interference power is random and
% we perform the joint MAP estimation of $\overline{\mathbf a}_0$ and $\overline{\mathbf x}_0$.
%The ML estimator of $\overline{\mathbf a}_0$ after simplification is
The maximization of \textcolor{black}{the conditional joint density $\overline{f}_{\overline{\mathbf a}_0,\overline{\mathbf x}_0|\overline{\mathbf Y}_0}\left(\overline{\mathbf a}_0,\overline{\mathbf x}_0,\overline{\mathbf Y}_0\right)$} is equivalent to the minimization of the negative logarithm of \textcolor{black}{the conditional joint density} $-\log \textcolor{black}{\overline{f}_{\overline{\mathbf a}_0,\overline{\mathbf x}_0|\overline{\mathbf Y}_0}\left(\overline{\mathbf a}_0,\overline{\mathbf x}_0,\overline{\mathbf Y}_0\right)}\propto \overline{f}_{\rm MAP}(\overline{\mathbf a}_0,\overline{\mathbf x}_0)$, where
% $\overline{f}_{\rm MAP}(\overline{\mathbf a}_0,\overline{\mathbf x}_0) $, where
\begin{align}
% &-\log p\left(\overline{\mathbf a}_0,\overline{\mathbf x}_0|\overline{\mathbf Y}_0\right)\notag\\
&\overline{f}_{\rm MAP}(\overline{\mathbf a}_0,\overline{\mathbf x}_0) \triangleq \overline{f}_{\rm ML}(\overline{\mathbf a}_0,\overline{\mathbf x}_0)+\frac{1}{M}\sum_{j=0}^6\sum_{\ell=1}^L\frac{(x_{j,\ell}-\mu_j)^2}{2\sigma_j^2}-\frac{1}{M}\sum_{j=0}^6\sum_{\omega\in\Psi_j}\left(c_{\omega}\prod\limits_{i\in\omega}a_i\right).\notag
\end{align}
Note that $\frac{1}{M}\sum_{j=0}^6\sum_{\ell=1}^L\frac{(x_{j,\ell}-\mu_j)^2}{2\sigma_j^2}$ is from the p.d.f. of $\overline{\mathbf x}_0$, and $-\frac{1}{M}\sum_{j=0}^6\sum_{\omega\in\Psi_j}(c_{\omega}\prod_{i\in\omega}a_i)$ is from the p.m.f. of $\overline{\mathbf a}_0$.
% Here, $\overline{f}_{\rm ML}(\overline{\mathbf a}_0,\mathbf x_j)$ is the negative log-likelihood of $\mathbf Y_j$ under activity detection with AP cooperation.
The joint MAP estimate of $\overline{\mathbf a}_0$ and $\overline{\mathbf x}_0$ with AP cooperation can be formulated as follows.
\begin{Prob}[Joint MAP Estimation with AP Cooperation]\label{Prob:MAP_coop}
% \vspace{-2mm}
\begin{align}
\min_{\overline{\mathbf a}_0,\overline{\mathbf x}_0} &\quad \overline{f}_{\rm MAP}(\overline{\mathbf a}_0,\overline{\mathbf x}_0)\notag\\
s.t. &\quad  \eqref{eqn:coop_cstrt_a}, \eqref{eqn:coop_cstrt_x}.\notag%1\geq a_i\geq 0,\quad i\in\overline{\Phi}_0,\notag\\
% &\quad x_{j,\ell}\geq 0,\quad j\in\{0,1\cdots,6\}, \ell\in \mathcal L.\notag
\end{align}
\vspace{-2mm}
Let $(\overline{\mathbf a}_0^{\dagger},\overline{\mathbf x}_0^\dagger)$ denote an optimal solution of Problem~\ref{Prob:MAP_coop}.
\end{Prob}

% As $\frac{1}{M}\sum_{j=0}^6\sum_{\ell=1}^L\frac{(x_{j,\ell}-\mu_j)^2}{2\sigma_j^2}$ is a convex function of $\overline{\mathbf x}_0$, and $-\frac{1}{M}\log \frac{p_a}{1-p_a}\sum_{i\in\overline{\Phi}_0}a_i$ is a liner function of $\mathbf a$,

By comparing $\overline{f}_{\rm MAP}(\overline{\mathbf a}_0,\overline{\mathbf x}_0)$ with $\overline{f}_{\rm ML}(\overline{\mathbf a}_0,\overline{\mathbf x}_0)$, we can draw the following conclusions. The \textcolor{black}{incorporating} of prior distribution $g_j(\mathbf x_j)$ pushes the estimate of $x_{j,\ell}$ towards its mean $\mu_j$ for all $\ell\in\mathcal L$. \textcolor{black}{Incorporating} prior distribution $p_j(\mathbf a_j)$ pushes the estimate of $a_i$, $i\in\Phi_j$ to \textcolor{black}{the activity states} with high probabilities.
As $\overline{f}_{\rm MAP}(\overline{\mathbf a}_0,\overline{\mathbf x}_0)-\overline{f}_{\rm ML}(\overline{\mathbf a}_0,\overline{\mathbf x}_0)$ decreases with $M$, the impacts of the prior
distributions of $\overline{\mathbf a}_0$ and $\overline{\mathbf x}_0$ reduce as $M$ increases.
% which diminishes to $0$ as $M\to \infty$. Therefore, we have the following result.
% \begin{Lem}[Property of Problem~\ref{Prob:MAP_coop}]
As $M\to \infty$, $\overline{f}_{\rm MAP}(\overline{\mathbf a}_0,\overline{\mathbf x}_0)\to \overline{f}_{\rm ML}(\overline{\mathbf a}_0,\overline{\mathbf x}_0)$, Problem~\ref{Prob:MAP_coop} reduces to Problem~\ref{Prob:ML_coop}, and $(\overline{\mathbf a}_0^\dagger,\overline{\mathbf x}_0^\dagger)$  becomes $(\overline{\mathbf a}_0^*,\overline{\mathbf x}_0^*)$.
% \end{Lem}

% Similarly, we can see that $\overline{f}_{\rm MAP}(\overline{\mathbf a}_0,\overline{\mathbf x}_0)$ is a DC function and Problem~\ref{Prob:MAP_coop} is a DC programming problem.
% From Problem~\ref{Prob:MAP_cooper}, we can see that the joint MAP estimation of device activity and the interference power is the one that minimizes the sum of the negative log-likelihood of $\mathbf Y_j$, $j\in\{0,1,\cdots,6\}$ minus the logarithm of the prior distribution of $\overline{\mathbf a}_0$ and $\overline{\mathbf x}_0$.
As $\overline{f}_{\rm ML}(\overline{\mathbf a}_0,\overline{\mathbf x}_0)$ is a DC function and $-\frac{1}{M}\sum_{j=0}^6\sum_{\omega\in\Psi_j}\left(c_{\omega}\prod\limits_{i\in\omega}a_i\right)$ is a non-convex function, we can see that %$\overline{f}_{\rm MAP}(\mathbf a,\overline{\mathbf x}_0)$ is a DC function and
Problem~\ref{Prob:MAP_coop} is a challenging non-convex
% DC programming
problem with a complicated objective function.
We adopt the coordinate descent method to obtain a stationary point of Problem~\ref{Prob:MAP_coop}.
Specifically, given $\overline{\mathbf a}_0$ and $\overline{\mathbf x}_0$  obtained in the previous step, the coordinate descent optimization with respect to $a_i$ is  equivalent to the optimization of the increment $d$ in $a_i$:
\begin{align}
\min_{1-a_i\geq  d\geq -a_i} \ \overline{f}_{\rm MAP}(\overline{\mathbf a}_0 + d\mathbf e_i, \overline{\mathbf x}),\label{eqn:MAP_a_coop}
\end{align}
and the coordinate descent optimization with respect to $x_{j,\ell}$ is equivalent to the optimization of the increment $d$ in $x_{j,\ell}$:
\begin{align}
\min_{d\geq -x_{j,\ell}} \ \overline{f}_{{\rm ML},j}(\overline{\mathbf a}_0, \mathbf x_j+d\mathbf e_{\ell})+ \frac{(x_{j,\ell}-\mu_j+d)^2}{2M\sigma_j^2}.\label{eqn:MAP_x_coop}
\end{align}
%The coordinate descend optimization of $\mathbf a_j$ in Problem~\ref{Prob:MAP} is the same as that in Problem~\ref{Prob:ML}. %The coordinate descend optimization of $x_\ell$ is conducted as follows.
Define
\begin{align}
&\widetilde{f}_{a,i}(d,\overline{\mathbf a}_0, \overline{\mathbf x}_0)\triangleq \overline{f}_{a,i}(d,\overline{\mathbf a}_0, \overline{\mathbf x}_0)-\frac{d}{M}\sum_{j=0}^6\sum_{\omega\in\Psi_j:i\in\omega}\left(c_{\omega}\prod_{i^{'}\in\omega,i^{'}\neq i}a_{i^{'}}\right),\notag\\
&\widetilde{f}_{x,j,\ell}(d,\overline{\mathbf a}_0, \overline{\mathbf x}_0)\triangleq \log(1+d\mathbf e_\ell^H\mathbf \Sigma_j^{-1}\mathbf e_\ell)-\frac{d\mathbf e_\ell^H\mathbf \Sigma_j^{-1}\widehat{\mathbf \Sigma}_{\mathbf Y_j}\mathbf \Sigma_j^{-1}\mathbf e_\ell}{1+d\mathbf e_\ell^H\mathbf \Sigma_j^{-1}\mathbf e_\ell} +\frac{(x_{j,\ell}-\mu_j+d)^2}{2M\sigma_j^2},\notag\\
% \end{align}
% \begin{align}
&\widetilde{h}_{a,i}(d,\overline{\mathbf a}_0, \overline{\mathbf x}_0)\triangleq \overline{h}_{a,i}(d,\overline{\mathbf a}_0, \overline{\mathbf x}_0) %-\frac{1}{M}\log \frac{p_a}{1-p_a}
-\frac{1}{M}\sum_{j=0}^6\sum_{\omega\in\Psi_j:i\in\omega}\left(c_{\omega}\prod_{i^{'}\in\omega,i^{'}\neq i}a_{i^{'}}\right),\notag\\
% \end{align}
% \begin{align}
&\widetilde{h}_{x,j,\ell}(d,\overline{\mathbf a}_0, \overline{\mathbf x}_0) \triangleq  \frac{\mathbf e_\ell^H\mathbf \Sigma_j^{-1}\mathbf e_\ell}{1+d\mathbf e_\ell^H\mathbf \Sigma_j^{-1}\mathbf e_\ell}-\frac{\mathbf e_\ell^H\mathbf \Sigma_j^{-1}\widehat{\mathbf \Sigma}_{\mathbf Y_j}\mathbf \Sigma_j^{-1}\mathbf e_\ell}{(1+d\mathbf e_\ell^H\mathbf \Sigma_j^{-1}\mathbf e_\ell)^2} +\frac{d+x_{j,\ell}-\mu_j}{M\sigma_j^2}.\notag
\end{align}
We write $\widetilde{f}_{a,i}(d,\overline{\mathbf a}_0, \overline{\mathbf x}_0)$, $\widetilde{f}_{x,j,\ell}(d,\overline{\mathbf a}_0, \overline{\mathbf x}_0)$, $\widetilde{h}_{a,i}(d,\overline{\mathbf a}_0, \overline{\mathbf x}_0)$, and $\widetilde{h}_{x,j,\ell}(d,\overline{\mathbf a}_0, \overline{\mathbf x}_0)$ as functions of $\overline{\mathbf a}_0$ and $\overline{\mathbf x}_0$, as $\mathbf \Sigma_j$, $j\in\{0,1,\cdots,6\}$ are functions of $\overline{\mathbf a}_0$ and ${\mathbf x}_j$.
Note that $\widetilde{h}_{a,i}(d,\overline{\mathbf a}_0, \overline{\mathbf x}_0)$ and $\widetilde{h}_{x,j,\ell}(d,\overline{\mathbf a}_0, \overline{\mathbf x}_0)$ are the derivative functions of $\widetilde{f}_{a,i}(d,\overline{\mathbf a}_0, \overline{\mathbf x}_0)$ and $\widetilde{f}_{x,j,\ell}(d,\overline{\mathbf a}_0, \overline{\mathbf x}_0)$ with respect to $d$, respectively. Denote $\widetilde{\mathcal A}_{i}(\overline{\mathbf a}_0, \overline{\mathbf x}_0)\triangleq \{d\in[-a_i,1-a_i]:\widetilde{h}_{a,i}(d,\overline{\mathbf a}_0, \overline{\mathbf x}_0)=0\}$ as the set of roots of equation $\widetilde{h}_{a,i}(d,\overline{\mathbf a}_0, \overline{\mathbf x}_0)=0$ that lie in $[-a_i,1-a_i]$, for given $\overline{\mathbf a}_0$ and $\overline{\mathbf x}_0$.
Denote $\widetilde{\mathcal X}_{j,\ell}(\overline{\mathbf a}_0, \overline{\mathbf x}_0)\triangleq \{d>-x_{j,\ell}:\widetilde{h}_{x,j,\ell}(d,\overline{\mathbf a}_0, \overline{\mathbf x}_0) =0\}$ as the set of roots of equation $\widetilde{h}_{x,j,\ell}(d,\overline{\mathbf a}_0, \overline{\mathbf x}_0) =0$ that are no smaller than $-x_{j,\ell}$, for given $\overline{\mathbf a}_0$ and $\overline{\mathbf x}_0$.
Based on the structure properties of the coordinate optimization problems in \eqref{eqn:MAP_a_coop} and \eqref{eqn:MAP_x_coop}, we have the following results.

\begin{Thm}[Optimal Solutions of Coordinate Descent Optimizations in \eqref{eqn:MAP_a_coop} and \eqref{eqn:MAP_x_coop}]\label{Thm:Step_APss}
Given $\overline{\mathbf a}_0$ and $\overline{\mathbf x}_0$  obtained in the previous step, the optimal solution to the coordinate optimization with respect to \textcolor{black}{the increment in} $a_i$ in \eqref{eqn:MAP_a_coop} is given by
\begin{align}
\mathop{\arg\min}\limits_{ d\in\widetilde{\mathcal A}_{i}(\overline{\mathbf a}_0, \overline{\mathbf x}_0)\cup\{-a_i,1-a_i\}}\widetilde{f}_{a,i}(d,\overline{\mathbf a}_0, \overline{\mathbf x}_0),\label{eqn:d_MAP_a_coop}
\end{align}
and the optimal solution to the coordinate optimization with respect to \textcolor{black}{the increment in} $x_{j,\ell}$ in~\eqref{eqn:MAP_x_coop} is given by
\begin{align}
\mathop{\arg\min}\limits_{d\in\widetilde{\mathcal X}_{j,\ell}(\overline{\mathbf a}_0, \overline{\mathbf x}_0)\cup\{-x_{j,\ell}\}}\widetilde{f}_{x,j,\ell}(d,\overline{\mathbf a}_0, \overline{\mathbf x}_0).\label{eqn:d_MAP_x_coop}
\end{align}
\end{Thm}
\begin{IEEEproof}
\textcolor{black}{Theorem~\ref{Thm:Step_APss} can be proved in a similar way to Theorem~\ref{Thm:Step_APs}. The details are omitted due to page limitation.}
\end{IEEEproof}

The roots of equation $\widetilde{h}_{a,i}(d,\overline{\mathbf a}_0, \overline{\mathbf x}_0)=0$ can be obtained by solving \textcolor{black}{a univariate polynomial equation of degree\footnote{Note that the degree of the univariate polynomial equation is $2u$, where $u$ denotes the number of cooperative APs.}  14} using mathematical tools, e.g., MATLAB.
The roots of equation $\widetilde{h}_{x,j,\ell}(d,\overline{\mathbf a}_0, \overline{\mathbf x}_0) =0$ can be obtained by solving
a cubic equation with one variable, which has closed-form solutions. From Theorem~\ref{Thm:Step_APss}, we can see that in the coordinate descent optimizations, prior information on $\overline{\mathbf a}_0$ and $\overline{\mathbf x}_0$ affects the updates of $a_i$, $i\in\overline{\Phi}_0$ and $x_{j,\ell}$, $j\in\{0,1,\cdots,6\}$, $\ell\in\mathcal L$, respectively.
The details of the coordinate descent algorithm for solving Problem~\ref{Prob:MAP_coop} are also summarized in Algorithm~\ref{alg:ML_descend_coop}.
% Note that
\textcolor{black}{As shown in Table~\ref{tab:computation_complexity},} the computational complexities for solving the coordinate optimizations in \eqref{eqn:MAP_a_coop} and \eqref{eqn:MAP_x_coop} per iteration %are $O(\sum_{j=0}^6N_j2^{N_j}+\overline{N}_0L^2+L^3)$, and
are higher than those for solving the coordinate optimizations in \eqref{eqn:ML_a_coop} and \eqref{eqn:ML_x_coop}, as the objective functions incorporating the prior distributions of $\overline{\mathbf a}_0$ and $\overline{\mathbf x}_0$ are more complex. %in the worst case, where an $N_j$-order interaction exists in $\mathbf a_j$. %In the group activity case given by the first instance in Section~\ref{Sec:prior_dist}, the computational complexity per iteration is $O(N_0L^2+L^3)$. In the group activity case given by the second instance in Section~\ref{Sec:prior_dist}, the computational complexity per iteration is $O(N_0L^2+L^3+\frac{N_0^2}{K}2^\frac{N_0}{K})$ when $|\mathcal N_k|=\frac{N_0}{K}$ for $k\in\mathcal K$. In addition, as $\mathbf a_0$ is a sparse vector, the real computational complexity is much lower than that in the worst-case.
Similarly, \textcolor{black}{as $\overline{f}_{\rm MAP}(\overline{\mathbf a}_0,\overline{\mathbf x}_0)$ is continuously differentiable}, we can obtain a stationary point of Problem~\ref{Prob:MAP_coop} by Algorithm~\ref{alg:ML_descend_coop} for the \textcolor{black}{joint} MAP estimation under a mild condition that each of the coordinate optimizations in \eqref{eqn:MAP_a_coop} and \eqref{eqn:MAP_x_coop} has a unique optimal solution~\cite[Proposition 2.7.1]{Bertsekas99}.
% From Theorem~\ref{Thm:Step_APss}, we can see that the negative log-likelihood of $\mathbf Y_j$ affects the update the each entry in $\{a_i$: $i\in\{1,2,\cdots,N_6\}\}$ in the coordinate descent optimization.

\section{Numerical Results}\label{sec:simulation}

In this section, we evaluate the performance of the proposed activity detection designs via numerical results.
We compare the proposed designs with four \textcolor{black}{state-of-the-art designs}, i.e., AMP (non-cooperative) in~\cite{Liu18TSP,Chen18TSP}, ML, and Group-Lasso in~\cite{Caire18ISIT}, %\textcolor{black}{and Group-Lasso in~\cite{Caire18ISIT}},
which do not consider inter-cell interference or AP cooperation, and AMP (cooperative) in~\cite{Chen19TWC} which considers inter-cell interference and AP cooperation. %\footnote{\textcolor{black}{Note that the baseline ML achieves order gains over  Multiple Measurement Vector (MMV) and Non-Negative Least Squares (NNLS) schemes as shown in \cite{Caire18ISIT}, and we do not shown the performance of MMV and NNLS schemes in the simulation.}}
Note that the proposed designs, %ML (non-cooperative), proposed MAP (non-cooperative), proposed ML (cooperative), proposed MAP (cooperative)
ML, \textcolor{black}{and Group-Lasso} are optimization-based device activity detection designs, \textcolor{black}{and the computational complexities of ML and Group-Lasso per iteration are $\mathcal O(N_0L^2)$}. AMP (non-cooperative) and AMP (cooperative) are based on the AMP algorithm with the minimum mean squared error (MMSE) denoiser, \textcolor{black}{and their computational complexities per iteration are $\mathcal O(N_0LM)$}. \textcolor{black}{ML has been recognized as the best design among all existing designs in most cases\textcolor{black}{, and the} AMP-based designs are also treated as very competitive \textcolor{black}{designs} among existing ones.}
% \textcolor{black}{The complexity of each iteration of the AMP algorithms for the non-cooperative case in \cite{Shao19IoTJ,Senel18TCOM,Liu18TSP,Chen18TSP} is $\mathcal O(N_0LM)$. The complexity of each iteration of the coordinate descent algorithm for the ML estimation problem %for the non-cooperative case
% in \cite{Caire18ISIT} is $\mathcal O(N_0L^2)$.
% The complexity of each iteration of the proposed coordinate descent algorithm for the joint ML estimation problem %in the non-cooperative case
% is $\mathcal O(N_0L^2+L^3)$.}

In the simulation, $N_0$ devices are uniformly distributed in  cell \textcolor{black}{$0$,} and the active devices out of cell \textcolor{black}{$0$} are distributed according to a homogeneous PPP with \textcolor{black}{density} $\lambda$.\footnote{\textcolor{black}{Note that under a homogeneous PPP, points are uniformly distributed in any given area, and the number of points in any given area is a random variable and follows a Poisson distribution. Thus, the model for the devices in cell $0$ does not contradict with that for the devices outside of  cell \textcolor{black}{$0$}.}} The number of devices in cell $0$ $N_0$ is deterministic, and the number of active devices in \textcolor{black}{any other cell} is random and has average $\frac{3\sqrt{3}}{2}R^2\lambda$. We treat the devices in cell $0$ and outside of it differently for the purpose of separating the impacts of $N_0$ and the intensity of inter-cell interference. We assume that each device  is active with probability $p_a$. %and consider two cases of random device activities, i.e., the i.i.d. case and correlated case. In the i.i.d. case, the devices are active  in an i.i.d. manner. In the correlated case, $N_0$ devices are divided into $\frac{N_0}{2}=250$ groups each containing two devices; device activities in different groups are independent; and every two devices in a group are correlated with correlation coefficient $\eta$. More specifically, for two devices $i_1$ and $i_2$ in one group, $\Pr(a_{i_1}=1,a_{i_2}=1)=\eta p_a +(1-\eta)p^2$, $\Pr(a_{i_1}=0,a_{i_2}=1)=\Pr(a_{i_1}=1,a_{i_2}=0)=(1-\eta)(p_a-p_a^2)$ and $\Pr(a_{i_1}=0,a_{i_2}=0)=1+(\eta-2)p_a+(1-\eta)p_a^2$.
% In the simulation,  %we consider that the $N$ devices in $\Phi_0$ are uniformly and randomly distributed in the typical cell.
We independently generate $2000$ realizations for the locations of devices, \textcolor{black}{$\mathbf p_i$, $i\in\mathcal I$}, $a_i$, $i\in\mathcal I$, and $\mathbf h_{i,j}$, $i\in\mathcal I$, $j\in\{0,1,\cdots,6\}$, perform device activity detection in each realization, and evaluate the average error probability over all $2000$ realizations.
For the proposed designs, ML, \textcolor{black}{and Group-Lasso}, %each design,
%refers to the activity detection design in~\cite{Caire18ISIT}, which performs ML estimation of $\mathbf a$ and sets $\mathbf x$ to $0$. ``AMP'' refers to the AMP-based activity detection designs in~\cite{Bockelmann13} which solves the random access problem from compressed sensing perspective.
% \subsection{Performance Metric}
let $\hat{a}_i\triangleq \mathbf 1[a_i^{*}\geq \theta]$ denote the estimate of the activity state of device $i$, where $\mathbf 1[\cdot]$ is the indicator function, and $\theta>0$ is a threshold. \textcolor{black}{As $a_i\in[0,1]$, the threshold $\theta$ should be chosen from $[0,1]$. It is obvious that the error probability is $1-p_a$ when $\theta\leq 0$ and is $p_a$ when $\theta\geq 1$. For each of the proposed designs, ML, and Group-Lasso, we evaluate the average error probability for $\theta\in\{0.01,0.02,\cdots,\textcolor{black}{1}\}$ and choose the optimal threshold, i.e., the one achieving the minimum as its average error probability.}\footnote{As the optimal threshold varies with the network parameters and the distributions of device activities and interference powers, it is not reasonable to fix a threshold for all the schemes and compare their performances in different cases. Note that \textcolor{black}{choosing} a threshold based on \textcolor{black}{prior knowledge} are widely considered~\cite{Shao19IoTJ,Senel18TCOM,Chen18TSP}.}
For AMP (non-cooperative) and AMP (cooperative),  let $\hat{a}_i\triangleq \mathbf 1[{\rm LLR}_i\geq 0]$ denote the estimate of the activity state of device $i$ \cite{Liu18TSP}, where ${\rm LLR}_i$ is the log-likelihood ratio for device $i$ and is given by~\textcolor{black}{\cite[Eq. (13)]{Chen19TWC} and \cite[Eq. (15)]{Chen19TWC}} for AMP (non-cooperative) and AMP (cooperative), respectively.
A detection error happens when $\hat{a}_i\neq a_i$.
% There are two types of error detection events: false alarm and missed detection. A false alarm event happens when a device is inactive but declared active. A missed detection event happens when a device is active but declared inactive.
% There are two types of error detection events: a device is inactive but declared active, or a device is active but declared inactive.
% We adopt the error probability as the performance metric, which is defined as follows:
% Therefore, the error probability is defined as
% \begin{align}
% \min_{\theta}\mathbb E_{\Phi,\mathbf h_i,a_i}\left(\frac{\sum_{i\in\Phi_0} \mathbf 1[\hat{a}_i\neq a_i]}{N}\right).\notag
% \end{align}
% The error probability is averaged over the location distributions of devices, channel fading between devices and the AP and device activity patterns.
% We select the threshold $\theta$ such that the average error probability achieves its minimum.
% Note that the selection of $\theta$ can be done in practical applications by historical data.
% The probability of error corresponds to two types of error events: a device is inactive but declared active or a device is active but declared inactive.
% \subsection{Performance Comparison}
In the simulation, %the error probability is obtained by averaging over 2000 trials.
unless otherwise stated, we choose $R=200$, \textcolor{black}{$\lambda=0.00025$}, $p_a=0.05$, $N_0=500$, $\alpha=3$, $L=40$, $M=60$ and $\delta^2=\frac{R^{-\alpha}}{10}$.\footnote{\textcolor{black}{$\lambda=0.00025$ represents that there are on average $50$ devices  in a square of $100$m$\times 100$m.}}

\begin{figure}[t]
\begin{center}
 \includegraphics[width=7cm]{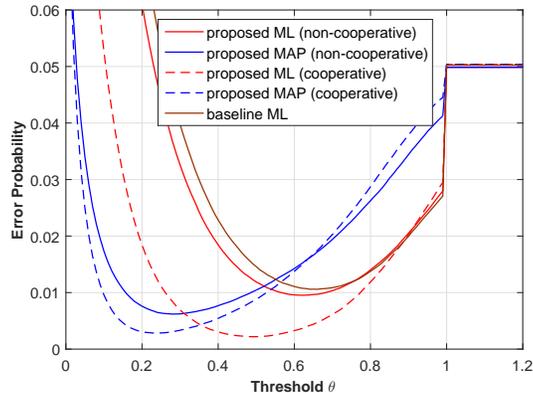}
  \end{center}
  % \vspace{-6mm}
  \caption{\small{\textcolor{black}{Error probability versus threshold $\theta$ in the i.i.d. case.
  % $R=200$, $\lambda=0.005$, $p_a=0.05$, $N_0=500$, $\alpha=3$, $L=40$, $M=60$ and $\delta^2=\frac{R^{-\alpha}}{10}$.
  }}}
\label{fig:error_prob_vs_theta}
\end{figure}

\textcolor{black}{%First, we illustrate the optimal threshold for different schemes.
Fig.~\ref{fig:error_prob_vs_theta} plots the error probabilities  of the proposed ML (non-cooperative), proposed MAP (non-cooperative), proposed ML (cooperative), proposed MAP (cooperative), and
ML versus the threshold $\theta$.
Note that when $\theta$ increases, more devices are detected as inactive.
From Fig.~\ref{fig:error_prob_vs_theta}, we can see that the error probability of each detection design first decreases with $\theta$ due to the decrease of false alarm and then increases with $\theta$ due to the increase of missed detection.
In addition, we observe that the optimal thresholds corresponding to the minimum error probabilities for the MAP-based detection designs are smaller than those for the corresponding ML-based detection designs.
This is because the incorporation of the prior distribution of device activities (e.g., $p_a=0.05$ in the i.i.d. case) makes the estimated \textcolor{black}{activity states} of most devices under the MAP-based detection designs smaller than those under the ML-based detection designs.
% From Fig.~\ref{fig:error_prob_vs_theta}, we can see that the error probability of each detection scheme first decreases with $\theta$ and then increases with $\theta$.
% This is because when $\theta$ increases, more devices are declared as inactive.
% In addition, we observe that the optimal thresholds corresponding to the minimum error probabilities for the MAP-based detection schemes are smaller than those for the ML-based detection schemes.
% This is because in the MAP-based detection schemes, the incorporation of the prior distribution of device activities pushes the estimate of $\mathbf a_0$ or $\overline{\mathbf a}_0$ to the activities with high probabilities.
}

% \begin{figure*}
\begin{figure}[t]
\begin{center}
\subfigure[\small{Length of pilot sequences $L$.}]
{\resizebox{7 cm}{!}{\includegraphics{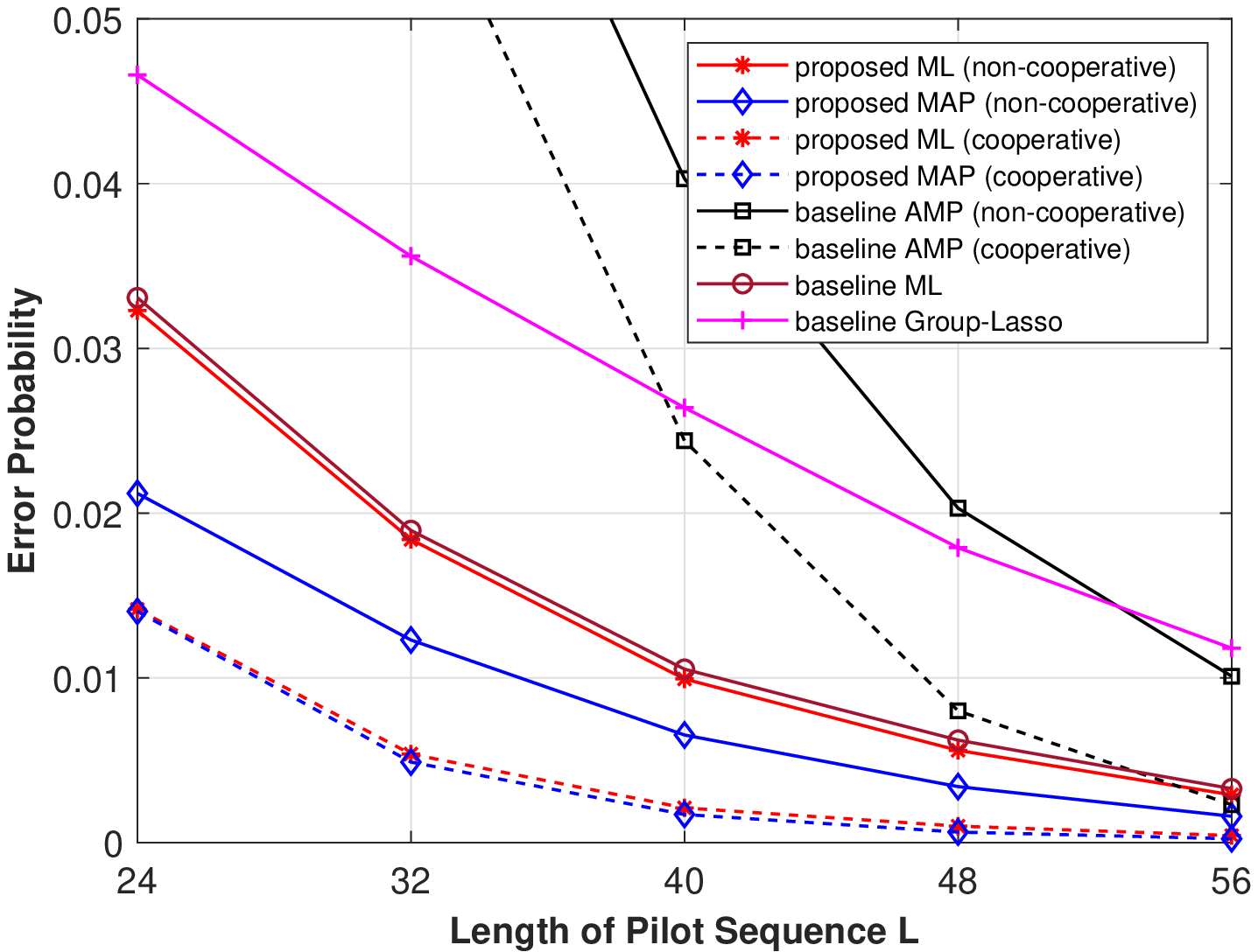}}}\quad\quad
\subfigure[\small{Number of antennas $M$.}]
{\resizebox{7cm}{!}{\includegraphics{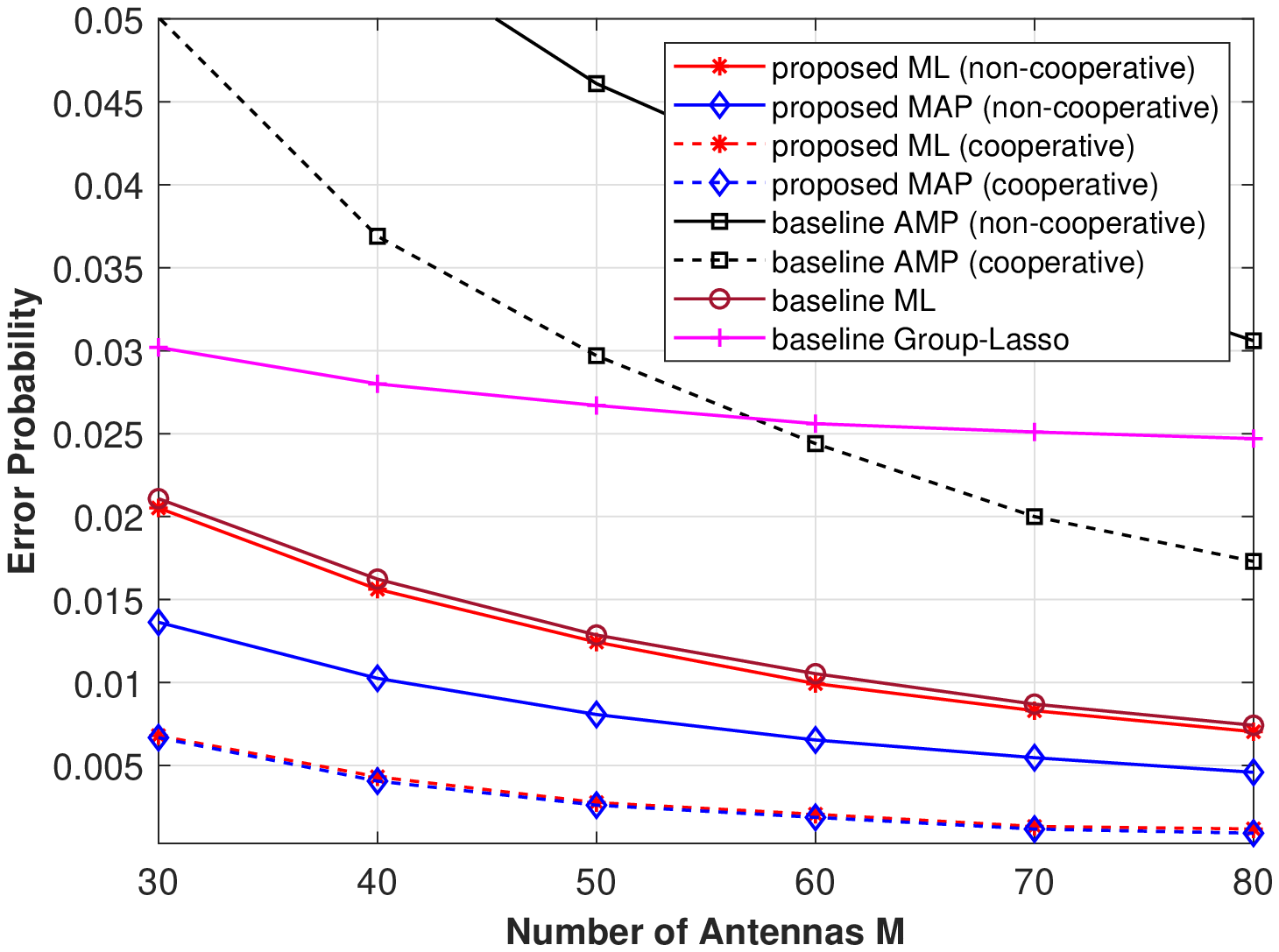}}}
\subfigure[\small{Density of active devices out of typical cell $\lambda$.}]
{\resizebox{7cm}{!}{\includegraphics{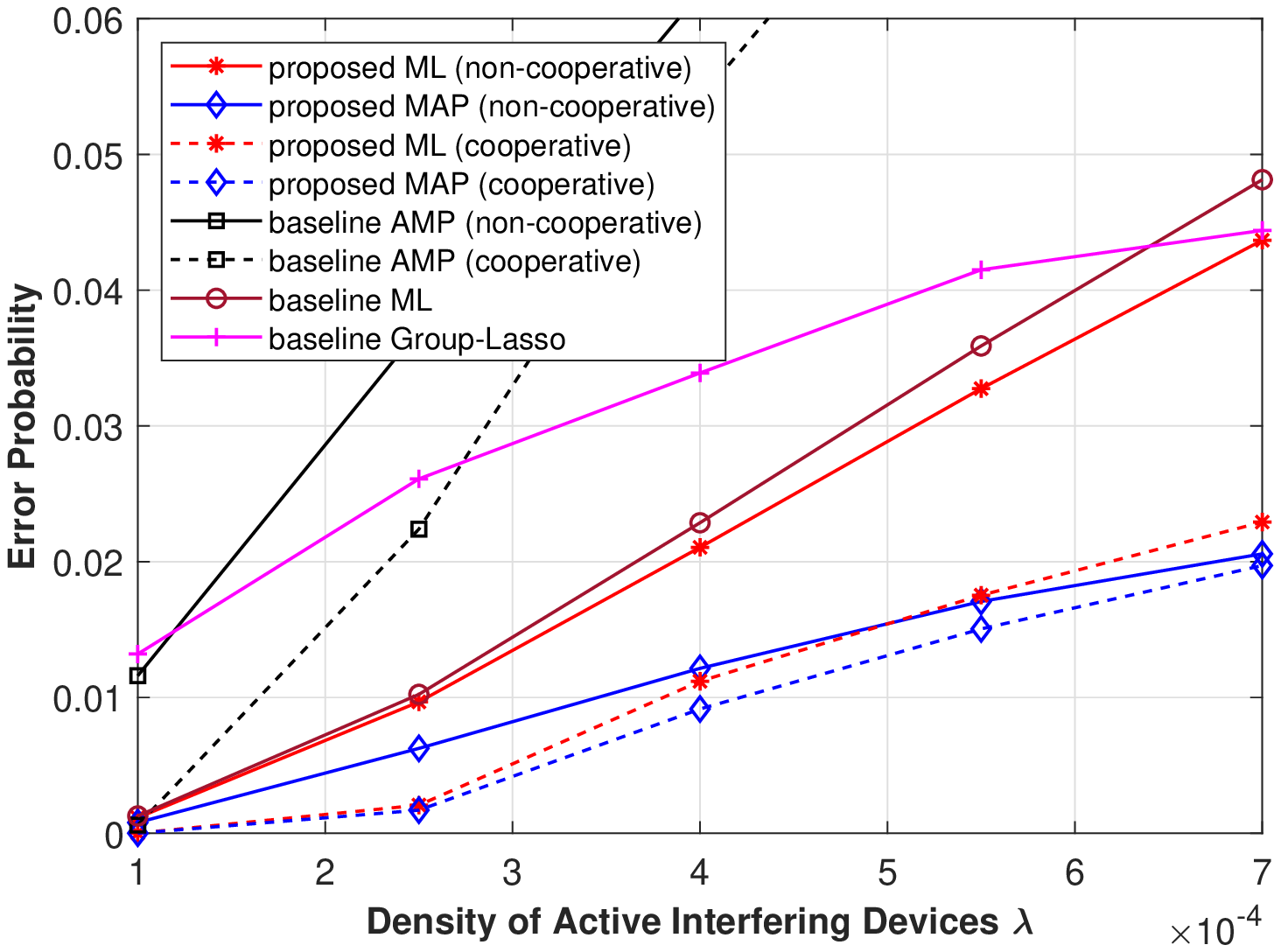}}}%\quad
% \subfigure[\small{Correlation Coefficient $\eta$ in the correlation case.}]
% {\resizebox{8cm}{!}{\includegraphics{Fig_Jan17_iid_vs_perfect_v5.eps}}}
\end{center}
\vspace{-2mm}
\caption{\small{\textcolor{black}{Error probability versus length of pilot sequences $L$, number of antennas $M$ and density of active devices out of typical cell $\lambda$ in the i.i.d. case.}}}
\vspace{-2mm}
\label{fig:performance_comparison}
\end{figure}
% \end{figure*}

Fig.~\ref{fig:performance_comparison} plots the error probability versus the length of pilot sequences $L$, the number of antennas $M$, and the density of active devices outside cell \textcolor{black}{$0$} $\lambda$ \textcolor{black}{in the i.i.d. case, where the devices \textcolor{black}{activate}  in an i.i.d. manner}. From Fig.~\ref{fig:performance_comparison}, we observe that  \textcolor{black}{the proposed designs %the optimization-based designs
% significantly
outperform the AMP-based designs and Group-Lasso \textcolor{black}{(the error probability of the proposed MAP (non-cooperative) is smaller than $\frac{1}{5}$ of that of AMP (non-cooperative), and the error probability of the proposed MAP (cooperative) is about $\frac{1}{10}$ of that of  AMP (cooperative))};} %at the cost of computational complexity increase;
the proposed ML (non-cooperative) outperforms ML, especially in the high interference regime; the proposed MAP (non-cooperative) can reduce the error probability by nearly a half, compared to the proposed ML (non-cooperative); %the proposed ML (cooperative) and the proposed MAP (cooperative) nearly coincide and outperform the proposed ML (non-cooperative) and the proposed MAP (non-cooperative)
the proposed cooperative designs significantly outperform their respective non-cooperative counterparts; the performance of the proposed MAP (cooperative) is similar to that of the proposed ML (cooperative). Note that the performance gain of the proposed ML (non-cooperative) over ML comes from the explicit consideration of \textcolor{black}{inter-cell interference}; the performance gain of the proposed MAP (non-cooperative) over the proposed ML (non-cooperative) derives from the incorporation of prior knowledge of the interference powers and device activities;  the performance gain of each proposed cooperative design over its non-cooperative counterpart is due to the exploitation of more observations from neighbor APs and the utilization of more network parameters; similar performance of the proposed cooperative designs indicates that exploiting prior knowledge of the interference powers and device activities brings a relatively small gain under AP cooperation.
% This is because the proposed designs take the interference into consideration and the proposed joint MAP estimation also considers the prior distribution of interference and device activity pattern.
% Specifically, from Fig.~\ref{fig:performance_comparison}~(a), we can see that the error probability of each design increases with the density of interfering devices, demonstrating the impact of interference in device activity detection. In addition, we can see that the gap between the proposed joint MAP estimation and the proposed joint ML estimation increases with $\lambda$, which shows that the value of prior knowledge of interference increases with its strength. %the proposed joint MAP estimation can achieve significant performance gains when the interference is severe.
%Specifically, from Fig.~\ref{fig:performance_comparison}~(a) and (b),
Specifically, from Fig.~\ref{fig:performance_comparison}~(a) and (b), we observe that the error probability of each design decreases with $L$ and $M$; and %. In addition, we can see that
the gap between the proposed MAP (non-cooperative) and the proposed ML (non-cooperative) increases as $L$ and
$M$ decrease, which highlights the benefit of prior information
at small $L$ and $M$ %shows that the proposed MAP estimation can achieve superior performance
% when the length of pilot sequences and the number of antennas are limited
under non-cooperative device activity detection.
% \begin{figure}[t]
% \begin{center}
%  \includegraphics[width=7cm]{fig_Dec_27_error_vs_lambda.eps}
%   \end{center}
%   \vspace{-2mm}
%   \caption{\small{Error probability versus density of interfering devices $\lambda$.}}
%   \vspace{-4mm}
% \label{fig:performance_comparison_vs_lambda}
% \end{figure}
% Fig.~\ref{fig:performance_comparison} plots the error probability versus the density of interfering devices $\lambda$.
From Fig.~\ref{fig:performance_comparison}~(c), we can see that the error probability of each design increases with $\lambda$,
% the density of interfering devices,
demonstrating the influence of \textcolor{black}{inter-cell interference} in device activity detection. In addition, we can see that the gap between the proposed MAP \textcolor{black}{(non-cooperative)}
and the proposed  ML \textcolor{black}{(non-cooperative)}
increases with $\lambda$, which shows that the value of prior knowledge of the interference powers increases with their strengths under non-cooperative device activity detection. %and the gap between each proposed cooperative design and the respective non-cooperative design increases with $\lambda$, which shows that the value of gathering more observations from the neighbor APs and obtaining more network parameters increases with the interference powers. %the proposed joint MAP estimation can achieve significant performance gains when the interference is severe.

% \begin{figure}[t]
% \begin{center}
%  \includegraphics[width=8cm]{Fig_Jan17_iid_vs_perfect_v5.eps}
%   \end{center}
%   \vspace{-2mm}
%   \caption{\small{Comparison between the p.d.f. of $x_\ell$ and its corresponding Gaussian approximation. $R=196$, $\lambda=0.01$, $p_a=0.05$ and $\alpha=4$.}}
%   \vspace{-4mm}
% \label{fig:aaa}
% \end{figure}

\begin{figure}[t]
\begin{minipage}[t]{0.45\linewidth}
\centering
\includegraphics[width=7 cm]{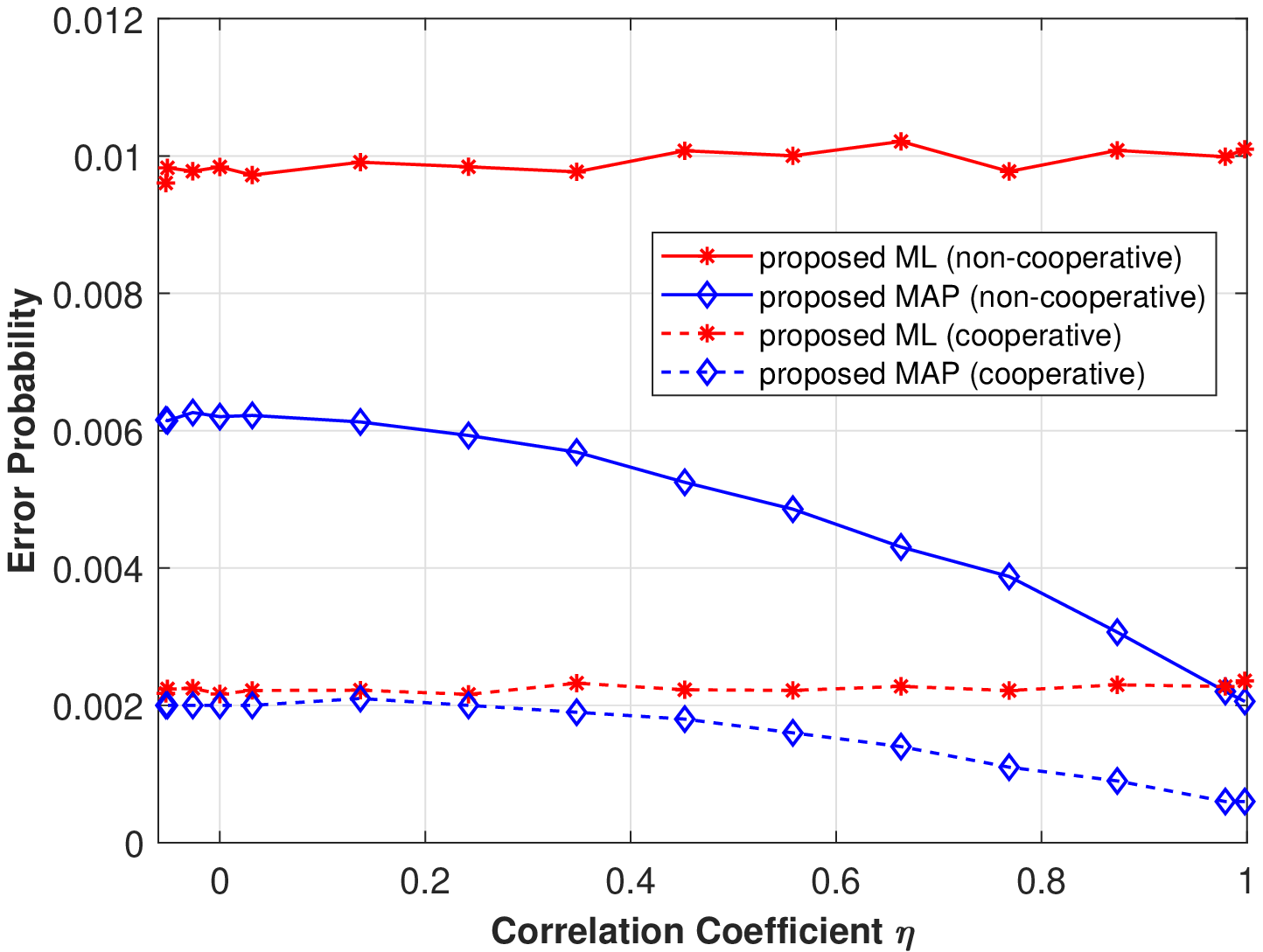}%Fig_Jan17_iid_vs_perfect_v5
\caption{Error probability versus correlation coefficient $\eta$ in the 1st instance in Section \ref{Sec:prior_dist}}
\label{fig:error_vs_eta}
\end{minipage}\quad
\begin{minipage}[t]{0.45\linewidth}
\centering
\includegraphics[width=7 cm]{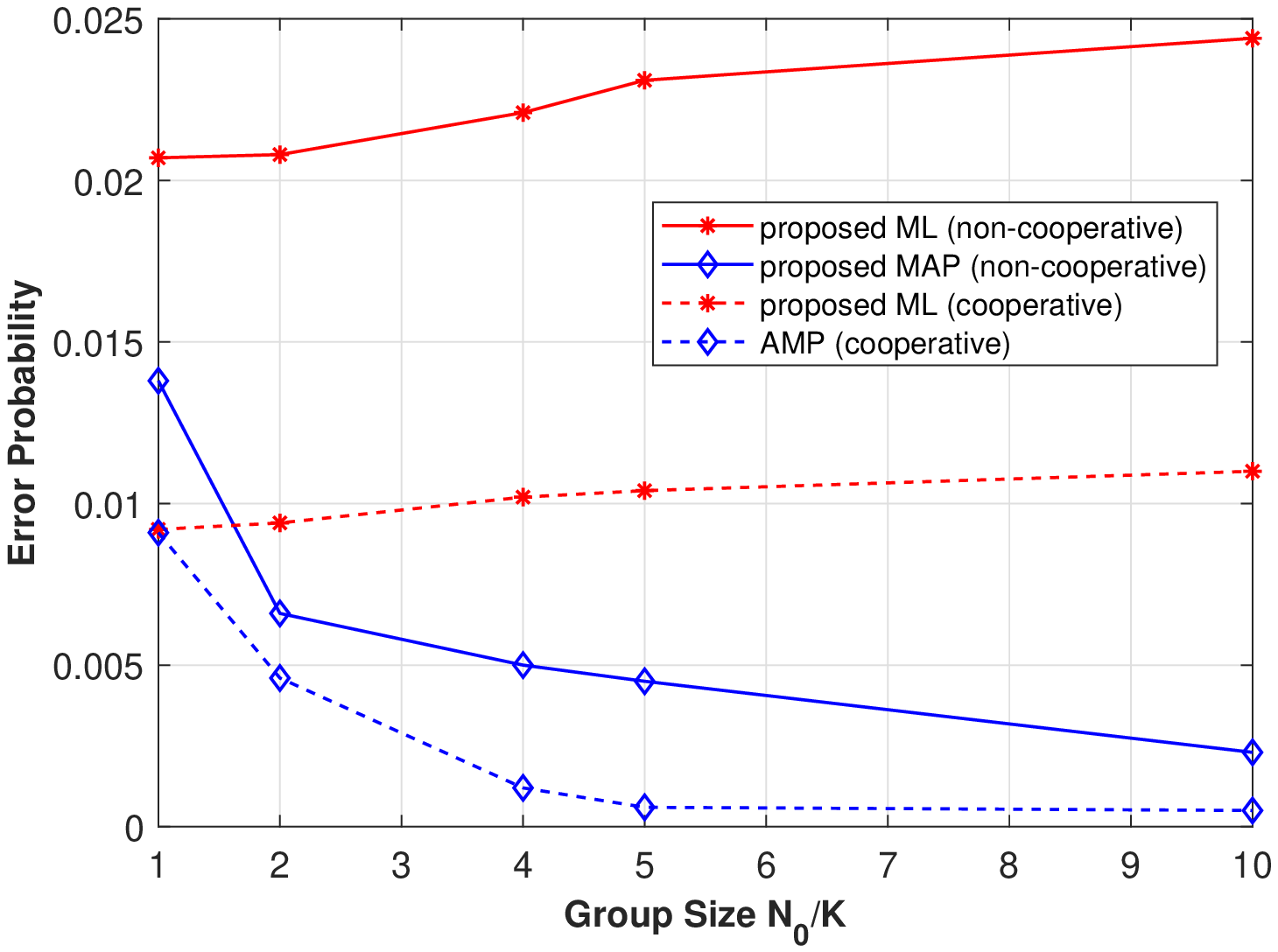}%Fig_Feb15_grp_error_prob_vs_S
\caption{Error probability versus group size $\frac{N_0}{K}$ in the 2nd instance in Section \ref{Sec:prior_dist}. $|\mathcal G_1|=\cdots=|\mathcal G_K|$, $L=30$, and $N_0= N_1=\cdots=N_6$.}
\label{fig:error_vs_group_size}
\end{minipage}
\vspace{-2mm}
\end{figure}

Fig.~\ref{fig:error_vs_eta} plots the error probability \textcolor{black}{of each proposed design} versus the correlation coefficient $\eta$ in the group activity case given by the first instance in Section~\ref{Sec:prior_dist}. As the baseline \textcolor{black}{designs} cannot exploit general sparsity patterns of device activities, we do not show their error probabilities in Fig.~\ref{fig:error_vs_eta} and Fig.~\ref{fig:error_vs_group_size}.
From Fig.~\ref{fig:error_vs_eta}, we can observe that the error probabilities of the proposed MAP (non-cooperative) and the proposed MAP (cooperative) decrease with $\eta$, \textcolor{black}{whereas} the error probabilities of the other designs nearly do not change with $\eta$. In addition, the error probabilities of the proposed MAP (non-cooperative) and the proposed MAP (cooperative) at $\eta=1$ reduce to about $1/3$ of the corresponding ones at $\eta=0$, which demonstrates the value of exploiting correlation among device activities.

\textcolor{black}{Fig.~\ref{fig:error_vs_group_size} plots the error probability versus the group size $\frac{N_0}{K}$ \textcolor{black}{in the group activity case given by the second instance in Section~\ref{Sec:prior_dist}}.
From Fig.~\ref{fig:error_vs_group_size}, we can see that the error probabilities of the proposed ML (non-cooperative) and the proposed ML (cooperative) increase with $\frac{N_0}{K}$, as the device activity detection is more challenging when the number of active devices is larger, and the correlation among device activities is not utilized. In contrast, the error probabilities of the proposed MAP (non-cooperative) and the proposed MAP (cooperative) decrease with $\frac{N_0}{K}$, as the exploitation of correlation successfully narrows down the set of \textcolor{black}{possible activity states}.
Note that in the second instance in Section~\ref{Sec:prior_dist}, when the group size $\frac{N_0}{K}$ increases, the variance of the number of active devices in each cell increases, the probability of having a larger number of active devices increases, and the sample space of device activities becomes smaller.}

% \vspace{-2mm}
\section{Conclusion}

\textcolor{black}{This} paper considered non-cooperative and cooperative device activity detection in grant-free \textcolor{black}{massive} access in a \textcolor{black}{multi-cell} network with interfering devices. Under each activity detection mechanism, we formulated the problems for the joint ML estimation and joint MAP estimation of both the device activities and interference powers. \textcolor{black}{Furthermore}, for each challenging non-convex problem, we proposed a coordinate descent algorithm to obtain a stationary point. Both analytical and numerical results demonstrated the importance of explicit consideration of \textcolor{black}{inter-cell interference}, the values of prior information and network parameters, and the advantage of AP cooperation in device activity detection. To our knowledge, this is the first time that techniques from probability, stochastic geometry, and optimization are jointly utilized in device activity detection for \textcolor{black}{grant-free} massive access. Furthermore, this is the first work that considers joint estimation of the device activities and interference powers to improve the accuracy of device activity detection.

\section*{Appendix A: Proof of Theorem~\ref{Thm:Step_one_AP}}

\textcolor{black}{First,} consider the coordinate optimization with respect to $a_i$. \textcolor{black}{By \eqref{eqn:f_ml},} \textcolor{black}{we have}
\begin{align}
&f_{\rm ML}(\mathbf a_0+d\mathbf e_i, \mathbf x)%\notag\\
\eqla \log|\mathbf \Sigma+d\gamma_{i,0}\mathbf p_i\mathbf p_i^H|+{\rm tr}((\mathbf \Sigma+d\gamma_{i,0}\mathbf p_i\mathbf p_i^H)^{-1}\widehat{\mathbf \Sigma}_{\mathbf Y_0}) \notag\\
&\eqlb \log\left(|\mathbf \Sigma|(1+d\gamma_{i,0}\mathbf p_i^H\mathbf \Sigma^{-1}\mathbf p_i)\right)+{\rm tr}\left(\left(\mathbf \Sigma^{-1}-\frac{d\gamma_{i,0}\mathbf \Sigma^{-1}\mathbf p_{i}\mathbf p_{i}^H\mathbf \Sigma^{-1}}{1+d\gamma_{i,0}\mathbf p_{i}^H\mathbf \Sigma^{-1}\mathbf p_{i}}\right)\widehat{\mathbf \Sigma}_{\mathbf Y_0} \right)\notag\\
&\textcolor{black}{\eqlc} f_{\rm ML}(\mathbf a_0, \mathbf x)+\log\left(1+d\gamma_{i,0}\mathbf p_i^H\mathbf \Sigma^{-1}\mathbf p_i\right)-\frac{d\gamma_{i,0}\mathbf p_{i}^H\mathbf \Sigma^{-1}\widehat{\mathbf \Sigma}_{\mathbf Y_0}\mathbf \Sigma^{-1}\mathbf p_{i} }{1+d\gamma_{i,0}\mathbf p_{i}^H\mathbf \Sigma^{-1}\mathbf p_{i}},\label{eqn:ML_expansion_a}
\end{align}
where $(a)$ is due to \textcolor{black}{$\mathbf \Sigma = \mathbf P_0\mathbf A_0\bm \Gamma_{0}\mathbf P_0^H+\mathbf X+\delta^2\mathbf I_L$ and} $\mathbf P_0\mathbf A_0\bm \Gamma_0\mathbf P_0^H = \sum_{i\in\Phi_0}a_i\gamma_{i,0}\mathbf p_i\mathbf p_i^H$, (b) is due to the fact that for any positive definite matrix $\mathbf \Sigma$, $(\mathbf \Sigma+d\gamma_{i,0}\mathbf p_i\mathbf p_i^H)^{-1}=\mathbf \Sigma^{-1}-\frac{d\gamma_{i,0}\mathbf \Sigma^{-1}\mathbf p_{i}\mathbf p_{i}^H\mathbf \Sigma^{-1}}{1+d\gamma_{i,0}\mathbf p_{i}^H\mathbf \Sigma^{-1}\mathbf p_{i}}$ and $|\mathbf \Sigma+d\gamma_{i,0}\mathbf p_i\mathbf p_i^H|=|\mathbf \Sigma|(1+d\gamma_{i,0}\mathbf p_i^H\mathbf \Sigma^{-1}\mathbf p_i)$ hold \cite{Caire18ISIT}, and (c) is due to the cyclic property of trace. %Taking the derivative of $f_{\rm ML}(\mathbf a_0+d\mathbf e_i, \mathbf x)$ with respect to $d$,
\textcolor{black}{Note that $f_{\rm ML}(\mathbf a_0+d\mathbf e_i, \mathbf x)$ is well-defined only when $d>d_0\triangleq -\frac{1}{\gamma_{i,0}\mathbf p_{i}^H\mathbf \Sigma^{-1}\mathbf p_{i}}$}.
By \eqref{eqn:ML_expansion_a}, we have
\begin{align}
\frac{\partial f_{\rm ML}(\mathbf a_0+d\mathbf e_i, \mathbf x)}{\partial d}
=\frac{\gamma_{i,0}\mathbf p_i^H\mathbf \Sigma^{-1}\mathbf p_i}{1+d\gamma_{i,0}\mathbf p_i^H\mathbf \Sigma^{-1}\mathbf p_i}-\frac{\gamma_{i,0}\mathbf p_{i}^H\mathbf \Sigma^{-1}\widehat{\mathbf \Sigma}_{\mathbf Y_0}\mathbf \Sigma^{-1}\mathbf p_{i} }{(1+d\gamma_{i,0}\mathbf p_{i}^H\mathbf \Sigma^{-1}\mathbf p_{i})^2}.\label{eqn:partial_a}
\end{align}
Thus, the solution of $\frac{\partial f_{\rm ML}(\mathbf a+d\mathbf e_i, \mathbf x)}{\partial d}=0$ is $d_a^*\triangleq\frac{\mathbf p_{i}^H\mathbf \Sigma^{-1}\widehat{\mathbf \Sigma}_{\mathbf Y_0}\mathbf \Sigma^{-1}\mathbf p_{i}-\mathbf p_{i}^H\mathbf \Sigma^{-1}\mathbf p_{i}}{\gamma_{i,0}(\mathbf p_{i}^H\mathbf \Sigma^{-1}\mathbf p_{i})^2}$.
% \begin{align}
% d_a^*=\frac{\mathbf p_{i}^H\mathbf \Sigma^{-1}\widehat{\mathbf \Sigma}_{\mathbf Y_0}\mathbf \Sigma^{-1}\mathbf p_{i}-\mathbf p_{i}^H\mathbf \Sigma^{-1}\mathbf p_{i}}{\gamma_{i,0}(\mathbf p_{i}^H\mathbf \Sigma^{-1}\mathbf p_{i})^2}.\notag
% \end{align}
As $\widehat{\mathbf \Sigma}_{\mathbf Y_0} = \frac{1}{M}\mathbf Y_0\mathbf Y_0^H$ and $\mathbf \Sigma$ is a Hermitian matrix, we have $\mathbf p_{i}^H\mathbf \Sigma^{-1}\widehat{\mathbf \Sigma}_{\mathbf Y_0}\mathbf \Sigma^{-1}\mathbf p_{i}\geq 0$ and %, we have
$d_a^*\geq -\frac{1}{\gamma_{i,0}\mathbf p_{i}^H\mathbf \Sigma^{-1}\mathbf p_{i}}$.
\textcolor{black}{In addition, we note that $\lim_{\epsilon\to 0^+}f_{\rm ML}(\mathbf a_0+(d_0+\epsilon)\mathbf e_i, \mathbf x)=+\infty$ and $\lim_{d\to +\infty}f_{\rm ML}(\mathbf a_0+d\mathbf e_i, \mathbf x)=+\infty$.}
% Recall that we require $a_i\in [0,1]$.
Therefore, \textcolor{black}{combining with $a_i\in [0,1]$}, we can obtain the optimal solution of \eqref{eqn:ML_a} as in \eqref{eqn:d_ML_a}.

Next, consider the coordinate optimization with respect to $x_\ell$. Similarly, by~\eqref{eqn:f_ml}, we have
\begin{align}
&f_{\rm ML}(\mathbf a_0, \mathbf x+d\mathbf e_\ell)%\notag\\
= \log|\mathbf \Sigma+d\mathbf e_\ell\mathbf e_\ell^H|+{\rm tr}((\mathbf \Sigma+d\mathbf e_\ell\mathbf e_\ell^H)^{-1}\widehat{\mathbf \Sigma}_{\mathbf Y_0}) \notag\\
&= \log\left(|\mathbf \Sigma|(1+d\mathbf e_\ell^H\mathbf \Sigma^{-1}\mathbf e_\ell)\right)+{\rm tr}\left(\left(\mathbf \Sigma^{-1}-\frac{d\mathbf \Sigma^{-1}\mathbf e_\ell\mathbf e_\ell^H\mathbf \Sigma^{-1}}{1+d\mathbf e_\ell^H\mathbf \Sigma^{-1}\mathbf e_\ell}\right)\widehat{\mathbf \Sigma}_{\mathbf Y_0} \right)\notag\\
&= f_{\rm ML}(\mathbf a_0, \mathbf x)+\log\left(1+d\mathbf e_\ell^H\mathbf \Sigma^{-1}\mathbf e_\ell\right)-\frac{d\mathbf e_\ell^H\mathbf \Sigma^{-1}\widehat{\mathbf \Sigma}_{\mathbf Y_0}\mathbf \Sigma^{-1}\mathbf e_\ell }{1+d\mathbf e_\ell^H\mathbf \Sigma^{-1}\mathbf e_\ell},\label{eqn:ML_expansion_x}
\end{align}
% where $(a)$ is due to $\mathbf P\mathbf A\bm \Gamma_0\mathbf P^H = \sum_{i\in\mathcal N}a_i\gamma_{i,0}\mathbf p_i\mathbf p_i^H$, (b) is due to the fact that for any positive definite matrix $\mathbf \Sigma$, $(\mathbf \Sigma+d\gamma_{i,0}\mathbf p_i\mathbf p_i^H)^{-1}=\mathbf \Sigma^{-1}-\frac{d\gamma_{i,0}\mathbf \Sigma^{-1}\mathbf p_{i}\mathbf p_{i}^H\mathbf \Sigma^{-1}}{1+d\gamma_{i,0}\mathbf p_{i}^H\mathbf \Sigma^{-1}\mathbf p_{i}}$ and $|\mathbf \Sigma+d\gamma_{i,0}\mathbf p_i\mathbf p_i^H|=|\mathbf \Sigma|(1+d\gamma_{i,0}\mathbf p_i^H\mathbf \Sigma^{-1}\mathbf p_i)$ hold.
% Taking the derivative of $f_{\rm ML}(\mathbf a_0, \mathbf x+d\mathbf e_\ell)$ with respect to $d$,
\textcolor{black}{By~\eqref{eqn:ML_expansion_x},} we have
\begin{align}
\frac{\partial f_{\rm ML}(\mathbf a_0, \mathbf x+d\mathbf e_\ell)}{\partial d}
=\frac{\mathbf e_\ell^H\mathbf \Sigma^{-1}\mathbf e_\ell}{1+d\mathbf e_\ell^H\mathbf \Sigma^{-1}\mathbf e_\ell}-\frac{\mathbf e_\ell^H\mathbf \Sigma^{-1}\widehat{\mathbf \Sigma}_{\mathbf Y_0}\mathbf \Sigma^{-1}\mathbf e_\ell }{(1+d\mathbf e_\ell^H\mathbf \Sigma^{-1}\mathbf e_\ell)^2}.\notag
\end{align}
\textcolor{black}{Thus,} the solution of $\frac{\partial f_{\rm ML}(\mathbf a_0, \mathbf x+d\mathbf e_\ell)}{\partial d}=0$ is $d_x^*\triangleq\frac{\mathbf e_\ell^H\mathbf \Sigma^{-1}\widehat{\mathbf \Sigma}_{\mathbf Y_0}\mathbf \Sigma^{-1}\mathbf e_\ell-\mathbf e_\ell^H\mathbf \Sigma^{-1}\mathbf e_\ell}{(\mathbf e_\ell^H\mathbf \Sigma^{-1}\mathbf e_\ell)^2}$.
% \begin{align}
% d_x^*=\frac{\mathbf e_\ell^H\mathbf \Sigma^{-1}\widehat{\mathbf \Sigma}_{\mathbf Y_0}\mathbf \Sigma^{-1}\mathbf e_\ell-\mathbf e_\ell^H\mathbf \Sigma^{-1}\mathbf e_\ell}{(\mathbf e_\ell^H\mathbf \Sigma^{-1}\mathbf e_\ell)^2}.\notag
% \end{align}
% As $\mathbf p_{i}^H\mathbf \Sigma_0^{-1}\widehat{\mathbf \Sigma}_{\mathbf Y_0}\mathbf \Sigma_0^{-1}\mathbf p_{i}\geq 0$, we have $d_a^*\geq -\frac{1}{\gamma_{i,0}\mathbf p_{i}^H\mathbf \Sigma_0^{-1}\mathbf p_{i}}\triangleq d_0$.
% Recall that we require $x_\ell\geq 0$. Therefore,
\textcolor{black}{Similarly, combining with $x_\ell\geq 0$,} we can obtain the optimal solution of \eqref{eqn:ML_x} as in \eqref{eqn:d_ML_x}.

% \vspace{-2mm}
\section*{Appendix B: Proof of Lemma~\ref{Lem:distribution_X}}

\textcolor{black}{To show Lemma~\ref{Lem:distribution_X}, it is sufficient to calculate the mean and variance of $\sum_{i\in\mathcal I\setminus\Phi_0} a_i \gamma_{i,0}$.}
% According to independent thin of a PPP,
% Let $\Phi_{t,j}$ denote the set of indices of the active devices in cell $j$.
Let $\mathcal I_t$ denote %\triangleq \cup_{j\in\mathcal J\setminus\{0\}}\Phi_{t,j}
the set of indices of the active interfering devices out of the typical cell. Recall that the locations of active interfering devices follow a homogeneous PPP with density $\lambda$.  According to Campbell's theorem for sums,  we \textcolor{black}{have}
\begin{align}
\textcolor{black}{\mu=}\mathbb E\left[\sum_{i\in\mathcal I\setminus\Phi_0} a_i \gamma_{i,0}\right]&=\mathbb E\left[\sum_{i\in\mathcal I_t} d_{i,0}^{-\alpha}\right]=\lambda  \int_{\mathbb R^2\setminus S_0}d(t)^{-\alpha}{\rm d}t%\notag\\
=12\lambda \int_{\frac{\sqrt{3}}{2}R}^{\infty}
\int_{0}^{\frac{\sqrt{3}}{3}x} (x^2+y^2)^{-\frac{\alpha}{2}}{\rm d }y{\rm d}x,\notag
\end{align}
where $S_0$ denotes the area of the typical cell and $d(t)$ is the distance between point $t$ and the origin. According to the variance result for PPPs in\textcolor{black}{~\cite[Page 85]{haenggi2012stochastic}}, we have
\begin{align}
\textcolor{black}{\sigma^2=}{\rm var}\left[\sum_{i\in\mathcal I\setminus\Phi_0} a_i \gamma_{i,0}\right] &= {\rm var}\left[\sum_{i\in\mathcal I_t} d_{i,0}^{-\alpha}\right]=\lambda \int_{\mathbb R^2\setminus S_0}d(t)^{-2\alpha}{\rm d}t%\notag\\
=12\lambda \int_{\frac{\sqrt{3}}{2}R}^{\infty}
\int_{0}^{\frac{\sqrt{3}}{3}x} (x^2+y^2)^{-\alpha}{\rm d }y{\rm d}x.\notag
\end{align}

\section*{\textcolor{black}{Appendix C: Proof of Theorem~\ref{Thm:Step_APs}}}

\textcolor{black}{First, we consider the coordinate descent optimization with respect to $a_i$ in~\eqref{eqn:MAP_a}. As $f_{\rm MAP}(\mathbf a_0 + d\mathbf e_i, \mathbf x) = f_{\rm ML}(\mathbf a_0+d\mathbf e_i, \mathbf x) - C_id$, we have $\frac{\partial f_{\rm MAP}(\mathbf a_0+d\mathbf e_i, \mathbf x)}{\partial d} = \frac{\partial f_{\rm ML}(\mathbf a_0+d\mathbf e_i, \mathbf x)}{\partial d} - C_i$. By~\eqref{eqn:partial_a}, we have
\begin{align}
\frac{\partial^2 f_{\rm MAP}(\mathbf a_0+d\mathbf e_i, \mathbf x)}{\partial d^2}&=\frac{\partial^2 f_{\rm ML}(\mathbf a_0+d\mathbf e_i, \mathbf x)}{\partial d^2}\notag\\
&=\frac{\gamma_{i,0}^2\mathbf p_{i}^H\mathbf \Sigma^{-1}\mathbf p_{i} \left(2\mathbf p_{i}^H\mathbf \Sigma^{-1}\widehat{\mathbf \Sigma}_{\mathbf Y_0}\mathbf \Sigma^{-1}\mathbf p_{i} - \mathbf p_{i}^H\mathbf \Sigma^{-1}\mathbf p_{i}-d\gamma_{i,0}(\mathbf p_{i}^H\mathbf \Sigma^{-1}\mathbf p_{i})^2\right)}{(1+d\gamma_{i,0}\mathbf p_{i}^H\mathbf \Sigma^{-1}\mathbf p_{i})^3}.\notag
\end{align}
The solution of $\frac{\partial^2 f_{\rm MAP}(\mathbf a_0+d\mathbf e_i, \mathbf x)}{\partial d^2}=0$ is $d_2\triangleq\frac{2\mathbf p_{i}^H\mathbf \Sigma^{-1}\widehat{\mathbf \Sigma}_{\mathbf Y_0}\mathbf \Sigma^{-1}\mathbf p_{i}-\mathbf p_{i}^H\mathbf \Sigma^{-1}\mathbf p_{i}}{\gamma_{i,0}(\mathbf p_{i}^H\mathbf \Sigma^{-1}\mathbf p_{i})^2}$. As $\frac{\partial^2 f_{\rm MAP}(\mathbf a_0+d\mathbf e_i, \mathbf x)}{\partial d^2}>0$ in $d\in(d_0,d_2)$ and  $\frac{\partial^2 f_{\rm MAP}(\mathbf a_0+d\mathbf e_i, \mathbf x)}{\partial d^2}<0$ in $d\in(d_2,\infty)$, we know that  $\frac{\partial f_{\rm MAP}(\mathbf a_0+d\mathbf e_i, \mathbf x)}{\partial d}$ increases with $d$ in  $(d_0,d_2)$, decreases with $d$ in $(d_2,\infty)$, and achieves its maximum $\frac{\gamma_{i,0}(\mathbf p_{i}^H\mathbf \Sigma^{-1}\mathbf p_{i})^2}{4\mathbf p_{i}^H\mathbf \Sigma^{-1}\widehat{\mathbf \Sigma}_{\mathbf Y_0}\mathbf \Sigma^{-1}\mathbf p_{i}}-C_i$ at $d=d_2$. As $\lim_{\epsilon\to 0^+}\frac{\partial f_{\rm MAP}(\mathbf a_0+(d_0+\epsilon)\mathbf e_i, \mathbf x)}{\partial d}=-\infty$ and $\lim_{d\to \infty}\frac{\partial f_{\rm MAP}(\mathbf a_0+d\mathbf e_i, \mathbf x)}{\partial d}=-C_i$, the range of $\frac{\partial f_{\rm MAP}(\mathbf a_0+d\mathbf e_i, \mathbf x)}{\partial d}$ is $\left(-\infty, \frac{\gamma_{i,0}(\mathbf p_{i}^H\mathbf \Sigma^{-1}\mathbf p_{i})^2}{4\mathbf p_{i}^H\mathbf \Sigma^{-1}\widehat{\mathbf \Sigma}_{\mathbf Y_0}\mathbf \Sigma^{-1}\mathbf p_{i}}-C_i\right]$.}

\textcolor{black}{Now, we consider the following three cases.
When $C_i\leq 0$, equation $\frac{\partial f_{\rm MAP}(\mathbf a_0+d\mathbf e_i, \mathbf x)}{\partial d}=0$ has one solution $s_i(\mathbf a_0,\mathbf x) \triangleq \frac{1}{2C_i}\left(1-\sqrt{1-\frac{4C_i\mathbf p_i^H\mathbf \Sigma^{-1}\widehat{\mathbf \Sigma}_{\mathbf Y_0}\mathbf \Sigma^{-1}\mathbf p_i}{\gamma_{i}(\mathbf p_i^H\mathbf \Sigma^{-1}\mathbf p_i)^2}}\right)-\frac{1}{\gamma_{i}\mathbf p_i^H\mathbf \Sigma^{-1}\mathbf p_i}$. We know that $f_{\rm MAP}(\mathbf a_0+d\mathbf e_i, \mathbf x)$ decreases with $d$ in $(d_0,s_i(\mathbf a_0,\mathbf x))$, increases with $d$ in $(s_i(\mathbf a_0,\mathbf x),+\infty)$ and achieves its maximum at $d=s_i(\mathbf a_0,\mathbf x)$. Combining with the constraint $a_i\in[0,1]$, we have the optimal solution given in \eqref{eqn:d_MAP_a}.
When $C_i \geq \frac{\gamma_{i,0}(\mathbf p_{i}^H\mathbf \Sigma^{-1}\mathbf p_{i})^2}{4\mathbf p_{i}^H\mathbf \Sigma^{-1}\widehat{\mathbf \Sigma}_{\mathbf Y_0}\mathbf \Sigma^{-1}\mathbf p_{i}}$, $f_{\rm MAP}(\mathbf a_0+d\mathbf e_i, \mathbf x)$ decreases with $d$ in $(d_0,+\infty)$. Combining with the constraint $a_i\in[0,1]$, we have the optimal solution given in \eqref{eqn:d_MAP_a}. When $0<C_i < \frac{\gamma_{i,0}(\mathbf p_{i}^H\mathbf \Sigma^{-1}\mathbf p_{i})^2}{4\mathbf p_{i}^H\mathbf \Sigma^{-1}\widehat{\mathbf \Sigma}_{\mathbf Y_0}\mathbf \Sigma^{-1}\mathbf p_{i}}$, equation $\frac{\partial f_{\rm MAP}(\mathbf a_0+d\mathbf e_i, \mathbf x)}{\partial d}=0$ has two solutions $s_i(\mathbf a_0,\mathbf x)$ and $\overline{s}_i(\mathbf a_0,\mathbf x)$, where
\begin{align}
\overline{s}_i(\mathbf a_0,\mathbf x) \triangleq \frac{1}{2C_i}\left(1+ \sqrt{1-\frac{4C_i\mathbf p_i^H\mathbf \Sigma^{-1}\widehat{\mathbf \Sigma}_{\mathbf Y_0}\mathbf \Sigma^{-1}\mathbf p_i}{\gamma_{i,0}(\mathbf p_i^H\mathbf \Sigma^{-1}\mathbf p_i)^2}}\right)-\frac{1}{\gamma_{i,0}\mathbf p_i^H\mathbf \Sigma^{-1}\mathbf p_i}.\notag
\end{align}
We know that $f_{\rm MAP}(\mathbf a_0+d\mathbf e_i, \mathbf x)$ decreases with $d$ in $(d_0,s_i(\mathbf a_0,\mathbf x))$ and $(\overline{s}_i(\mathbf a_0,\mathbf x),+\infty)$, and increases with $d$ in $(s_i(\mathbf a_0,\mathbf x),\overline{s}_i(\mathbf a_0,\mathbf x))$. Combining with the constraint $a_i\in[0,1]$, we have the optimal solution given in \eqref{eqn:d_MAP_a}.}

\textcolor{black}{Next, we consider the coordinate descent optimization with respect to $x_\ell$ in~\eqref{eqn:MAP_x}. Note that $h_{x,\ell}(d,\mathbf a_0,\mathbf x)$ is the derivative function of $f_{x,\ell}(d,\mathbf a_0,\mathbf x)$ with respect to $d$. Combining with the constraint $x_\ell>0$, we know that $f_{x,\ell}(d,\mathbf a_0,\mathbf x)$ achieves its maximum at one point in $\mathcal X_{\ell}(\mathbf a_0,\mathbf x)\cup\{-x_\ell\}$, where $\mathcal X_{\ell}(\mathbf a_0,\mathbf x)\triangleq \{d\geq-x_\ell:h_{x,\ell}(d,\mathbf a_0,\mathbf x)=0\}$, which has the optimal objective value. Therefore, we complete the proof.}

% \vspace{-2mm}
\section*{Appendix D: Proof of Corollary~\ref{Cor:opt_a_iid_no_coop}}

\textcolor{black}{First, we derive $f_{\rm MAP}(\mathbf a_0, \mathbf x)$ in the i.i.d. case. Note that in the i.i.d. case,} %components of $\mathbf a_0$ are i.i.d. Bernoulli random variables and
\begin{align}
p(\mathbf a_0)\textcolor{black}{=\prod_{i\in\Phi_0}p_a^{a_i}(1-p_a)^{1-a_i}}=\exp\left(\log \frac{p_a}{1-p_a}\sum_{i\in\Phi_0}a_i+N_0\log(1-p_a)\right).\notag
\end{align}
% Combining with the results on the likelihood function of $\mathbf Y_0$ and the distribution of $\mathbf x$, the joint posterior distribution of $\mathbf a$ and $\mathbf x$, given the observation $\mathbf Y_0$,
\textcolor{black}{Based on~\eqref{eqn:f_map},} $f_{\rm MAP}(\mathbf a_0, \mathbf x)$ in the i.i.d. case is given by
\begin{align}
f_{\rm MAP}(\mathbf a_0,\mathbf x)= f_{\rm ML}(\mathbf a_0, \mathbf x)
+ \frac{1}{2M\sigma^2}\sum_{\ell\in\mathcal L}(x_{\ell}-\mu)^2-\frac{1}{M}\log \frac{p_a}{1-p_a}\sum_{i\in\overline{\Phi}_0}a_i.\label{eqn:f_map_iid}
\end{align}
\textcolor{black}{By \eqref{eqn:f_map_iid}, we know that $C_i=\frac{1}{M}\log \frac{p_a}{1-p_a}<0$ (as $p_a\ll 1$). Substituting $C_i=\frac{1}{M}\log \frac{p_a}{1-p_a}$ into \eqref{eqn:d_MAP_a}, we can obtain the optimal solution of \eqref{eqn:MAP_a} as in \eqref{eqn:d_MAP_a_iid} in the i.i.d. case.}


\begin{thebibliography}{10}
\providecommand{\url}[1]{#1}
\csname url@samestyle\endcsname
\providecommand{\newblock}{\relax}
\providecommand{\bibinfo}[2]{#2}
\providecommand{\BIBentrySTDinterwordspacing}{\spaceskip=0pt\relax}
\providecommand{\BIBentryALTinterwordstretchfactor}{4}
\providecommand{\BIBentryALTinterwordspacing}{\spaceskip=\fontdimen2\font plus
\BIBentryALTinterwordstretchfactor\fontdimen3\font minus
  \fontdimen4\font\relax}
\providecommand{\BIBforeignlanguage}[2]{{%
\expandafter\ifx\csname l@#1\endcsname\relax
\typeout{** WARNING: IEEEtran.bst: No hyphenation pattern has been}%
\typeout{** loaded for the language `#1'. Using the pattern for}%
\typeout{** the default language instead.}%
\else
\language=\csname l@#1\endcsname
\fi
#2}}
\providecommand{\BIBdecl}{\relax}
\BIBdecl

\bibitem{Jiang20WCNC}
D.~{Jiang} and Y.~{Cui}, ``Ml estimation and map estimation for device activity
  in grant-free massive access with interference,'' in \emph{Proc. IEEE WCNC},
  Apr. 2020, pp. 1--6.

\bibitem{Jiang20SPAWC}
------, ``Map-based pilot state detection in grant-free random access for
  mmtc,'' in \emph{2020 IEEE 21st International Workshop on Signal Processing
  Advances in Wireless Communications (SPAWC)}, 2020, pp. 1--5.

\bibitem{Popovski17Mag}
E.~d.~{Carvalho}, E.~{Bjornson}, J.~H. {Sorensen}, P.~{Popovski}, and E.~G.
  {Larsson}, ``Random access protocols for massive mimo,'' \emph{IEEE Commun.
  Mag.}, vol.~55, no.~5, pp. 216--222, May 2017.

\bibitem{Liu18Mag}
L.~{Liu}, E.~G. {Larsson}, W.~{Yu}, P.~{Popovski}, C.~{Stefanovic}, and E.~{de
  Carvalho}, ``Sparse signal processing for grant-free massive connectivity: A
  future paradigm for random access protocols in the internet of things,''
  \emph{IEEE Signal Process. Mag.}, vol.~35, no.~5, pp. 88--99, Sep. 2018.

\bibitem{Bockelmann16Mag}
C.~{Bockelmann}, N.~{Pratas}, H.~{Nikopour}, K.~{Au}, T.~{Svensson},
  C.~{Stefanovic}, P.~{Popovski}, and A.~{Dekorsy}, ``Massive machine-type
  communications in 5g: physical and mac-layer solutions,'' \emph{IEEE Commun.
  Mag.}, vol.~54, no.~9, pp. 59--65, Sep. 2016.

\bibitem{Chen18TCOM}
X.~{Chen}, Z.~{Zhang}, C.~{Zhong}, R.~{Jia}, and D.~W.~K. {Ng}, ``Fully
  non-orthogonal communication for massive access,'' \emph{IEEE Trans.
  Commun.}, vol.~66, no.~4, pp. 1717--1731, Apr. 2018.

\bibitem{Senel18TCOM}
K.~{Senel} and E.~G. {Larsson}, ``Grant-free massive mtc-enabled massive mimo:
  A compressive sensing approach,'' \emph{IEEE Trans. Commun.}, vol.~66,
  no.~12, pp. 6164--6175, Dec 2018.

\bibitem{Yu19ICC}
Z.~{Chen}, F.~{Sohrabi}, Y.~{Liu}, and W.~{Yu}, ``Covariance based joint
  activity and data detection for massive random access with massive mimo,'' in
  \emph{Proc. IEEE ICC}, May 2019, pp. 1--6.

\bibitem{Liu18TSP}
L.~{Liu} and W.~{Yu}, ``Massive connectivity with massive mimo—part i: Device
  activity detection and channel estimation,'' \emph{IEEE Trans. Signal
  Process.}, vol.~66, no.~11, pp. 2933--2946, Jun. 2018.

\bibitem{Cui20JSAC}
Y.~{Cui}, S.~{Li}, and W.~{Zhang}, ``Jointly sparse signal recovery and support
  recovery via deep learning with applications in mimo-based grant-free random
  access,'' \emph{IEEE Journal on Selected Areas in Communications}, pp. 1--1,
  2020.

\bibitem{Xu15ICC}
X.~{Xu}, X.~{Rao}, and V.~K.~N. {Lau}, ``Active user detection and channel
  estimation in uplink cran systems,'' in \emph{Proc. IEEE ICC}, Jun. 2015, pp.
  2727--2732.

\bibitem{Bockelmann13}
C.~Bockelmann, H.~F. Schepker, and A.~Dekorsy, ``Compressive sensing based
  multi-user detection for machine-to-machine communication,''
  \emph{Transactions on Emerging Telecommunications Technologies}, vol.~24,
  no.~4, pp. 389--400, 2013.

\bibitem{Zhang18TVT}
Y.~{Zhang}, Q.~{Guo}, Z.~{Wang}, J.~{Xi}, and N.~{Wu}, ``Block sparse bayesian
  learning based joint user activity detection and channel estimation for
  grant-free noma systems,'' \emph{IEEE Trans. Veh. Technol.}, vol.~67, no.~10,
  pp. 9631--9640, Oct. 2018.

\bibitem{Chen18TSP}
Z.~{Chen}, F.~{Sohrabi}, and W.~{Yu}, ``Sparse activity detection for massive
  connectivity,'' \emph{IEEE Trans. Signal Process.}, vol.~66, no.~7, pp.
  1890--1904, April 2018.

\bibitem{Shao19IoTJ}
X.~{Shao}, X.~{Chen}, C.~{Zhong}, J.~{Zhao}, and Z.~{Zhang}, ``A unified design
  of massive access for cellular internet of things,'' \emph{IEEE Internet of
  Things Journal}, vol.~6, no.~2, pp. 3934--3947, Apr. 2019.

\bibitem{Chen19TWC}
Z.~{Chen}, F.~{Sohrabi}, and W.~{Yu}, ``Multi-cell sparse activity detection
  for massive random access: Massive mimo versus cooperative mimo,'' \emph{IEEE
  Trans. Wireless Commun.}, vol.~18, no.~8, pp. 4060--4074, Aug. 2019.

\bibitem{Caire18ISIT}
S.~{Haghighatshoar}, P.~{Jung}, and G.~{Caire}, ``Improved scaling law for
  activity detection in massive mimo systems,'' in \emph{Proc. IEEE ISIT}, Jun.
  2018, pp. 381--385.

\bibitem{Alexander19}
\BIBentryALTinterwordspacing
A.~Fengler, G.~Caire, P.~Jung, and S.~Haghighatshoar, ``Massive {MIMO}
  unsourced random access,'' \emph{CoRR}, vol. abs/1901.00828, 2019. [Online].
  Available: \url{http://arxiv.org/abs/1901.00828}
\BIBentrySTDinterwordspacing

\bibitem{Chen19Asilomar}
Z.~{Chen} and W.~{Yu}, ``Phase transition analysis for covariance based massive
  random access with massive mimo,'' in \emph{Proc. ASILOMAR}, Nov. 2019, pp.
  1--5.

\bibitem{Liu19GLOBECOM}
B.~Liu, Z.~Wei, J.~Yuan, and M.~Pajovic, ``Deep learning assisted user
  identification in massive machine-type communications,'' in \emph{Proc. IEEE
  GLOBECOM}, Dec. 2019, pp. 1--6.

\bibitem{Zhang19TVT}
Z.~{Zhang}, Y.~{Li}, C.~{Huang}, Q.~{Guo}, C.~{Yuen}, and Y.~L. {Guan},
  ``Dnn-aided block sparse bayesian learning for user activity detection and
  channel estimation in grant-free non-orthogonal random access,'' \emph{IEEE
  Trans. Veh. Technol.}, vol.~68, no.~12, pp. 12\,000--12\,012, Dec. 2019.

\bibitem{Ye19TII}
N.~{Ye}, X.~{Li}, H.~{Yu}, A.~{Wang}, W.~{Liu}, and X.~{Hou}, ``Deep learning
  aided grant-free noma toward reliable low-latency access in tactile internet
  of things,'' \emph{IEEE Trans. Ind. Inf.}, vol.~15, no.~5, pp. 2995--3005,
  May 2019.

\bibitem{NIPS2011_4209}
S.~Ding, G.~Wahba, and J.~Zhu, ``Learning higher-order graph structure with
  features by structure penalty,'' in \emph{Advances in Neural Information
  Processing Systems 24}.\hskip 1em plus 0.5em minus 0.4em\relax Curran
  Associates, Inc., 2011, pp. 253--261.

\bibitem{FTNhaenggi09}
M.~Haenggi and R.~K. Ganti, ``Interference in large wireless networks,''
  \emph{Foundations and Trends in Networking}, vol.~3, no.~2, pp. 127--248,
  2009.

\bibitem{haenggi2012stochastic}
M.~Haenggi, \emph{Stochastic geometry for wireless networks}.\hskip 1em plus
  0.5em minus 0.4em\relax Cambridge University Press, 2012.

\bibitem{Andrews11TCOM}
J.~G. {Andrews}, F.~{Baccelli}, and R.~K. {Ganti}, ``A tractable approach to
  coverage and rate in cellular networks,'' \emph{IEEE Trans. Commun.},
  vol.~59, no.~11, pp. 3122--3134, 2011.

\bibitem{Andrews07WC}
J.~G. {Andrews}, W.~{Choi}, and R.~W. {Heath}, ``Overcoming interference in
  spatial multiplexing mimo cellular networks,'' \emph{IEEE Wireless Commun.},
  vol.~14, no.~6, pp. 95--104, Dec. 2007.

\bibitem{Choi19TCOM}
J.~{Choi}, ``Noma-based compressive random access using gaussian spreading,''
  \emph{IEEE Trans. Commun.}, vol.~67, no.~7, pp. 5167--5177, Jul. 2019.

\bibitem{Bertsekas99}
D.~Bertsekas, \emph{Nonlinear Programming}.\hskip 1em plus 0.5em minus
  0.4em\relax Athena Scientific, 1999.

\bibitem{banerjee2008model}
O.~Banerjee, L.~E. Ghaoui, and A.~d’Aspremont, ``Model selection through
  sparse maximum likelihood estimation for multivariate gaussian or binary
  data,'' \emph{Journal of Machine learning research}, vol.~9, no. Mar, pp.
  485--516, 2008.

\bibitem{Aljuaid10TVT}
M.~{Aljuaid} and H.~{Yanikomeroglu}, ``Investigating the gaussian convergence
  of the distribution of the aggregate interference power in large wireless
  networks,'' \emph{IEEE Trans. Veh Technology}, vol.~59, no.~9, pp.
  4418--4424, Nov. 2010.

\bibitem{Hasan07TWC}
A.~{Hasan} and J.~G. {Andrews}, ``The guard zone in wireless ad hoc networks,''
  \emph{IEEE Trans. Wireless Commun.}, vol.~6, no.~3, pp. 897--906, Mar. 2007.

\end{thebibliography}
\end{document}